\documentclass[longauth]{aa} 
\usepackage{graphicx}
\usepackage{siunitx}
\usepackage{ctable}
\usepackage{textcomp}
\usepackage{multirow}
\usepackage{verbatim}
\usepackage{txfonts}
\usepackage[colorlinks=true,allcolors=blue]{hyperref}
\usepackage{placeins}

\usepackage{xcolor}

\newcommand{\micron}{\textmu m}
\newcommand{\kms}{$\rm km ~s^{-1}$}

\newcommand{\msun}{$\rm M_{\odot}$}
\newcommand{\sfr}{$\rm M_{\odot} ~yr^{-1}$}
\newcommand{\ha}{$\rm H \alpha$}
\newcommand{\hb}{$\rm H \beta$}
\newcommand{\hg}{$\rm H \gamma$}
\newcommand{\hd}{$\rm H \delta$}
\newcommand{\nii}{$\rm [\ion{N}{II}]$}
\newcommand{\jlow}{J$\_$low}
\newcommand{\hlow}{H$\_$low}
\newcommand{\klow}{K$\_$low}
\newcommand{\gaia}{\emph{Gaia}}

\begin{document}

\title{Dual and lensed AGN at sub-arcsecond separations uncovered by ERIS}

\titlerunning{Dual and lensed AGN with ERIS}
\authorrunning{Zanchettin et al.}

\author{
M.~V.~Zanchettin\inst{\ref{inaf-oaa}}\thanks{Corresponding author, \email{maria.zanchettin@inaf.it}}
\and F.~Mannucci\inst{\ref{inaf-oaa}}
\and Q.~D'Amato\inst{\ref{iaps}}
\and M.~Scialpi\inst{\ref{unitn},\ref{iUNIFI},\ref{inaf-oaa}}
\and C.~Marconcini\inst{\ref{inaf-oaa}}
\and G.~Tozzi\inst{\ref{MPE}}
\and E.~Bertola\inst{\ref{inaf-oaa}}
\and A.~Feltre\inst{\ref{inaf-oaa}}
\and C.~Bracci\inst{\ref{iUNIFI},\ref{inaf-oaa}}
\and V.~Braito\inst{\ref{inaf-brera},\ref{unitn}, \ref{CUA}}
\and S.~Carniani\inst{\ref{sns}}
\and E.~Cataldi\inst{\ref{iUNIFI},\ref{inaf-oaa}}
\and M.~Ceci\inst{\ref{iUNIFI},\ref{inaf-oaa}}
\and A.~Chakraborty\inst{\ref{IIA}}
\and C.~Cicone\inst{\ref{oslo}}
\and G.~Cresci\inst{\ref{inaf-oaa}}
\and A.~De~Rosa\inst{\ref{iaps}}
\and M.~Ginolfi\inst{\ref{iUNIFI},\ref{inaf-oaa}}
\and I.~Lamperti\inst{\ref{CAB}}
\and A.~Marconi\inst{\ref{iUNIFI},\ref{inaf-oaa}}
\and B.~Moreschini\inst{\ref{iUNIFI},\ref{inaf-oaa}}
\and E.~Nardini\inst{\ref{inaf-oaa}}
\and M.~Perna\inst{\ref{CAB}}
\and E.~Portaluri\inst{\ref{oaab}}
\and K.~Rubinur\inst{\ref{oslo}}
\and P.~Saracco\inst{\ref{inaf-brera}}
\and P.~Severgnini\inst{\ref{inaf-brera}}
\and L.~Ulivi\inst{\ref{CAB}}
\and G.~Venturi\inst{\ref{inaf-oaa}}
\and C.~Vignali\inst{\ref{unibo},\ref{inaf-oas}}
\and G.~Agapito\inst{\ref{inaf-oaa}}
\and A.~Agudo~Berbel\inst{\ref{MPE}}
\and A.~Baruffolo\inst{\ref{padova}}
\and Y.~Cao\inst{\ref{MPE}}
\and R.~Davies\inst{\ref{MPE}}
\and M.~Deysenroth\inst{\ref{MPE}}
\and I.~Di~Antonio\inst{\ref{oaab}}
\and A.~Di Cianno\inst{\ref{oaab}}
\and G.~Di~Rico\inst{\ref{oaab}}
\and M.~Dolci\inst{\ref{oaab}}
\and F.~Eisenhauer\inst{\ref{MPE}}
\and S.~Esposito\inst{\ref{inaf-oaa}}
\and D.~Fantinel\inst{\ref{padova}}
\and D.~Ferruzzi\inst{\ref{inaf-oaa}}
\and H.~Feuchtgruber\inst{\ref{MPE}}
\and N.~M.~Förster Schreiber\inst{\ref{MPE}}
\and R.~Genzel\inst{\ref{MPE}}
\and S.~Gillessen\inst{\ref{MPE}}
\and M.~Hartl\inst{\ref{MPE}}
\and H.~Huber\inst{\ref{MPE}}
\and C.~Keller\inst{\ref{nso}}
\and K.~Kravchenko\inst{\ref{MPE}}
\and D.~Lutz\inst{\ref{MPE}}
\and M.~Macintosh\inst{\ref{atc}}
\and T.~Ott\inst{\ref{MPE}}
\and A.~Puglisi\inst{\ref{inaf-oaa}}
\and S.~Rabien\inst{\ref{MPE}}
\and C.~Rau\inst{\ref{MPE}}
\and A.~Riccardi\inst{\ref{inaf-oaa}}
\and B.~Salasnich\inst{\ref{padova}}
\and T.~Shimizu\inst{\ref{MPE}}
\and E.~Sturm\inst{\ref{MPE}}
\and L.~Tacconi\inst{\ref{MPE}}
\and A.~Valentini\inst{\ref{oaab}}
\and M.~Xompero\inst{\ref{inaf-oaa}}
}

\institute{ INAF -- Osservatorio Astrofisico di Arcetri, Largo E. Fermi 5, 50125 Firenze, Italy
\label{inaf-oaa}
\and INAF -- Istituto di Astrofisica e Planetologia Spaziali, Via Fosso del Cavaliere 100, 00133 Rome, Italy
\label{iaps}
\and Dipartimento di Fisica, Universit`a di Trento, Via Sommarive 14, Trento 38123, Italy \label{unitn}
\and Università di Firenze, Dipartimento di Fisica e Astronomia, Via G. Sansone 1, 50019 Sesto Fiorentino, Firenze, Italy
\label{iUNIFI}
\and Max Planck Institute for Extraterrestrial Physics, Giessenbachstrasse 1, 85748 Garching, Germany
\label{MPE}
\and Dipartimento di Fisica e Astronomia ‘Augusto Righi’, Universit\`a degli Studi di Bologna, via Gobetti 93/2, 40129 Bologna, Italy
\label{unibo}
\and INAF –- Osservatorio di Astrofisica e Scienza dello Spazio di Bologna, Via Gobetti 93/3, 40129 Bologna, Italy
\label{inaf-oas}
\and Indian Institute of Astrophysics, 2nd Block, Koramangala, Bangalore 560 034, India
\label{IIA}
\and Centro de Astrobiología (CAB),  CSIC-INTA,  Ctra. de Ajalvir Km. 4, 28850 Torrejón de Ardoz, Madrid, Spain
\label{CAB}
\and INAF – Osservatorio Astronomico d'Abruzzo, Via Mentore Maggini snc, 64100 Teramo, Italy
\label{oaab}
\and INAF -- Osservatorio Astronomico di Brera,
Milano: Via Brera 28, I-20121 Milano, Italy;
Merate: Via Bianchi 46, I-23807 Merate (LC), Italy 
\label{inaf-brera}
\and Institute for Astrophysics and Computational Sciences, Department of Physics, The Catholic University of America, Washington, DC 20064, USA \label{CUA}
\and Scuola Normale Superiore, Piazza dei Cavalieri 7, I-56126 Pisa, Italy \label{sns}
\and Institute of Theoretical Astrophysics, University of Oslo, PO Box 1029, Blindern, Oslo, 0315, Norway \label{oslo}
\and INAF -- Osservatorio Astronomico di Padova, Vicolo dell'Osservatorio 5, 35122 Padova, Italy
\label{padova}
\and STFC UK Astronomy Technology Centre, Royal Observatory Edinburgh, Blackford Hill, Edinburgh EH9 3HJ, UK
\label{atc}
\and National Solar Observatory, 3665 Discovery Drive, Boulder, CO 80303, USA
\label{nso}
}

\date{Accepted October 18, 2026}

\abstract{We present the first results of the ongoing INAF GTO program focused on the search of sub--arcsecond dual AGN, carried out with the near-infrared integral-field unit SPIFFIER of the Enhanced Resolution Imager and Spectrograph (ERIS) at the ESO Very Large Telescope, operating in adaptive-optics mode with a laser guide star. We report integral field spectroscopy with the low--resolution gratings ($R \sim 5000 - 5600$) of 34 multiple AGN candidates selected from the \gaia\ Data Release~3 source catalog via the \gaia\ Multi-Peak (GMP) method, which preferentially identifies luminous unobscured AGN at sub-arcsecond separations.
The targets span the redshift range $0.7 \lesssim z \lesssim 3.4$ and exhibit projected angular separations between ${\sim}0.2''$ and ${\sim}0.9''$, corresponding to physical separations of ${\sim}1.7$--$7.8$~kpc. Of the 34 observed systems,13 (${\sim}40\%$) are resolved into multiple AGN-like components; 18 (${\sim}53\%$) are classified as AGN plus foreground star superpositions, one as a single AGN, and two as systems with an AGN primary and an unclassified secondary component.
Among the multiple AGN systems, we confirm six dual AGN at projected separations of $0.26''$ -- $0.80''$ ($2.2$--$6.7$~kpc), in the redshift range $z~=~0.68$ -- $1.57$, and with velocity offsets in the range $\Delta v \sim 290$ -- $950$~\kms.
We also identify seven gravitationally lensed quasars, only one of which was previously known, including four doubly imaged and three quadruply imaged systems.
These results substantially enlarge the census of spectroscopically confirmed compact multiple AGN systems at $z > 0.5$ and demonstrate the efficiency of ERIS--SPIFFIER for the systematic characterization of sub-arcsecond dual and lensed AGN candidates.}


   \maketitle

\section{Introduction}\label{sec:intro}

In hierarchical models of galaxy formation, major mergers are expected to be frequent, especially at high redshift, and to strongly shape galaxy morphology, star formation activity, and chemical evolution \citep{1978MNRAS.183..341W}, while also involving the supermassive black holes (SMBHs) residing at the centers of the merging galaxies \citep{Begelman80, 2003ApJ...582..559V, White78}. Hydrodynamical simulations indicate that these mergers drive substantial gas inflows toward the central regions of galaxies, fueling both intense star formation and accretion onto SMBHs, likely triggering active galactic nucleus \citep[AGN;][]{2007Sci...316.1874M} activity.
Following the merger, the central SMBHs of the progenitor galaxies sink toward the center of the common remnant via dynamical friction, eventually forming a gravitationally bound pair \citep{Colpi14}. The subsequent evolution toward coalescence can require extended timescales, of the order of $\sim$1~Gyr \citep[e.g.][]{2021NatRP...3..732V}, before the final inspiral produces gravitational-wave (GW) emission. Consequently, pairs of SMBHs at kpc-scale separation within the same host are predicted to be a natural phase of this evolutionary path, particularly at early epochs \citep{2003ApJ...582..559V}.
When both SMBHs are actively accreting material from the surroundings, the system is observed as a dual AGN \citep{Steinborn16, Rosas-Guevara19, Volonteri22}, with projected separations ranging from tens of kpc down to sub-kpc scales \citep{derosa2019}.
Compact dual AGN systems are therefore direct observational tracers of late-stage mergers and SMBH--host galaxy co-evolution, offering a unique laboratory for studying merger-driven gas accretion, AGN feedback in disturbed environments, and the dynamical processes that drive two SMBHs toward coalescence. The final inspiral and merger of such SMBH pairs are expected to produce GW signals in the nHz regime, already probed by pulsar timing arrays \citep[PTA;][]{Arzoumanian2018, Agazie2023}, and at mHz frequencies detectable by the upcoming Laser Interferometer Space Antenna \citep[LISA;][]{AmaroSeoane2023, Colpi2024}.

Cosmological simulations predict that dual AGN at separations below 10~kpc represent a non-negligible fraction of the overall AGN population, reaching up to a few percent at $z > 0.5$, where the galaxy merger rate is significantly higher than in the local Universe \citep{2021NatRP...3..732V,Steinborn16, Rosas-Guevara19}.
Despite these theoretical expectations, and more than a decade of dedicated observational searches, the number of spectroscopically confirmed dual AGN at sub-kpc to few-kpc separations remains small \citep{chen2022, mannucci2022, mannucci2023,ciurlo23, scialpi24, scialpi2026}, the deficit being particularly severe beyond the local Universe \citep{derosa2019,chen2022, mannucci2022}.
A full list of all spectroscopically confirmed dual AGN at $z > 0.5$ will be presented by \citet{scialpi2026_bhmass}.
Resolving two AGN separated by less than one arcsecond requires either space-based imaging or ground-based adaptive-optics (AO)-assisted observations, both of which are observationally expensive and have historically been applied to small samples. Furthermore, late-stage mergers are often associated with increased column densities and dust obscuration \citep{Koss18, Blecha18, Ricci17, Barrows2023}, making it harder to detect both nuclei even when the spatial resolution is sufficient. As a result, wide-separation systems and low-redshift pairs dominate current observational samples \citep[e.g.][]{derosa2019}, leaving the compact merger regime largely unexplored.

Systematic searches for close dual AGN have employed a variety of indirect and direct techniques. 
Early approaches exploited double-peaked narrow emission lines in spatially integrated spectra as a signature of two kinematically distinct nuclei \citep[e.g.][]{Zhou2004,Comerford2009}, though such features can also arise from gas rotation, disturbed kinematics, or AGN-driven outflows, leading to significant contamination. 
Radio-selected samples \citep{Fu2015}, mid-infrared pre-selection \citep{Satyapal2017,Pfeifle2019b}, and quasar clustering studies \citep{Hennawi2006, Hennawi2010} have each contributed to the identification of dual AGN.
Recently, \citet{Jing2026_desi} identified 1020 quasar pairs with separations $>2\arcsec$ in the Dark Energy Spectroscopic Instrument (DESI) DR1 quasar sample \citep{DESI_Karim2026}, with a redshift distribution peaking at $z\sim1$--2.5.
These approaches often probe relatively wide separations or specific AGN populations, and therefore do not provide a complete census of the merging SMBH population. More recently, probabilistic pair-finding methods and serendipitous discoveries in lensing surveys have added individual systems \citep[e.g.][]{Inada2008, lemon_2020,lemon2023,Glikman2023}, and integral field unit (IFU) spectroscopy with JWST has begun to reveal dual AGN candidates even at high redshift \citep{ubler2024, Perna2025, ubler2025, zamora2025}.

The \gaia\ mission \citep{2016A&A...595A...1G} of the European Space Agency (ESA) has opened a new window for the systematic identification of compact dual AGN candidates through multiple complementary all-sky selection strategies.
Dual AGN candidates can be identified as multiple \gaia\ detections within a few arcseconds of a known AGN \citep{Lemon2017, lemon2023, chen2022, Shen2023}, via characteristic variability \citep{Kochanek2006, Krone-Martins2019}, or by exploiting the astrometric jitter induced in the photocenter position of unresolved pairs by AGN variability, a technique known as ``varstrometry'' \citep{Shen2019,Hwang2020,chen2022, Schwartzman2024, Schwartzman2025, Gross2025}.
A further approach exploits the detection of multiple peaks in the along-scan light profiles of otherwise point-like sources, which is the foundation of the \gaia\ Multi-Peak (GMP) method \citep{mannucci2022}. This technique identifies sub-arcsecond multiple AGN candidates down to separations of $\sim$0.15$\arcsec$ and is particularly effective at $z > 0.3$, where the host-galaxy contribution to the \gaia\ light profile is negligible. Since \gaia\ operates at optical wavelengths and identifies sources down to $G \sim 20.5$~mag (in the Vega system), all these approaches are most sensitive to unobscured, optically bright AGN, which are expected to be type~1 QSOs.
High-resolution imaging from ESA/\textit{Euclid} provides a powerful complementary approach, extending the search to fainter systems down to  $I_{\rm AB} \sim 24.5$ in the VIS filter through machine-learning methods applied to sub-arcsecond-resolution images \citep{Ulivi2025}.
Strongly lensed quasars are one of the main classes of objects selected, along with intrinsic dual AGN, in searches for sub-arcsecond multiple AGN candidates \citep[e.g.][]{scialpi2026, lemon2023}. Their multiple lensed images can generate the same multi-peak signatures used to identify close AGN pairs. In these systems, however, the observed components are not distinct accreting SMBHs, but multiple images of the same background quasar produced by the gravitational potential of an intervening galaxy or galaxy group.
Lensed AGN systems are especially interesting because they allow to measure the total mass enclosed within the Einstein radius, that can provide stringent constraints to stellar initial mass function and the relative contributions of baryons and dark matter of the lensing galaxy
\citep[e.g.][]{Bolton2008_slacs,Treu2010,Massey2010,Sonnenfeld2015,Newman2017,Shajib2021}.
In addition, strongly lensed AGN can provide an independent way to constrain cosmological parameters, in particular the Hubble parameter \citep[e.g.][]{Wong2018A}. At the same time, the lensing magnification makes it possible to study intrinsically faint or distant AGN in greater detail than would otherwise be feasible, including their feedback and host-galaxy properties at high redshift \citep[e.g.][]{Feruglio2017,Tozzi2021}.  
In this context, compact lenses selected through the GMP method probe a complementary and largely unexplored regime, with sub-arcsecond image separations and Einstein radii sampling the inner regions of lens galaxies, often below their effective radii and at higher redshifts than classical galaxy-scale lens surveys \citep[e.g.][]{Treu2010, Sonnenfeld2019, damato2026}.

Multiple systems selected with the GMP method can therefore correspond to three distinct physical configurations: intrinsic dual AGN, gravitationally lensed quasars producing multiple images at sub-arcsecond separation, or chance alignments between an AGN and a foreground star. Disentangling these scenarios requires spatially resolved spectroscopy at sub--arcsecond resolution and at sufficient spectral resolution, which provides redshifts, emission-line diagnostics, and continuum properties for the individual components.
The first spectroscopic follow-up observations of GMP-selected targets were carried out with the OH-Suppressing Infrared Imaging Spectrograph \citep[OSIRIS,][]{Larkin2006} at the W.~M.~Keck Observatory \citep{ciurlo23, mannucci2023}, confirming the first GMP-selected dual AGN systems and demonstrating the feasibility of AO-assisted classification. Subsequent dedicated observations with the Multi Unit Spectroscopic Explorer \citep[MUSE,][]{bacon2010} in narrow-field mode with AO at the VLT expanded the confirmed sample and established a systematic classification framework, revealing a population that includes both dual AGN and compact gravitationally lensed systems \citep{scialpi24, scialpi2026}.

In this paper, we present AO-assisted integral field spectroscopy of 34 GMP-selected multiple AGN candidates obtained with the near-infrared (NIR) integral field unit (IFU) of the Enhanced Resolution Imager and Spectrograph \citep[ERIS,][]{Davies2023} at the VLT.
Spatially resolved spectra allow us to classify each component independently and confirm six dual AGN, and seven gravitationally lensed quasars, substantially enlarging the census of spectroscopically confirmed compact multiple AGN systems at $z > 0.5$.
The paper is organized as follows. Section~\ref{sec:targets} describes the target selection, section~\ref{sec:method} presents the observation strategy, the data reduction, the image decomposition and spectral extraction. In Section~\ref{sec:classification} we present the classification scheme and classify the multiple systems in our sample. In Section~\ref{sec:discussion} we discuss our results, and in Section~\ref{sec:concl} we present our conclusions.
Throughout this work we adopt a flat $\Lambda$CDM cosmology with $H_0 = 67.7$~\kms~Mpc$^{-1}$ and $\Omega_{\rm m} = 0.31$ \citep{Planck2020}, and all magnitudes are reported in the Vega system unless explicitly stated otherwise.

\section{Target selection}\label{sec:targets}

Our sample of dual AGN candidates was built by applying the GMP technique \citep{mannucci2022, mannucci2023} to sources in the \gaia~Data Release~3 catalog \citep{Gaia_EDR3_1, Gaia_EDR3_2}. The GMP method exploits the \texttt{ipd\_frac\_multi\_peak} parameter, hereafter \texttt{FMP}, \citep{Gaia_EDR3_2}, which quantifies the percent fraction of \gaia~scans, over all scan orientations, in which a source is detected with multiple peaks in its light profile. The parameter ranges from 0, for sources never detected as multi-peaked, to 100, for sources detected as multi-peaked in all scans. We adopted the same selection thresholds used in previous GMP studies, \texttt{FMP}\,$\geq 8$, which has been demonstrated to reliably identify sub-arcsecond multiple systems \citep{mannucci2022, mannucci2023}.
In particular, the GMP method enables the selection of multiple systems with angular separations down to $\sim$\SI{0.15}{\arcsecond}, where the individual components are blended within the \gaia\ point-spread function (PSF). At separations larger than $\sim$\SI{0.7}{\arcsecond}, the components may be resolved as separate \gaia\ catalog entries. To focus on compact systems, we restricted the sample to angular separations smaller than \SI{0.9}{\arcsecond}, some targets near this upper limit are resolved in the \gaia\ catalog (see Table~\ref{tab:sample}).
We restricted the sample to $z > 0.5$ because at lower redshifts the brightness of the host galaxy can significantly reduce the GMP detection efficiency \citep{mannucci2022}. This selection allows us to probe the cosmic noon epoch ($z \simeq 1$--3), which corresponds to the peak of galaxy mergers, star formation, and AGN accretion activity \citep{madau2014,chen23b}.
At these redshifts, the angular-separation range probed by the GMP method corresponds to projected physical separations of approximately 2--8 kpc, consistent with the regime in which the AGN are expected to reside within the same merging system \citep[e.g.,][]{Volonteri22, chen23b}.

We further required the primary source, i.e., the brightest component, to satisfy $G < 20.5$ mag, which marks the faint limit of the GMP method \citep{mannucci2023}.
Indeed, \citet{mannucci2023} showed that if the primary component has $G > 17$ mag, the GMP method can efficiently select systems with a secondary as faint as $G \sim 20.5$--21 mag.
To minimize stellar contamination, we excluded sources at Galactic latitudes $|b| < 20^{\circ}$, sources located in the vicinity of prominent Local Group galaxies, namely M31, the LMC, and the SMC, and sources in crowded fields, defined as regions with more than 12 \gaia~detections per square arcminute within 3 arcmin of the target position.

The parent AGN sample to which the GMP method was applied is composed by i) AGN from spectroscopic surveys, including the Sloan Digital Sky Survey (SDSS; \citealt{SDSS_lyke2020}) and DESI \citep{DESI_Karim2026} survey, compiled in version~8 of Milliquas \citep{flesch2023}, and ii) photometric AGN candidates from wide-field catalogs \citep{flesch2021, wise_ps1_2022,delchambre2022, quaia_Storey-Fisher2024, fu2024, salvato2025}.
The former provides robust classifications and accurate redshifts of the primary sources up to $z \sim 4$, while the latter extends the GMP selection to sky regions not yet covered by spectroscopic surveys. Applying the GMP method to this parent sample resulted in a catalog of approximately 2,300 multiple-AGN candidates.

Approximately 700 photometric AGN candidates identified by the GMP method were followed up with dedicated long-slit spectroscopic observations. Since these observations do not spatially resolve the individual components of the system, they cannot confirm the dual AGN nature of the systems. Instead, they are used to confirm the presence of at least one AGN and to obtain a robust spectroscopic redshift.
The observations were carried out under several observing programs with three instruments: the ESO Faint Object Spectrograph and Camera 2 (EFOSC2) mounted at the Nasmyth~B focus of the 3.58m ESO New Technology Telescope (NTT; \citealt{efosc2_buzzoni84}; Program IDs: 109.22W4, 112.25CT, 114.27FC, PI: Mannucci; 113.26TB, PI: Scialpi), the Device Optimized for the LOw RESolution spectrograph (DOLORES) on the Telescopio Nazionale \textit{Galileo} (TNG; Program IDs: \texttt{A47TAC\_28}, \texttt{A48TAC\_67}, \texttt{AOT51\_24}, PI: Scialpi), and the FOcal Reducer/low dispersion Spectrograph~2 (FORS2) on the VLT in Long-Slit Spectroscopy (LSS) mode (Program IDs: \texttt{115.28CQ}, \texttt{116.28PX}, PI: Scialpi).
Observations were carried out under varying seeing conditions, with a typical image quality better than $1''$. A $1''$ slit was adopted as the default configuration.
Under poorer seeing conditions, the slit width was increased to minimize flux losses by enclosing a larger fraction of the PSF and to ensure that the light from all components of the GMP-selected systems was included in the extracted spectrum.
Integration times were chosen to reach a continuum S/N of $\sim$10--15 per spectral resolution element at $G \simeq 20$ mag, as estimated with the ESO and TNG Exposure Time Calculators, yielding total on-source times of 20--60\,min per target depending on target brightness and weather conditions. In this work, we focus on the properties of objects observed with ERIS (see below for details on the selection of the ERIS sample); the full spectroscopic catalog of GMP-selected AGN candidates will be presented in Scialpi \& Marconcini, in prep.

The resulting sample of $\sim$700 spectra, for which the primary component was confirmed to be an AGN, was further inspected for possible stellar contamination through a spectral decomposition analysis (\citealp{scialpi2026}; Scialpi \& Marconcini, in prep.). This step is necessary because long-slit spectroscopy alone cannot distinguish between AGN pairs and projected AGN+star systems, both of which may appear as multiple-peaked sources in \gaia\ light profile. Each spectrum was modeled with \texttt{pPXF} \citep{Cappellari2004} using a linear combination of quasar \citep{temple2021} and stellar (G, K, M types, \citealt{yan2019}) templates.
The fit simultaneously accounts for the quasar redshift, dust attenuation, and relative template normalizations, while allowing for a velocity shift of the stellar component. Candidate AGN+star systems were identified by comparing the quality of the best-fitting quasar+star model with that of a pure quasar model. 
The presence of a stellar component was assessed by quantifying the improvement of the fit around prominent stellar absorption features relative to a pure quasar model. Sources satisfying the adopted contamination criteria ($\Delta\chi^2_{\rm R} > 10\%$ or $\Delta{\rm BIC} > 6$) were classified as likely AGN+star contaminants.
We note that the spectral-decomposition analysis was primarily used to rank candidates according to their probability of stellar contamination. To optimize the use of the allocated observing time, some lower-priority candidates were observed when higher-priority targets were unavailable. Consequently, a fraction of AGN+star contaminants remains in the final IFU sample.

Targets selected for ERIS--SPIFFIER follow-up observations were required to satisfy some observational constraints. The sub-arcsecond angular separations of our targets ($\sim 0.15''$--$0.8''$) require diffraction-limited observations on an 8m-class telescope with AO correction via a laser guide star (LGS), allowing for tip-tilt correction using reference stars located at relatively large angular separations from the science target. We therefore required the presence of a suitable tip-tilt (TT) star, with $R_p < 18.0$ and $B_p < 19.0$ within 60\arcsec~from the target. Additional observability constraints included a declination of $\delta < +15^{\circ}$ to ensure accessibility from Paranal.
A further constraint was imposed by the spectral coverage of the ERIS--SPIFFIER gratings used in this program. We required that at least one prominent emission line, namely \ha, \hb+[\ion{O}{III}]$\lambda5007$, \hg, \hd, or \ion{Mg}{II}, {falls within one of the three SPIFFIER bands employed here: \jlow\ (1.09--1.42~\micron), \hlow\ (1.45--1.87~\micron), or \klow\ (1.93--2.48~\micron). The presence of these lines in the IFU data cube is essential to study the nature of both sources, measure possible velocity offsets between the two components, investigate the presence of differences in the line profiles, and measure single-epoch black-hole masses.
We have a starting sample of $\sim$530 spectra, for which the primary component is a spectroscopically confirmed AGN at $z>0.5$, $\sim$350 of them are observable from Paranal. After applying the additional ERIS--SPIFFIER selection criteria described above, including the requirements on AO tip-tilt stars and spectral coverage, approximately 160 sources remained and were ranked according to their probability of stellar contamination. Among these, $\sim$40 sources showed clear evidence of stellar contamination.
The final target selection was further driven by the right ascension and visibility of the sources during the allocated GTO nights, as well as by the atmospheric conditions and other observing constraints at the time of the observations.
Therefore, we observed 34 systems with ERIS--SPIFFIER, see Table~\ref{tab:sample}. These targets span the redshift range $0.68 \lesssim z \lesssim 3.43$ and exhibit projected angular separations between $\sim 0.21''$ and $\sim 0.93''$, corresponding to physical separations of approximately $1.7$--$7.8$~kpc at these redshifts.

\section{Methods}\label{sec:method}

In this paper, we present the first results of the VLT/ERIS follow-up of the GMP-selected multiple AGN candidates. Table~\ref{tab:sample} summarizes the main properties of the full sample. 
We report the \texttt{FMP} parameter, as well as the G-band magnitudes of the primary ($G_A$) and secondary components ($G_B$, when multiple components are present in the \gaia\ catalog), as listed in the \gaia\ archive. By definition, the primary component (A) is the brightest source in the \gaia\ G band, i.e. $G_A < G_B$.
The redshift of each system is derived from the long-slit spectra, as described in Section~\ref{sec:targets}.
Because the selection is based on optically bright AGN, the targets are expected to be type~1 QSOs. The ERIS--SPIFFIER observations therefore allow us to classify the systems by comparing the emission-line properties of the different components (see Section~\ref{sec:classification}) and, for confirmed dual AGN, to derive black-hole mass estimates from the broad-line emission.
\citet{scialpi2026_bhmass} will present the black-hole mass estimates for all the GMP dual AGN confirmed so far, i.e. including from this work; \citet{mannucci2023, ciurlo23, scialpi24, scialpi2026}.
In Table~\ref{tab:sample}, we also report the classification of each system as: (i) AGN+AGN, corresponding to the 13 systems resolved into multiple AGN-like components, either dual AGN or gravitationally lensed systems; (ii) AGN+star, corresponding to 18 systems that are chance alignments between an AGN and a foreground star; (iii) AGN, for the system in which only a single AGN component is detected (see Section~\ref{app:singleAGN}); and (iv) AGN+Unclass., for the systems in which the secondary remains unclassified (see Section~\ref{subsec:unclass}).

\subsection{Observations}\label{sec:obs}

The VLT/ERIS observations presented in this paper are part of the INAF guaranteed-time observations (GTO) program ``The properties of the elusive population of dual AGNs with sub-arcsec separations'' (P.I. F.~Mannucci).
The targets were observed using the SPIFFIER IFU spectrograph \citep{George2016}, the upgraded version of the SPIFFI spectrograph of SINFONI \citep{Eisenhauer2003}, between April 2023 and March 2026 (run IDs: 111.24QJ, 112.25GT, 113.26BW, 114.26ZT, 115.27WA, 116.28MD).
Each target was observed in an individual observing block, split into short exposures because of the strong and rapid variation of the infrared sky emission, following an on-source jitter pattern for the science target and an object-sky pattern for the telluric star used for flux calibration and telluric correction. In Table~\ref{tab:sample}, the total integration time is reported as the number of exposures (N$_{\rm exp}$) multiplied by the integration time of each exposure (T$_{\rm exp}$).
All observations were carried out using the 100~mas plate scale, which corresponds to a spaxel size of 50~$\times$~100~mas and a field of view (FoV) of \SI{3.2}{\arcsecond}~$\times$~\SI{3.2}{\arcsecond}.
Thanks to the dithering strategy with half-spaxel shifts, the reduced data cubes have square pixels with a size of 50~mas.
The targets were observed using one of the low-resolution gratings in the \jlow, \hlow, or \klow\ band, which provide spectral resolving powers of $R \sim 5000$, $R \sim 5200$, and $R \sim 5600$ corresponding to a velocity resolution of $\sim$ 60 \kms.
The telluric stars were observed in seeing-limited mode with the same pixel scale and grating, during the same night as the science targets.

\subsection{Data reduction}\label{sec:reduction}

The data were reduced with the ERIS pipeline (version 1.8.1) using the ESO Recipe Execution Tool \href{https://www.eso.org/sci/software/cpl/esorex.html}{\textsc{EsoRex}} (version 3.3.10), together with our own \texttt{python} routines for telluric correction and flux calibration, see \citet{Lamperti2026} for a detailed description of the routine.
The pipeline applies the standard calibration steps, including dark subtraction, detector-linearity correction, geometrical-distortion correction, flat-fielding, and wavelength calibration, to both the science and telluric-star data.
Additionally, {individual frames were corrected for differential atmospheric refraction (DAR) by setting \texttt{--dar\_corr=TRUE} in the \texttt{eris\_ifu\_jitter} recipe.
For J0300-0532, the reduction was performed with \texttt{--dar\_corr=FALSE}, as applying the DAR correction introduced artifacts at the position of the line falling in a region of poor sky transmission (see Tozzi et al., in prep.).
Our observation strategy did not include dedicated sky frames for the science targets. Instead, we took advantage of the applied dithering offsets and of the point-like nature of the sources to estimate the sky background for each science exposure from the temporally adjacent object frames. Sky subtraction was performed using the Davies method \citep{Davies2007MNRAS.375.1099D} implemented in the ERIS pipeline, setting \texttt{--sky\_tweak=1} and allowing the sky spectrum to be stretched with \texttt{--stretch=TRUE}.
Individual frames were aligned and combined using the pipeline recipe \texttt{eris\_ifu\_combine\_hdrl}, with the dithering shifts computed from the WCS information in the header of each individual input cube.
Flux calibration and telluric correction of the combined cubes were applied using our own \texttt{python} routines and the dedicated observations of the telluric stars. The response functions were obtained from the spectra of the stars normalized to their J, H, and K magnitudes from the Two Micron All Sky Survey \citep[2MASS;][]{skrutskie2006} catalog.
The ERIS \jlow, \hlow, and \klow\ bands are fully contained within the 2MASS J, H, and K bands, respectively, allowing us to recover the spectral shape of the flux calibration.

\begingroup
\setlength{\tabcolsep}{6pt} 
\renewcommand{\arraystretch}{0.9} 
\begin{table*}
\caption[]{Main properties of the GMP targets observed with VLT/ERIS and of the observations.}\label{tab:sample}
\centering                          
\begin{tabular}{l c c c c c c c c c}        
\hline  
\hline
Target & Coordinates (J2000) & $z$ & \texttt{FMP} & $G_A$ & $G_B$ & $\rm N_{exp}\times T_{exp}$ & Obs. & Band & Class.\\
 &  &  &  &  & & [s]  & date & &\\
\hline
J0008--3655 & 00:08:29.40 $-$36:55:23.28 & 1.715 & 16 & 18.56 & 19.71 & 5$\times$450 & 22/08/2025 & Jlow & AGN+star \\
J0018--4815 & 00:18:41.36 $-$48:15:19.29 & 2.104 & 9 & 19.74 & - & 6$\times$450 & 02/10/2024 & Klow & AGN+star \\
J0025+0750 & 00:25:53.35 +07:50:34.41 & 3.107 & 17 & 20.55 & - & 6$\times$450 & 04/09/2024 & Jlow & AGN+star \\
J0109--3529 & 01:09:46.30 $-$35:29:55.65 & 1.575 & 10 & 20.45 & 21.05 & 5$\times$450 & 06/10/2024 & Hlow & AGN+AGN \\
\multirow{2}{*}{J0150+0809}  & \multirow{2}{*}{01:50:02.83 +08:09:11.88} & \multirow{2}{*}{1.592} & \multirow{2}{*}{8} & \multirow{2}{*}{20.40} & \multirow{2}{*}{-} & 6$\times$450 & 22/08/2025 & Hlow & \multirow{2}{*}{AGN+Unclass.}\\
 &  &  &  &  &  & 12$\times$450 & 12/10/2025 & Jlow & \\
J0216+1430 & 02:16:36.62 +14:30:46.06 & 0.977 & 13 & 19.83 & 19.84 & 8$\times$300 & 22/09/2023 & Jlow & AGN+AGN\\
J0300--0532 & 03:00:12.27 $-$05:32:33.70 & 0.682 & 26 & 19.98 & - & 6$\times$450 & 04/10/2024 & Jlow & AGN+AGN\\
J0434+0846 & 04:34:36.68 +08:46:30.21 & 0.913 & 29 & 17.31 & - & 7$\times$300 & 21/09/2023 & Jlow & AGN+star  \\
J0459--0413 & 04:59:07.63 $-$04:13:21.52 & 1.277 & 33 & 20.41 & - & 4$\times$600 & 03/02/2024 & Hlow & AGN+AGN\\
J0533--2303 & 05:33:54.46 $-$23:03:20.84 & 1.372 & 8 & 18.05 & - & 4$\times$600 & 02/02/2024 & Hlow  & AGN+AGN\\
J0602--3118 & 06:02:04.12 $-$31:18:50.86 & 1.300 & 34 & 20.18 & - & 6$\times$450 & 25/03/2025 & Hlow & AGN+AGN\\
\multirow{2}{*}{J0602--3852} & \multirow{2}{*}{06:02:11.66 --38:52:13.15} & \multirow{2}{*}{3.041} & \multirow{2}{*}{9} & \multirow{2}{*}{20.41} & \multirow{2}{*}{-} & 6$\times$450 & 11/04/2025 & Jlow & \multirow{2}{*}{AGN+AGN} \\
 &  &  &  &  &  & 11$\times$450 & 13/10/2025 & Klow \\
J0917--0223 & 09:17:15.27 $-$02:23:07.06 & 1.287 & 54 & 19.65 & - & 9$\times$300 & 01/05/2023 & Hlow & AGN+star  \\
J1245--0908 & 12:45:17.21 $-$09:08:36.04 & 0.956 & 14 & 18.54 & - & 10$\times$300 & 15/04/2023 & Jlow & AGN+star \\
J1251--2857 & 12:51:22.01 $-$28:57:53.76 & 2.711 & 22 & 18.99 & - & 6$\times$450 & 24/03/2025 & Klow & AGN+star \\
J1305--0232 & 13:05:05.68 $-$02:32:19.19 & 1.373 & 9 & 18.90 & 20.30 & 3$\times$300 & 04/02/2024 & Hlow & AGN+AGN \\
J1318--0136$^{*}$ & 13:18:58.73 $-$01:36:42.50 & 1.489 & 17 & 20.03 & - & 8$\times$300 & 14/04/2023 & Hlow & AGN+star  \\
\multirow{2}{*}{J1335+0527$^{**}$} & \multirow{2}{*}{13:35:12.13 +05:27:32.48} & \multirow{2}{*}{1.954} & \multirow{2}{*}{10} & \multirow{2}{*}{19.53} & \multirow{2}{*}{-} & \multirow{2}{*}{6$\times$450} &  22/03/2025 & \multirow{2}{*}{Jlow} & \multirow{2}{*}{AGN+AGN} \\
&  &  &  &  &  & & 12/04/2025 & & \\
J1423+1544 & 14:23:22.93 +15:44:37.57 & 1.328 & 13 & 19.53 & 20.75 & 8$\times$300 & 15/04/2023 & Hlow & AGN+AGN \\
J1453+0529 & 14:53:06.95 +05:29:28.06 & 2.826 & 54 & 19.41 & - & 8$\times$300 & 14/04/2023 & Hlow & AGN+AGN\\
J1502+0257 & 15:02:16.66 +02:57:19.85 & 1.481 & 23 & 19.78 & - & 6$\times$300 & 14/04/2023 & Hlow & AGN+star  \\
\multirow{2}{*}{J1518+0605} & \multirow{2}{*}{15:18:02.17 +06:05:24.75} &\multirow{2}{*}{0.990} & \multirow{2}{*}{8} & \multirow{2}{*}{20.42} & \multirow{2}{*}{-} & 2$\times$450 & 11/04/2025 & \multirow{2}{*}{Jlow} &  \multirow{2}{*}{AGN} \\
& & & & & & 9$\times$450 & 24/03/2026 & & \\
J2013--1306 & 20:13:24.89 $-$13:06:35.09 & 1.240 & 11 & 20.28 & - & 6$\times$450 & 22/08/2025 & Jlow & AGN+star  \\ 
J2018--0351 & 20:18:34.62 $-$03:51:52.65 & 0.750 & 27 & 19.83 & 20.37 & 6$\times$450 & 22/08/2025 & Jlow & AGN+star  \\
J2019--2525 & 20:19:44.00 $-$25:25:05.20 & 2.009 & 16 & 17.37 & 19.75 & 8$\times$300 & 22/09/2023 & Klow & AGN+AGN \\
J2054--2900 & 20:54:15.07 $-$29:00:40.99 & 1.483 & 10 & 20.06 & 20.77 & 6$\times$450 & 03/10/2024 & Hlow & AGN+star  \\
\multirow{2}{*}{J2056--1656} & \multirow{2}{*}{20:56:09.09 $-$16:56:01.83} & \multirow{2}{*}{1.441} & \multirow{2}{*}{33} & \multirow{2}{*}{18.08} & \multirow{2}{*}{-} & 9$\times$300 & 15/04/2023 & \multirow{2}{*}{Hlow} & \multirow{2}{*}{AGN+star} \\
 & &  & &  &  &  8$\times$300 & 23/09/2023 & & \\
J2104$-$0215 & 21:04:50.38 $-$02:15:03.97 & 1.997 & 10 & 19.51 & 19.68 &  6$\times$450 & 02/09/2024 & Klow & AGN+star \\
J2117+1057 & 21:17:47.65 +10:57:22.67 & 1.607 & 10 & 19.49 & - &  6$\times$450 & 01/10/2024 & Hlow & AGN+AGN \\
J2118$-$1640 & 21:18:39.28 $-$16:40:52.85 & 3.428 & 10 & 19.57 & 20.61 &  6$\times$450 & 03/10/2024 & Klow & AGN+star  \\
J2135$-$2025 & 21:35:33.03 $-$20:25:49.64 & 0.875 & 14 & 19.97 & 20.84 &  6$\times$450 & 01/10/2024 & Jlow & AGN+star  \\
J2208$-$3529 & 22:08:55.18 $-$35:29:26.24 & 2.515 & 26 & 19.85 & - &  6$\times$450 & 04/10/2024 & Klow & AGN+star \\
J2211+1310 & 22:11:58.07 +13:10:37.01 & 2.604 & 21 & 19.42 & - &  8$\times$300 & 22/09/2023 & Klow & AGN+star  \\
J2301$-$1906 & 23:01:34.37 $-$19:06:46.68 & 0.816 & 22 & 19.25 & - & 3$\times$300 & 14/04/2023 & Jlow & AGN+Unclass. \\
\hline
\end{tabular}
\tablefoot{The columns report the redshift $z$ of the systems measured from the long-slit spectra (see Section~\ref{sec:targets}), the \texttt{ipd$\_$frac$\_$multi$\_$peak} parameter (\texttt{FMP}) and the \gaia~G-band magnitudes of the primary ($G_A$) and of the secondary ($G_B$, if multiple components are present) from the \gaia~archive are reported for each system. The exposure time is reported as the number of frames (N$_{exp}$) per the exposure time of each frame (T$_{exp}$). $^{(*)}$: J1318--0136 is part of the sample of \citet{mannucci2023} and classified as AGN+star. $^{(**)}$: J1335+0527 was observed in March 2025 and in April 2025 with identical observational set-up because of instrumental issues of the first observation, the PSF of March 2025 data has an elliptical not physical shape.}
\end{table*}
\endgroup

\subsection{Image decomposition}\label{subsec:image}

For each target, we computed the broad-band image by integrating the flux over the widest spectral range encompassing both the emission line and the continuum emission in both components, while excluding wavelength regions affected by poor sky transmission.
Figure~\ref{fig:eris_maps} shows the maps of the multiple AGN systems, including dual and lensed AGN. The target name, classification, and projected separation in arcsecond between the primary (i.e. most luminous) and secondary (B) AGN are reported in each map. 
Figure~\ref{fig:eris_maps_quad} shows the three systems classified as quadruply lensed AGN. For J0459--0413 and J0602--3852 we report the radius $r$ of the circle enclosing all four AGN images, which approximates the Einstein radius, while for J1453+0529 we report the Einstein radius $r_E$ calculated by \citet{damato2026} through lens reconstruction.
Following \citet{damato2026} approach, in absence of an independent PSF model, the position of each component is determined by modeling the broad band map with a combination of two two-dimensional Moffat (\texttt{2D-Moffat}) functions, or four functions for quadruply lensed AGN, of the form:
\begin{equation}\label{eq:moffat}
I(x,y) = I_0 \left[1 + \left(\frac{x'}{\alpha_x}\right)^2 + \left(\frac{y'}{\alpha_y}\right)^2 \right]^{-\beta},
\end{equation}
where $I_0$ is the peak intensity, $\beta$ controls the power-law slope of the profile wings, and $\alpha_x$ and $\alpha_y$ are the scale parameters along the two principal axes. The rotated coordinates $(x',y')$ are defined as:
\begin{align}
x' &= (x-x_0)\cos\theta + (y-y_0)\sin\theta, \\
y' &= -(x-x_0)\sin\theta + (y-y_0)\cos\theta,
\end{align}
with $(x_0,y_0)$ the centroid position of the component and $\theta$ the position angle measured counter-clockwise from the $x$ axis.
The full width at half maximum (FWHM) along each axis is related to the scale parameters by:
\begin{equation}
\mathrm{FWHM}_{x,y} = 2~\alpha_{x,y}~\sqrt{2^{1/\beta}-1}.
\end{equation}
The modeling was performed using a custom Python routine, where each \texttt{2D-Moffat} profile accounts for the spatial light distribution of an individual component.
To reduce the number of free parameters and improve the stability of the fit, the secondary component, and the third and fourth components in quadruply lensed systems, were constrained to share the same shape parameters ($\alpha_x$, $\alpha_y$, and $\beta$) as the primary, allowing only the centroid positions and amplitudes to vary independently. This approach is based on the expectation that the PSF is constant across the central $\sim1\arcsec\times1\arcsec$ region containing all the components, and provides robust estimates of the relative astrometry between components.
\newline The resulting separations are reported in {Tables~\ref{tab:classified} and~\ref{tab:classified_app} for the systems classified as dual AGN, lensed AGN, dual AGN candidates, and AGN+star, respectively.

\begin{figure*}
\resizebox{\hsize}{!}{
\includegraphics{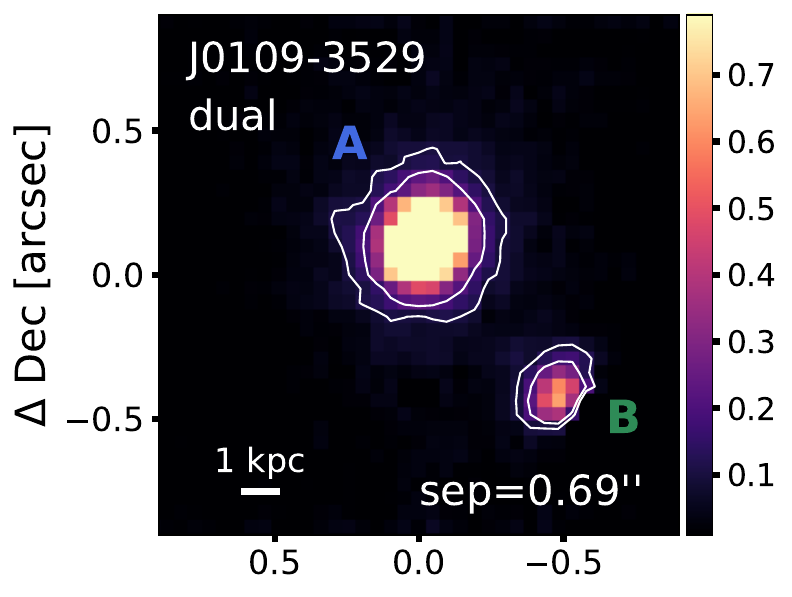}
\includegraphics{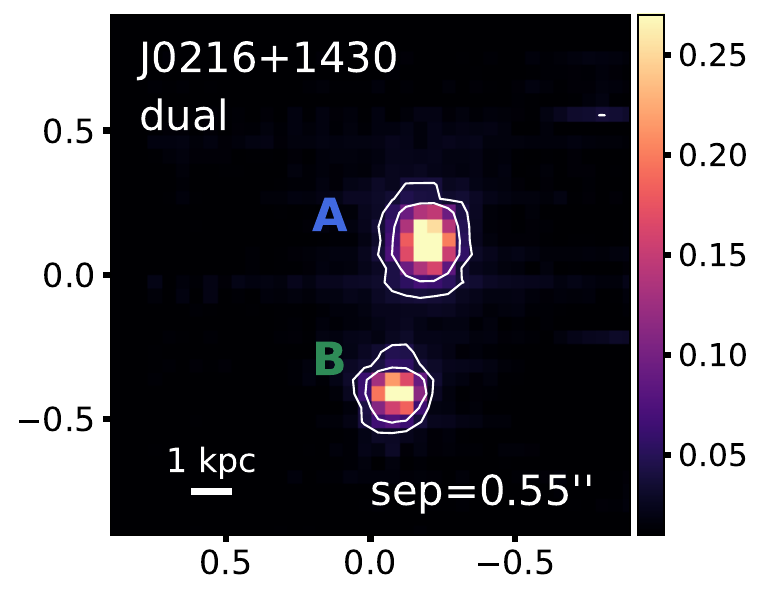}
\includegraphics{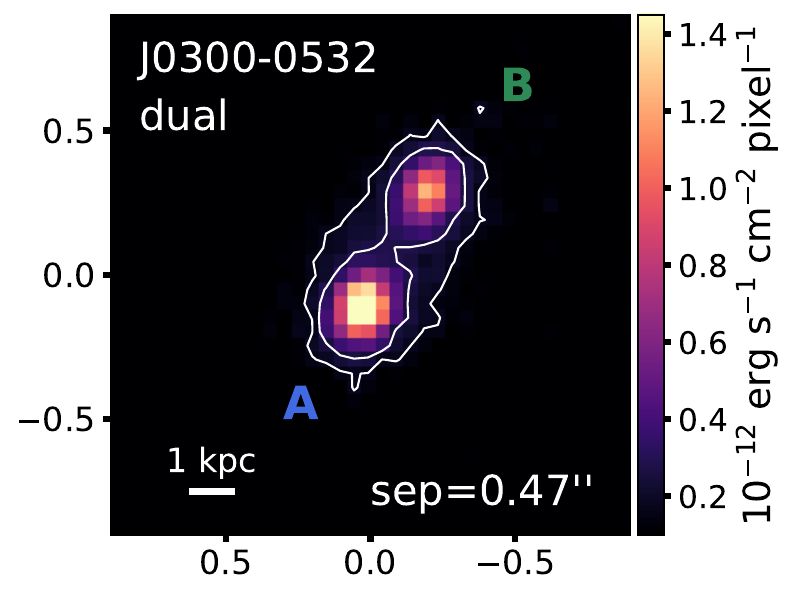}}
\resizebox{\hsize}{!}{
\includegraphics{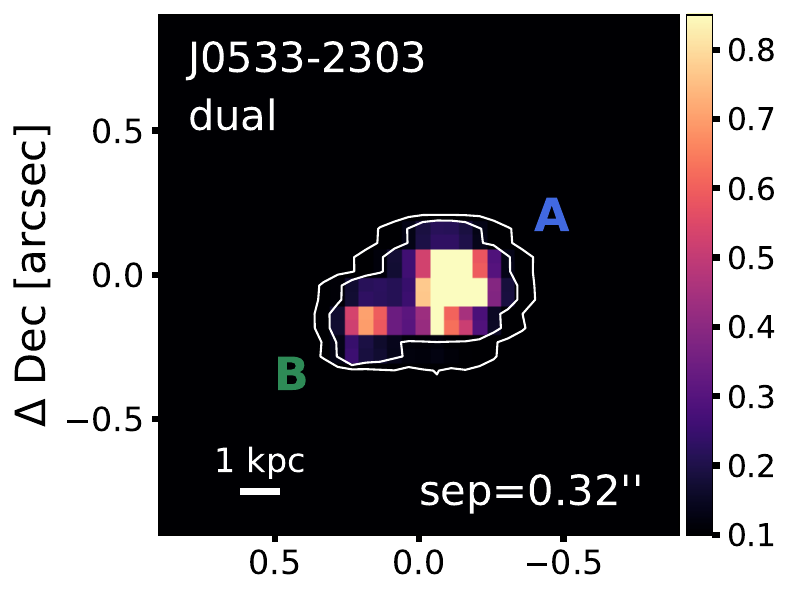}
\includegraphics{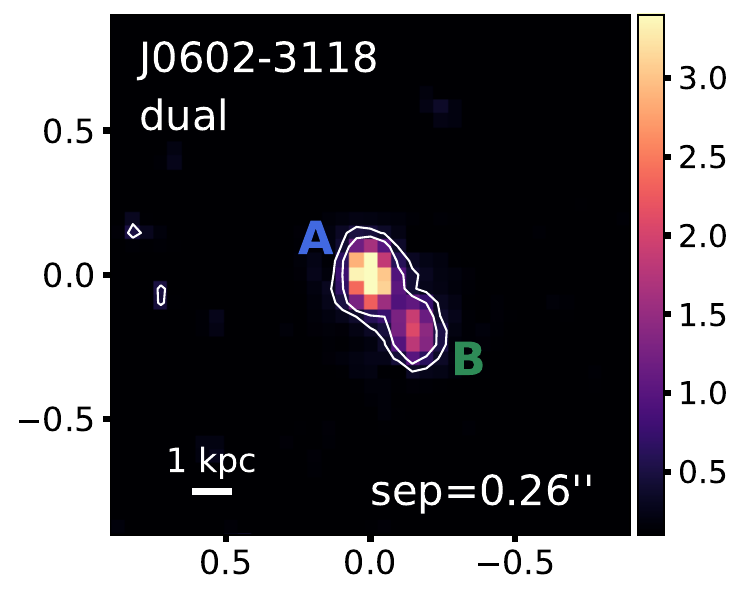}
\includegraphics{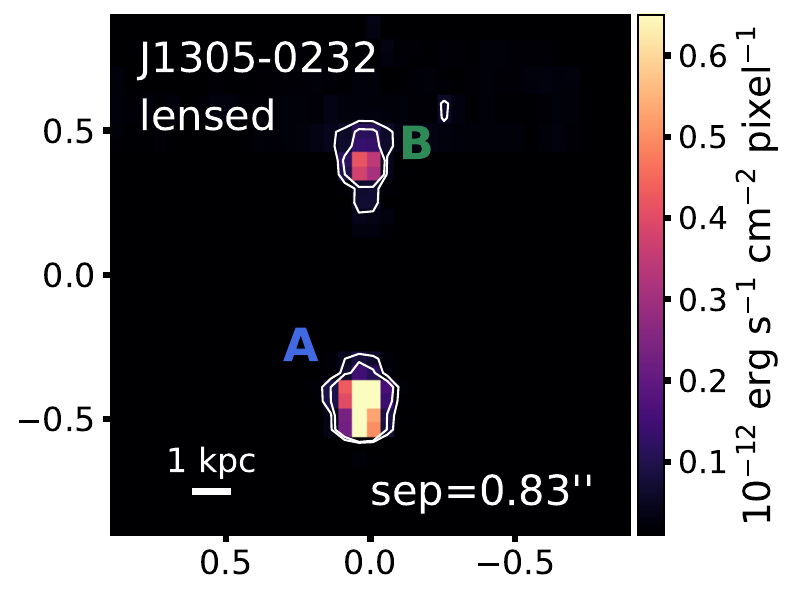}}
\resizebox{\hsize}{!}{
\includegraphics{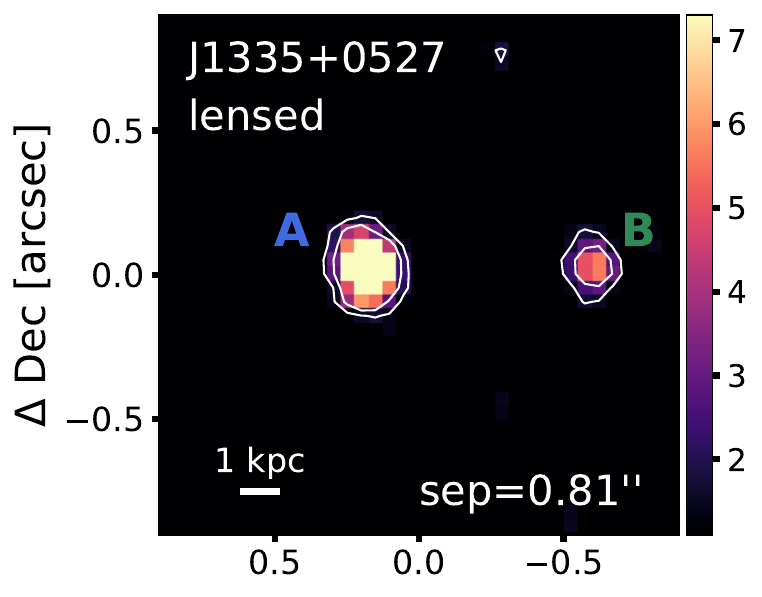}
\includegraphics{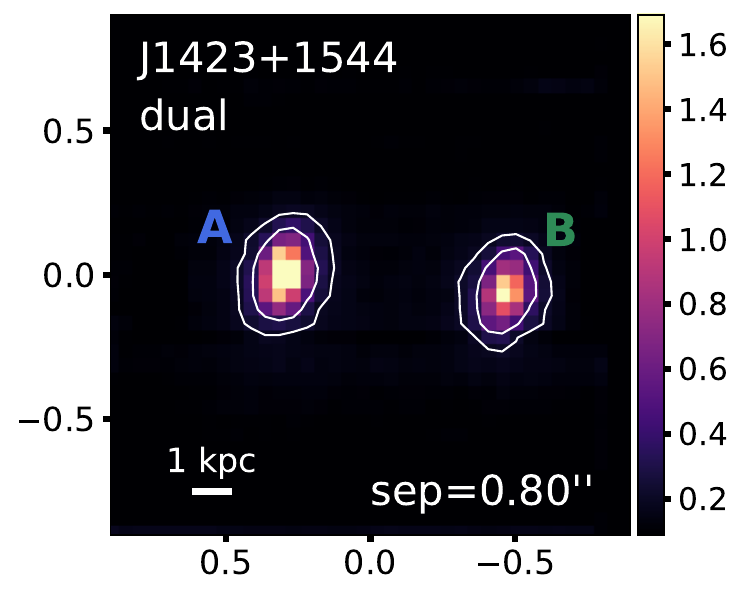}
\includegraphics{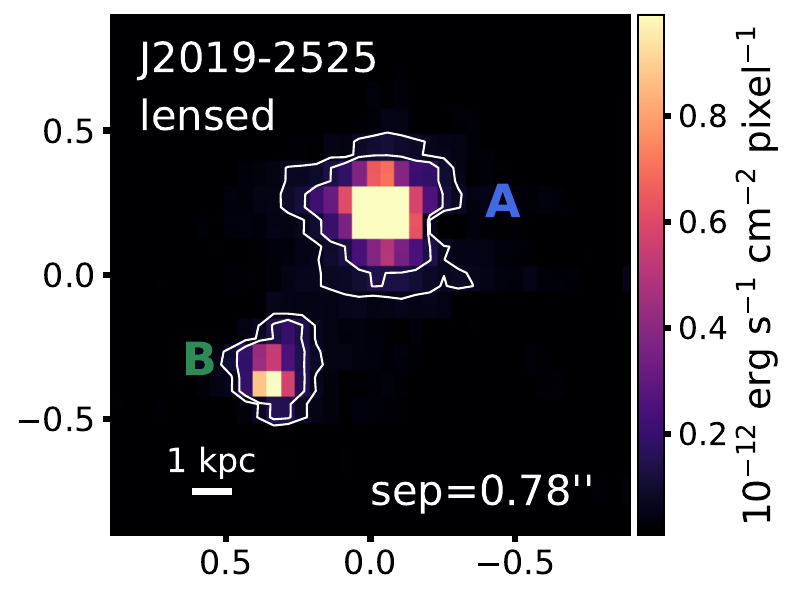}}
\begin{center}
\resizebox{0.7\hsize}{!}{
\includegraphics{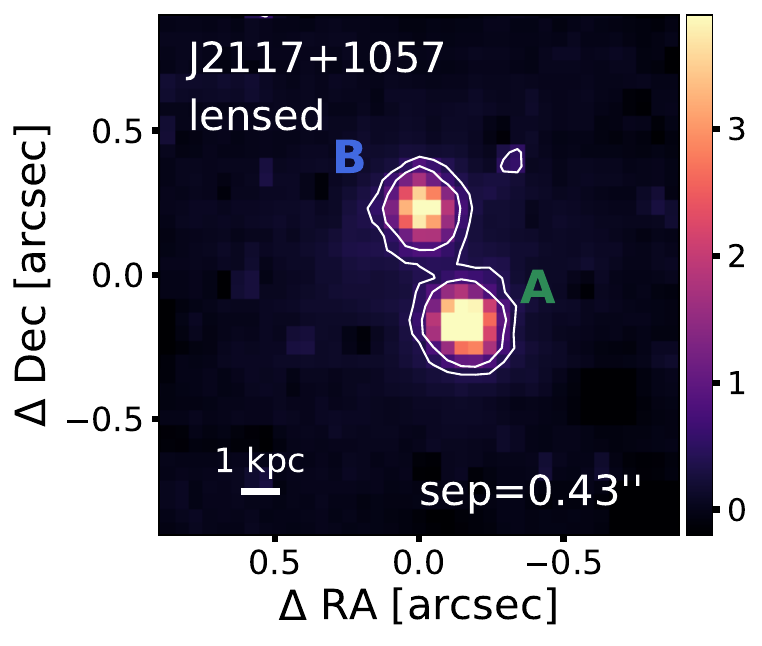}
\includegraphics{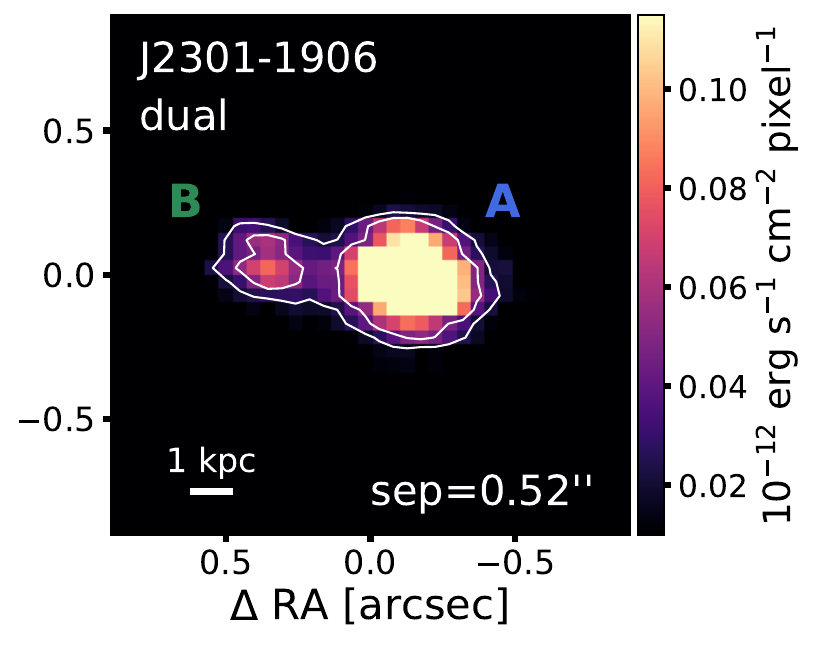}}
\end{center}
\caption{Broad band images of the AGN+AGN systems and J2301--1906 observed with ERIS--SPIFFIER: target name, classification and projected separation are reported on the figure. The maps are oriented with north at the top and west to the right.  The labels plotted over the maps mark the position of the two components: the primary A in blue and the secondary B in green. White contours are drawn at (5, 10) times the r.m.s. noise of the map.}
\label{fig:eris_maps}
\end{figure*}

\begin{figure*}
\resizebox{\hsize}{!}{
\includegraphics{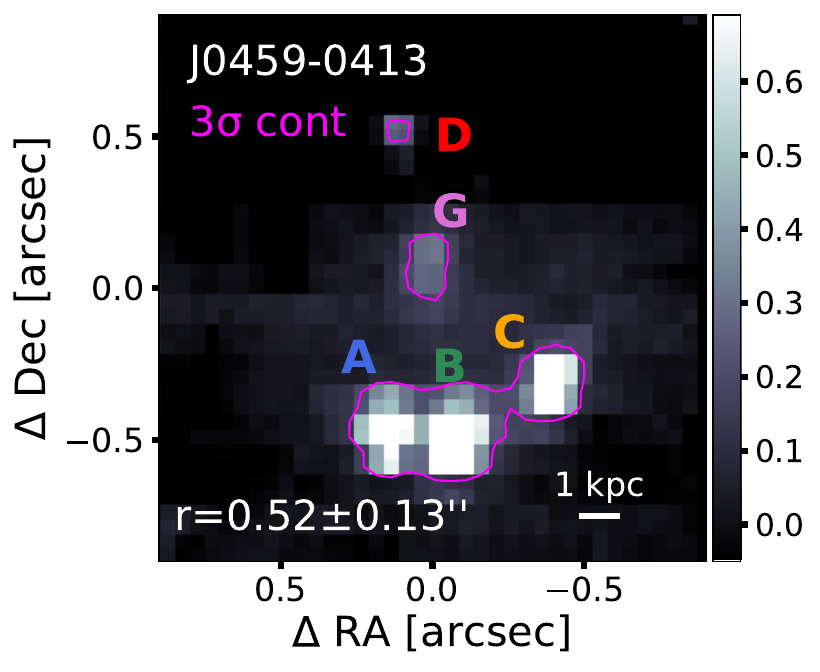}
\includegraphics{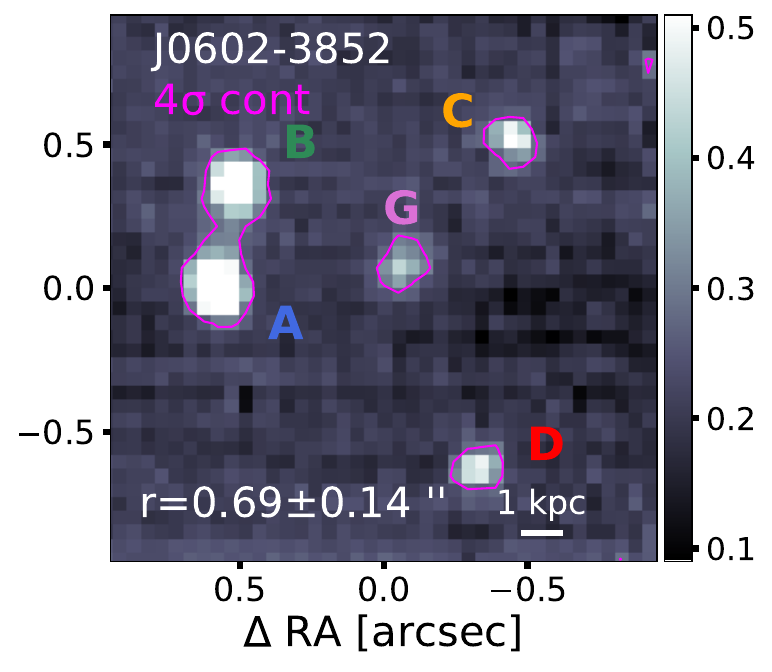}
\includegraphics{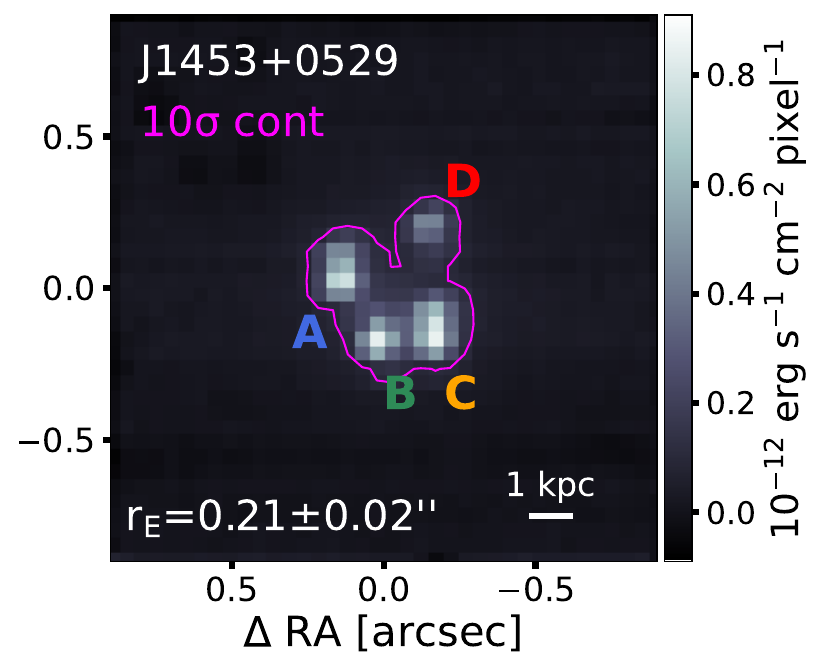}}
\caption{Same as Fig.~\ref{fig:eris_maps} but for the systems classified as quadruply lensed AGN. The confirmed AGN images are labeled from A to D.
In J0602--3852 and J0459-0413 the lensing galaxy is labeled as G and $r$ indicates the radius of the circle enclosing all four AGN images. For J1453+0529 we report the Einstein ring $r_E$ radius from \citet{damato2026}. Magenta contours level is reported on the maps.}
\label{fig:eris_maps_quad}
\end{figure*}

\subsection{Spectra extraction}\label{subsec:spectra}

The spectra extraction was performed by modeling the data cube on a channel-by-channel basis using a combination of \texttt{2D-Moffat} functions, as in Eq.~(\ref{eq:moffat}), with two components for double systems and four components for quadruply lensed AGN. 
This approach was adopted instead of a standard aperture extraction to minimize mutual contamination between the spectra of the different components. While the modeling procedure may introduce additional statistical uncertainties, it substantially reduces the systematic effects associated with flux cross-contamination between nearby sources. Given the small angular separations of the targets in our sample, we consider the latter to be the dominant source of uncertainty in a standard aperture extraction.
For each system, the centroid positions $(x_0,y_0)$ of all components were fixed to the values determined as described in Section~\ref{subsec:image}, after verifying that the relative astrometry was consistent within a fraction of a pixel when modeling the full cube and the blue- and red-side spectral ranges independently.
Following the approach adopted for the image analysis (Section~\ref{subsec:image}), the shape parameters of the secondary components ($\alpha_x$, $\alpha_y$, $\beta$, and $\theta$) were constrained to be identical to those of the primary component. Under these assumptions, only the amplitudes $I_0$ of the individual components and the common PSF shape were allowed to vary as a function of wavelength.
The flux of each component (A or B) at a given wavelength channel $\lambda_i$ is then given by:
\begin{equation}
F^{A,B}(\lambda_i) = I^{A,B}_0(\lambda_i)~\frac{\pi~\alpha_x(\lambda_i)~\alpha_y(\lambda_i)}{\beta(\lambda_i) - 1},
\end{equation}
where $I_0(\lambda_i)$ is the wavelength-dependent amplitude of the best-fit \texttt{2D-Moffat} function, while $\alpha_x(\lambda_i)$, $\alpha_y(\lambda_i)$, and $\beta(\lambda_i)$ describe the common PSF shape at the same wavelength channel.

To assess the potential impact of host-galaxy contamination on the \texttt{2D-Moffat} decomposition, we consider the limiting case in which up to 10\% of the total \ha\ flux of the primary AGN originates from the host galaxy rather than from the QSO, for the lowest-redshift dual AGN in our sample (J0300--0532, $z_A=0.679$; see Table~\ref{tab:classified}). This 10\% \ha\ flux would correspond to a star formation rate (SFR) of $\sim$7 \sfr\ \citep{kennicutt2012}, consistent with a star-forming galaxy of stellar mass $\sim1.3\times10^{10}$ \msun\ on the main sequence at $z=0.679$ \citep{speagle2014}. Conversely, if the host were an early-type, quiescent galaxy, the same \ha\ luminosity, if powered by old stellar populations including post-AGB stars, would require a stellar mass of $\sim9.6\times10^{12}$ \msun \citep{cidfernandes2011}. We therefore conclude that, if the host galaxy contributes significantly to the observed \ha\ luminosity, it would most plausibly be a star-forming galaxy.

AO-assisted observations typically achieve limited Strehl ratios, and therefore the extracted spectra capture only a fraction of the total flux. Assuming a common PSF among all components within each system, this fraction is expected to be the same for all sources. To account for this effect and for the absolute flux-calibration uncertainty of the observations, the spectra were normalized by matching the total flux of each system to the expected broadband magnitude. 
Specifically, the fluxes of the individual components were summed and convolved with the transmission curve of the corresponding ERIS grating to derive the measured magnitude. This value was then compared with the expected broadband magnitude, and the resulting scaling factor was applied to the extracted spectra.
The expected broadband magnitude was either taken directly from available VHS \citep{mcmahon2013} or 2MASS \citep{skrutskie2006} photometry, or derived using the empirical relations described in Appendix~\ref{app:mag}, based on ancillary photometry in other bands, e.g. $H$ inferred from $J$ and $K$, or from \gaia~$B_P$ and $R_P$ magnitudes. 
To determine the redshift of each component we modeled each ERIS spectrum independently, using empirical quasar templates from \citet{temple2021}, which incorporate an empirical dust attenuation law. The quasar template and dust attenuation were fixed to the values derived from the optical long-slit spectral model obtained during the deconvolution procedure (see Section~\ref{sec:targets}), leaving the redshift as the only free parameter. This provided the redshifts $z_A$ and $z_B$ and, for dual AGN systems, their relative velocity offset $\Delta v$ (see Table~\ref{tab:classified}).

Since the ERIS spectra cover only a limited wavelength range around either \ha\ or \hb, they do not provide sufficient information on their own to constrain the dust attenuation using standard methods (e.g. the Balmer decrement) or the continuum shape. The ERIS spectra were therefore combined with the available optical long-slit observations used for target selection (see Section~\ref{sec:targets}). While the optical spectrum provides broad wavelength coverage, it does not spatially resolve the individual components, as the slit encompasses the entire system. Conversely, the ERIS observations spatially resolve the individual sources but cover only a limited wavelength range.
We therefore performed a simultaneous fit of the optical long-slit spectrum and the ERIS spectra using the empirical quasar templates from \citet{temple2021}. Each ERIS near-infrared spectrum is modeled independently with a quasar template, while the sum of the two best-fit templates is required to reproduce the optical long-slit spectrum, see Appendix~\ref{app:temple} for the details.
The redshifts of the two templates were fixed to the values measured previously, while the normalization and dust attenuation $E(B-V)$ of each component were left as free parameters.
When the two components were spatially resolved by \gaia\ (see Table~\ref{tab:sample}), their flux ratio was used as an additional constraint on the relative normalization of the two templates.
The parameter space was explored using a Markov Chain Monte Carlo (MCMC) approach, from which we derived median values and associated uncertainties for the attenuation of each component, see Figure~\ref{fig:j0300_temple_mcmc} for an example of the fit.

\noindent Figures~\ref{fig:eris_spectra_dual} and~\ref{fig:eris_spectra_lens} show the spectra of the dual, dual candidates and lensed AGN systems, respectively, including both doubly and quadruply lensed systems. The spectra follow the same color coding as the labels in Figures~\ref{fig:eris_maps} and~\ref{fig:eris_maps_quad}. Vertical dotted lines mark the positions of the main expected emission lines at the redshifts reported in the panels. For the dual AGN systems, we report the redshifts of the primary ($z_A$) and secondary ($z_B$) components, and the expected emission-line positions are color-coded accordingly.

\begin{figure*}
\resizebox{\hsize}{!}{
\includegraphics{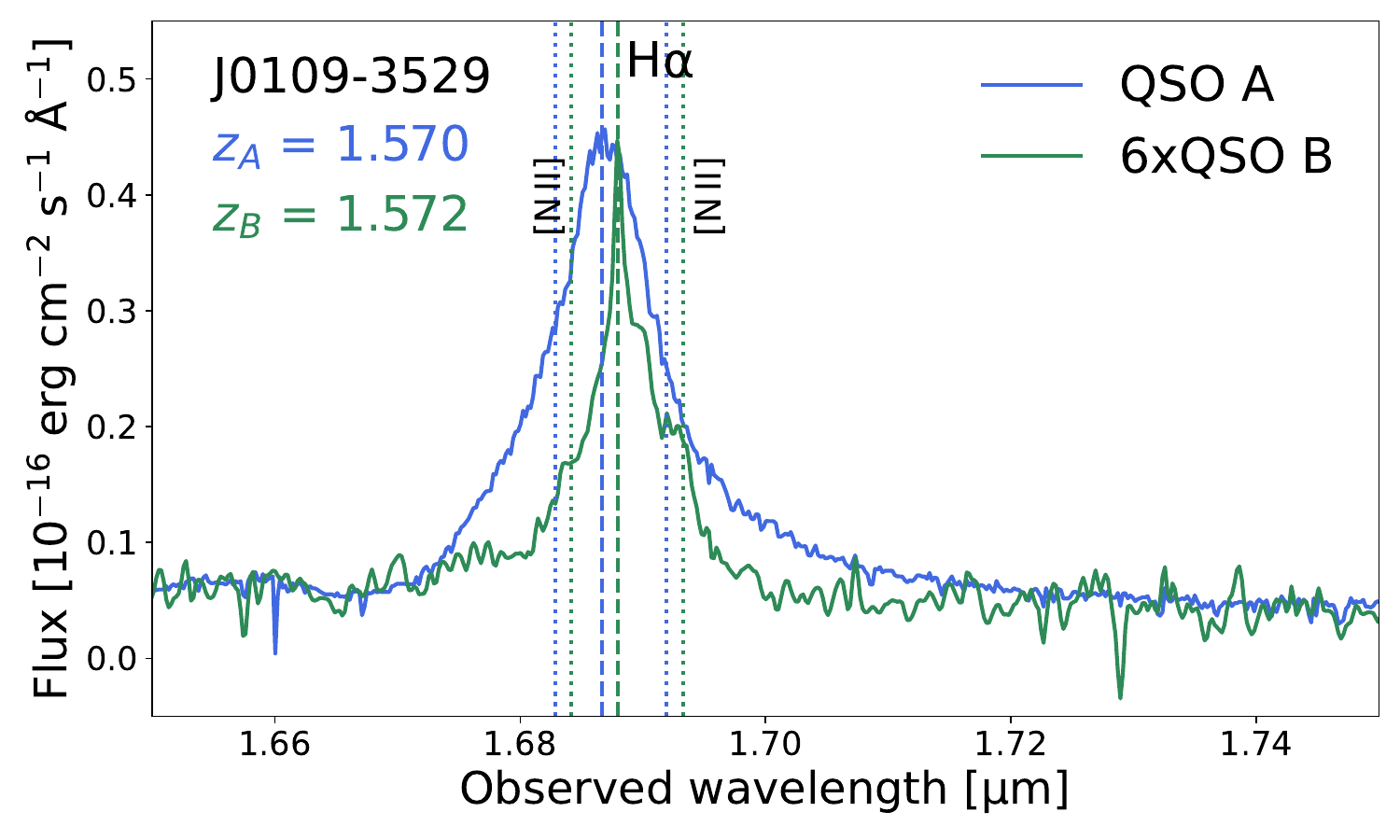}
\includegraphics{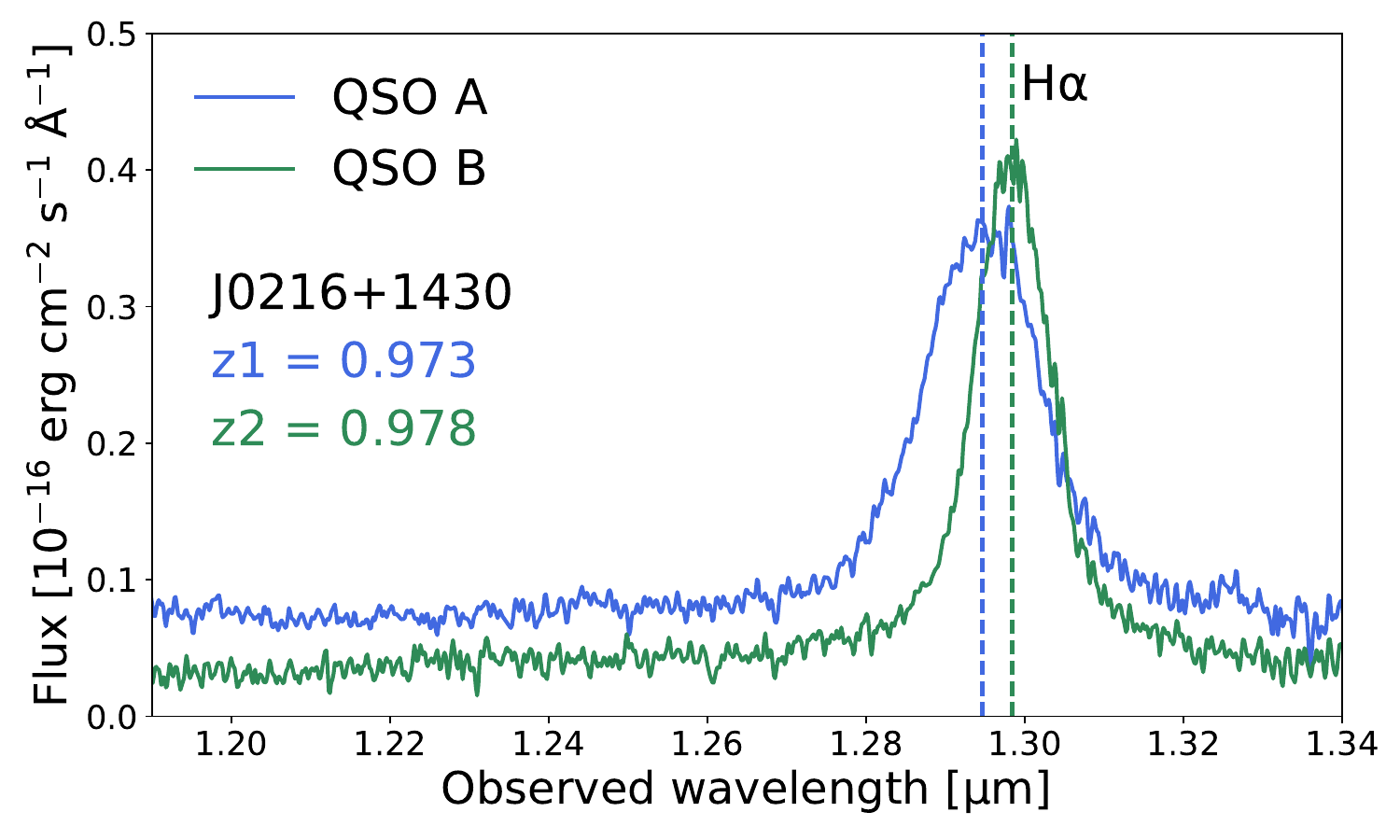}}
\resizebox{\hsize}{!}{
\includegraphics{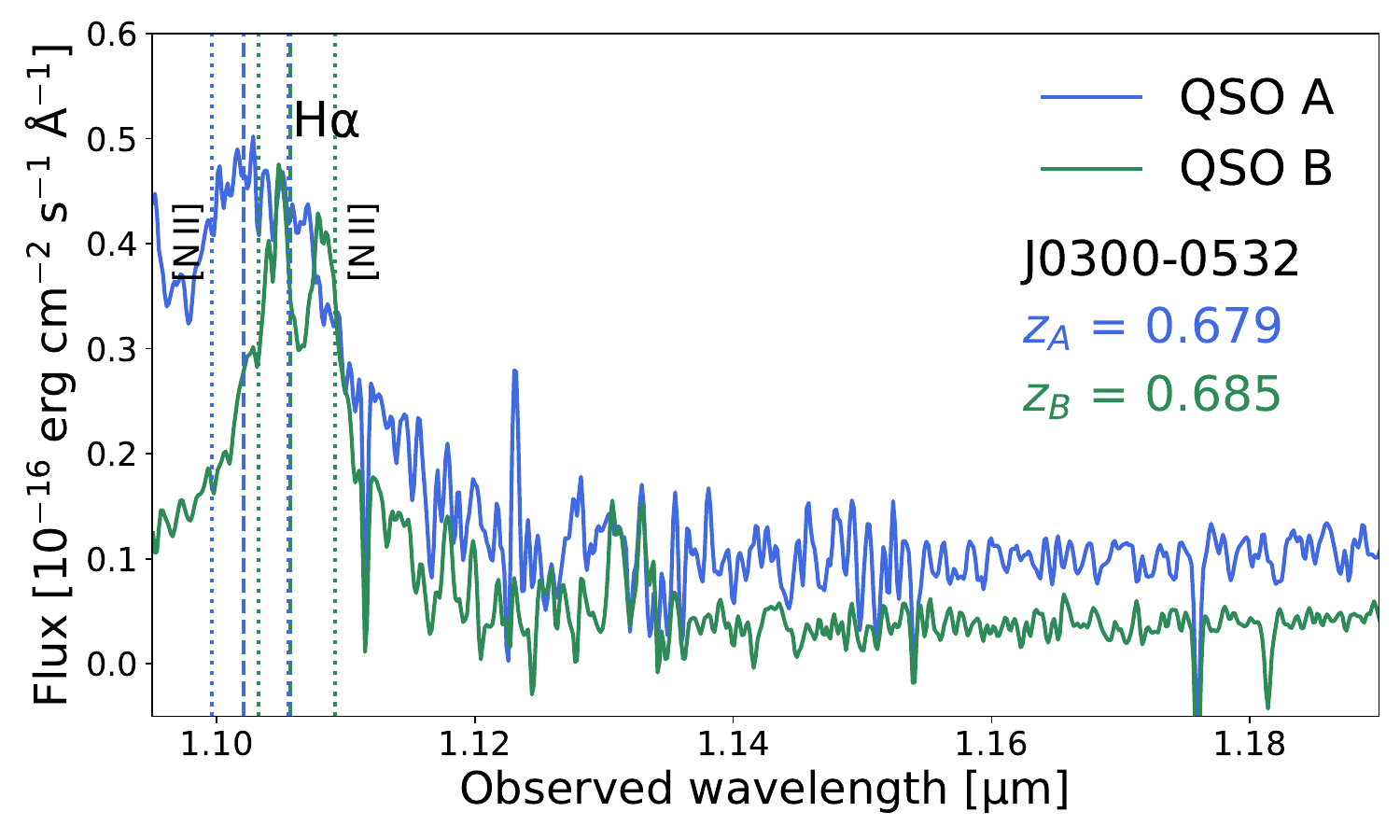}
\includegraphics{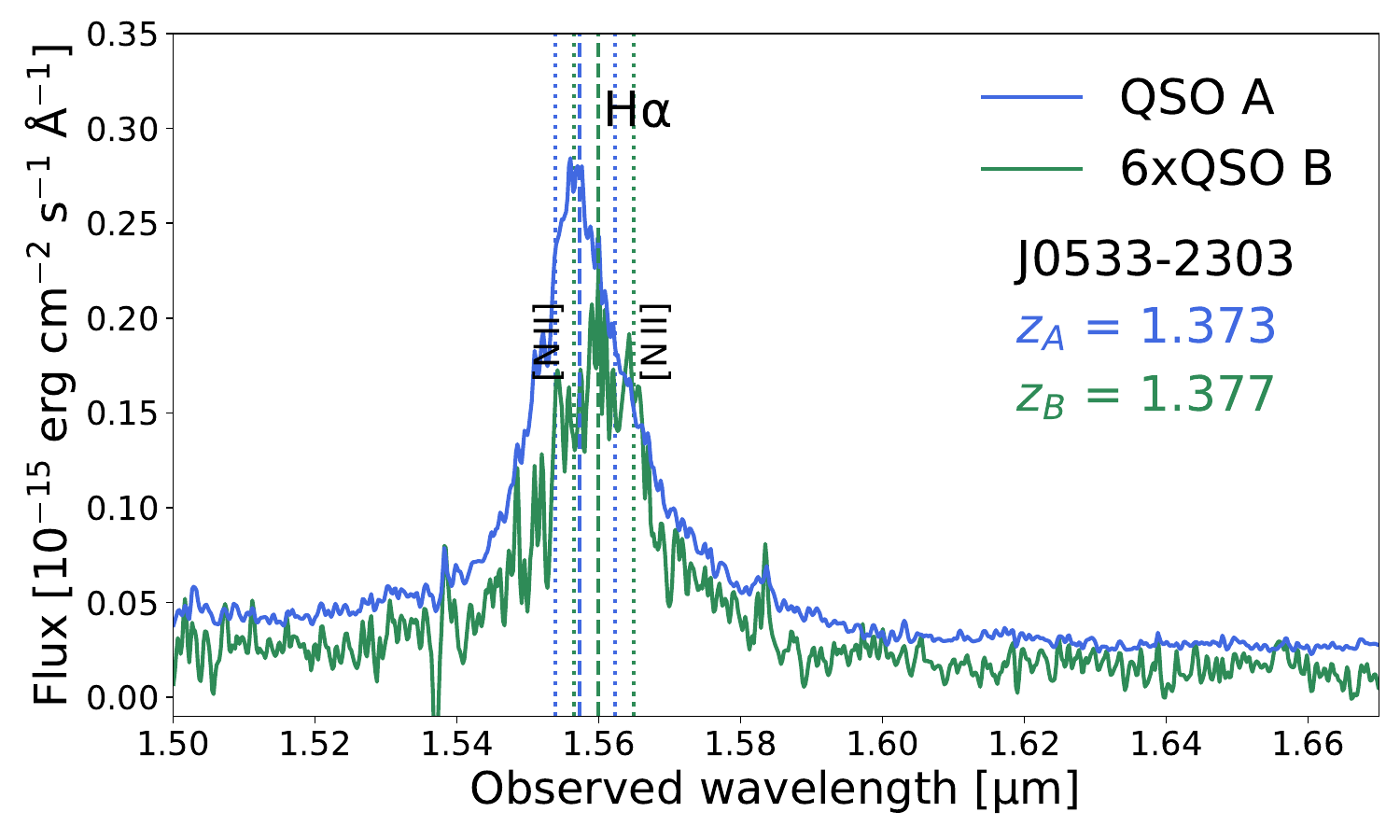}}
\resizebox{\hsize}{!}{
\includegraphics{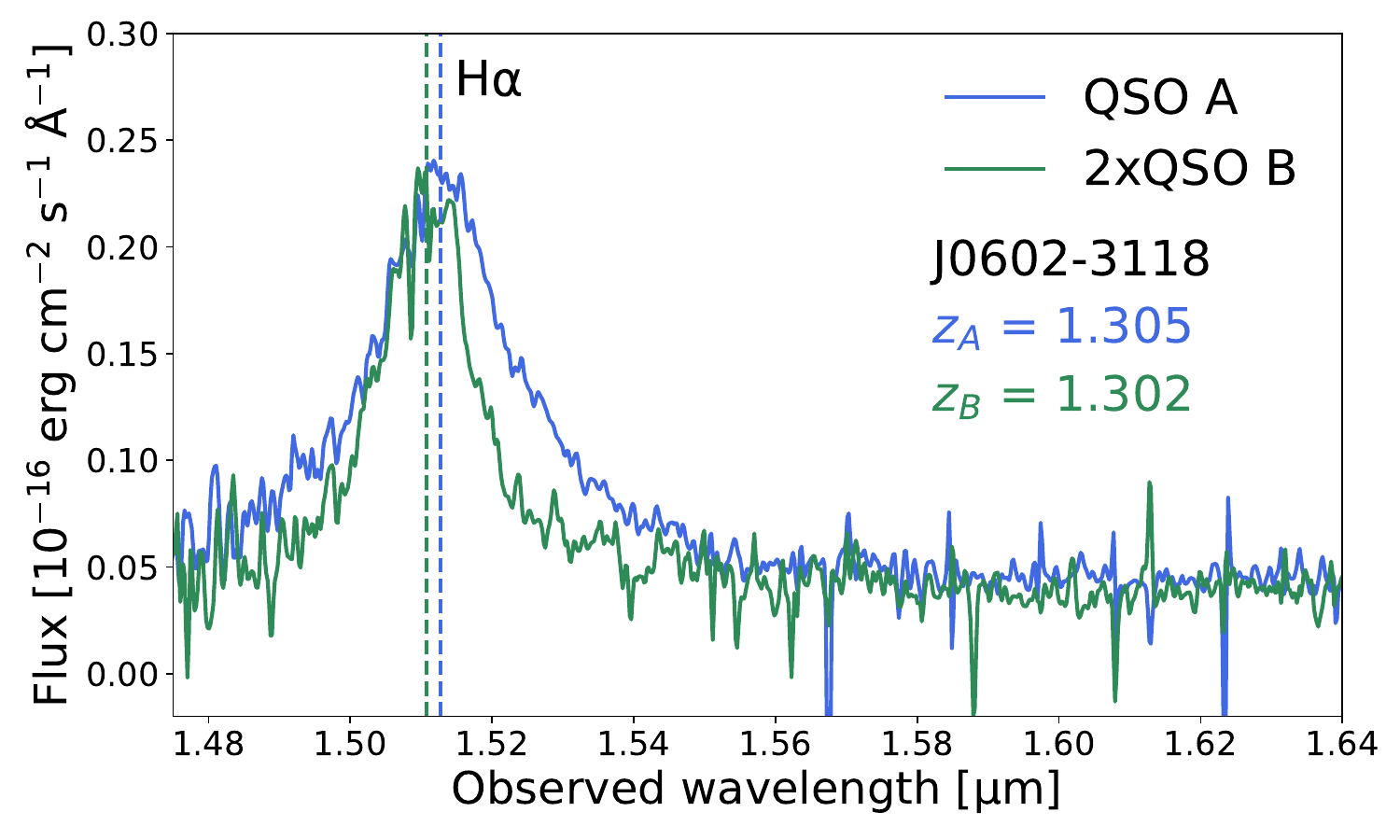}
\includegraphics{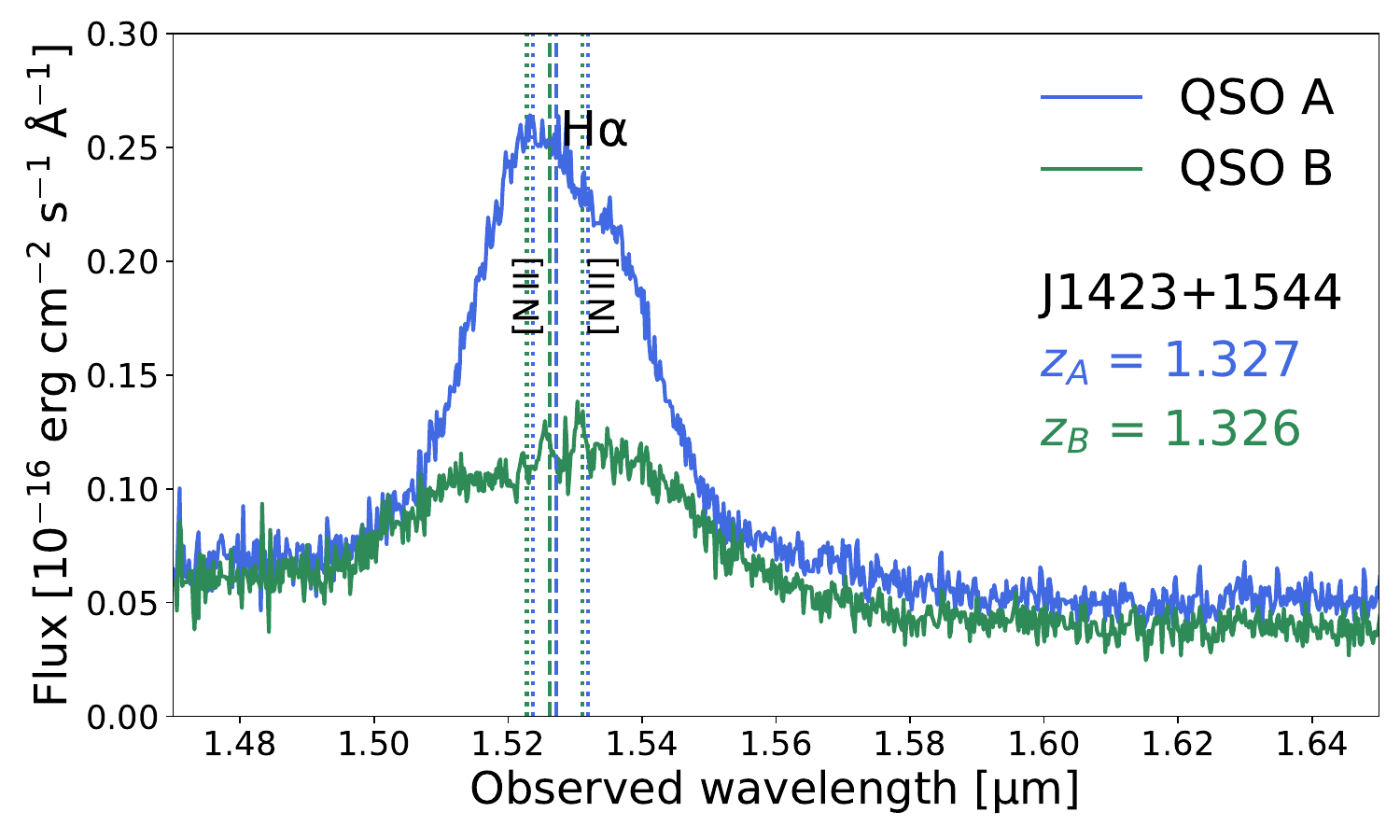}}
\begin{center}
\resizebox{0.5\hsize}{!}{\includegraphics{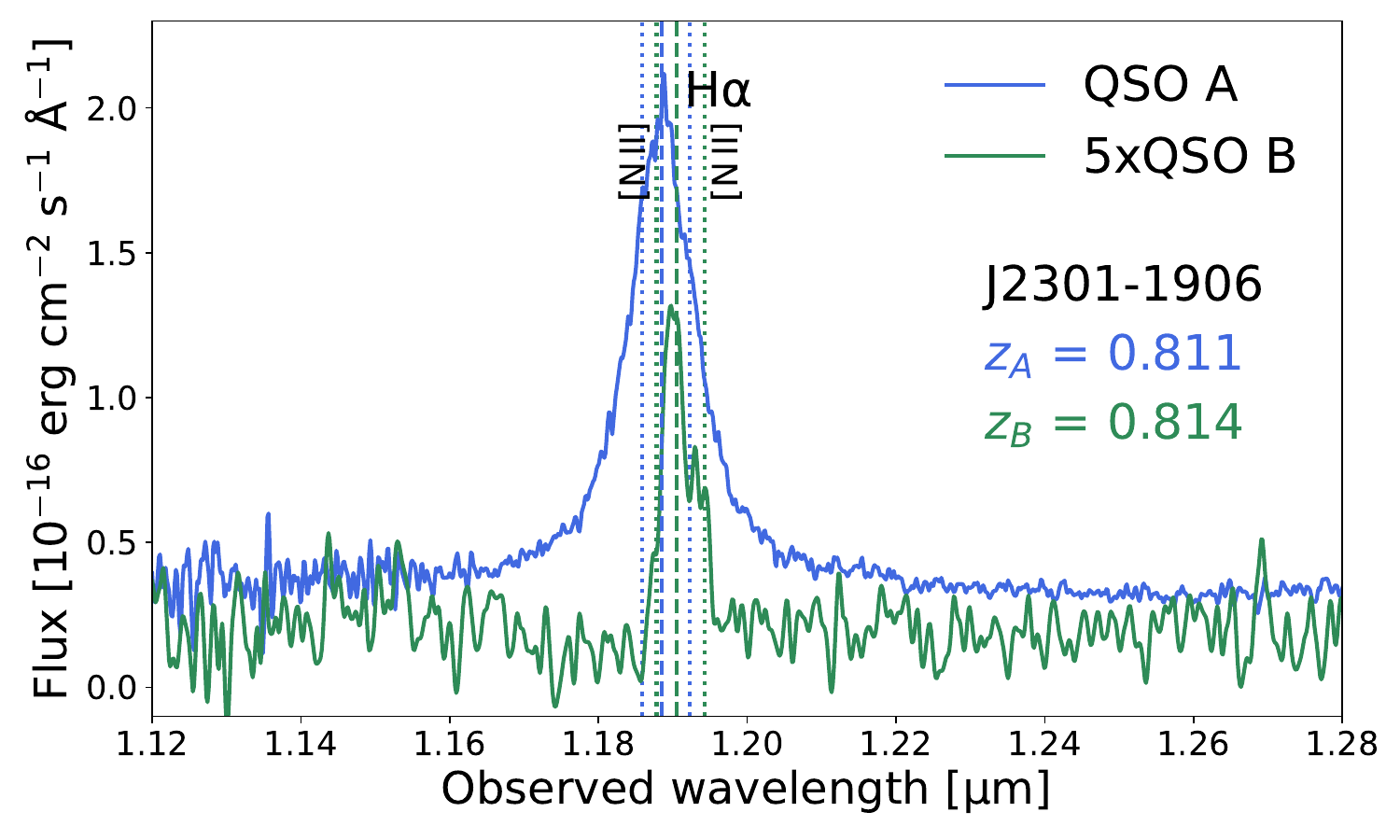}}
\end{center}
\caption{Spectra of the dual AGN systems and J2301--1906 observed with ERIS (target name and redshift are reported in the panels). They have been extracted using a combination of \texttt{2D-Moffat} functions in order to avoid contamination, as explained in the text. The spectra have the same color-coding of the labels of Figure \ref{fig:eris_maps}. Vertical dotted lines mark the position of the main expected emission lines.}
\label{fig:eris_spectra_dual}
\end{figure*}

\begin{figure*}
\resizebox{\hsize}{!}{
\includegraphics{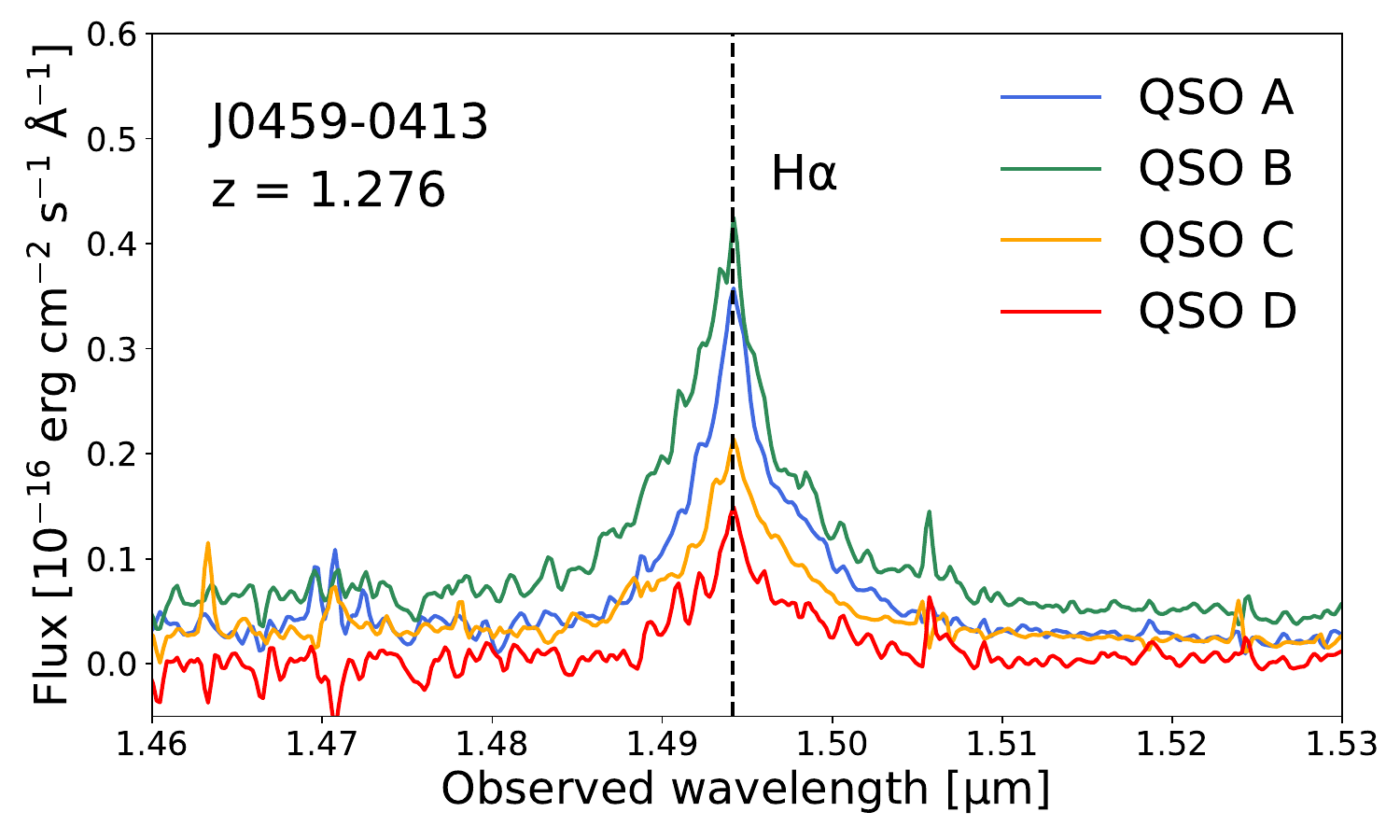}
\includegraphics{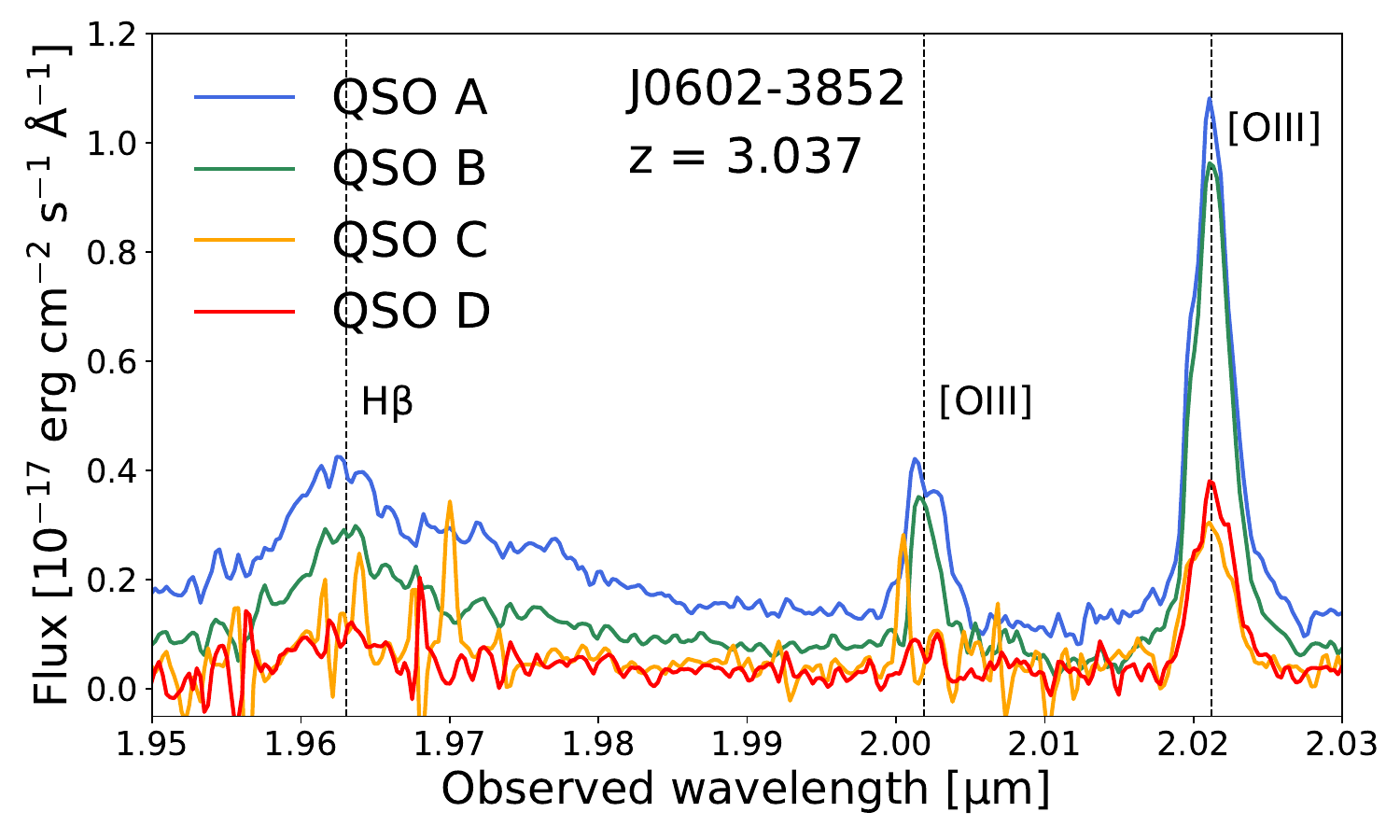}}
\resizebox{\hsize}{!}{
\includegraphics{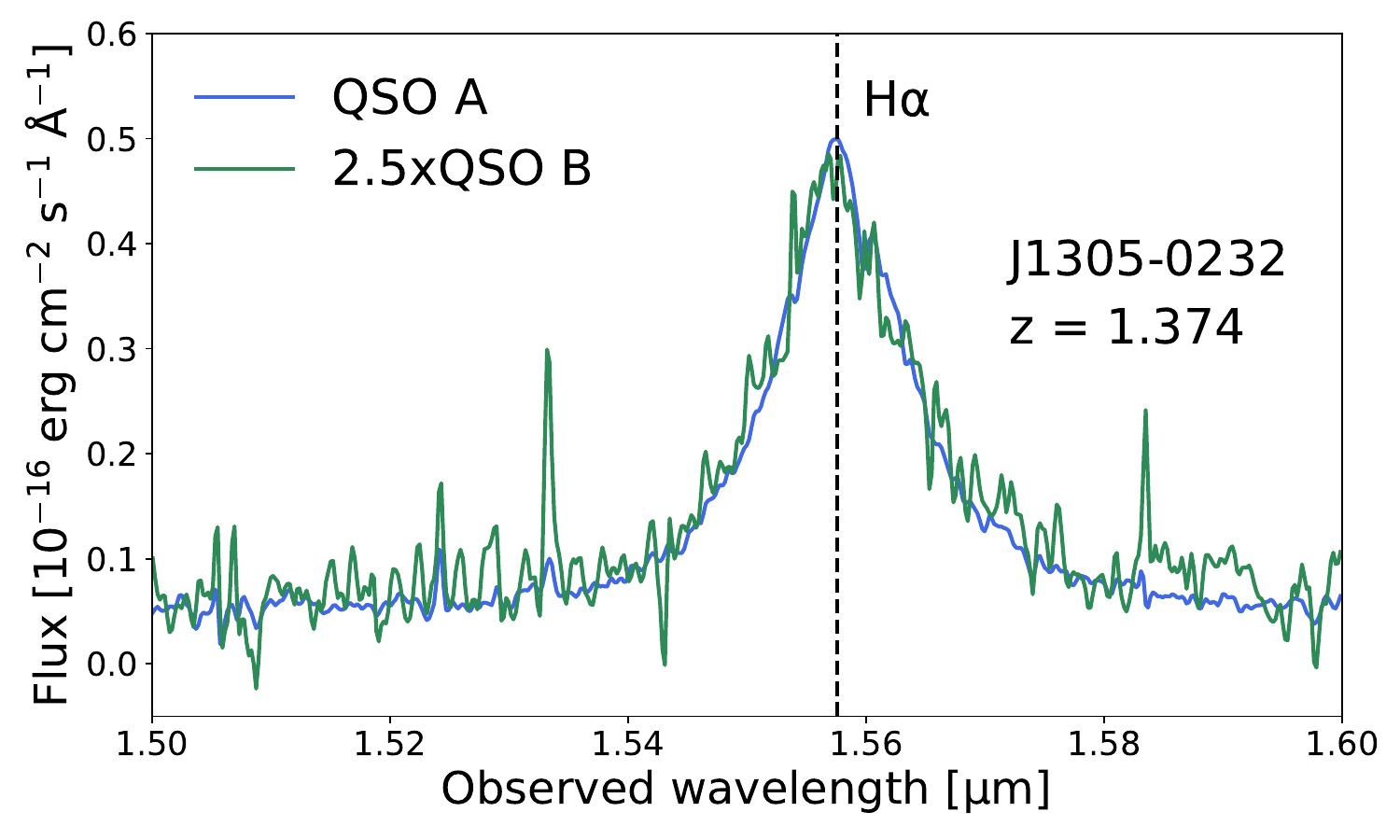}
\includegraphics{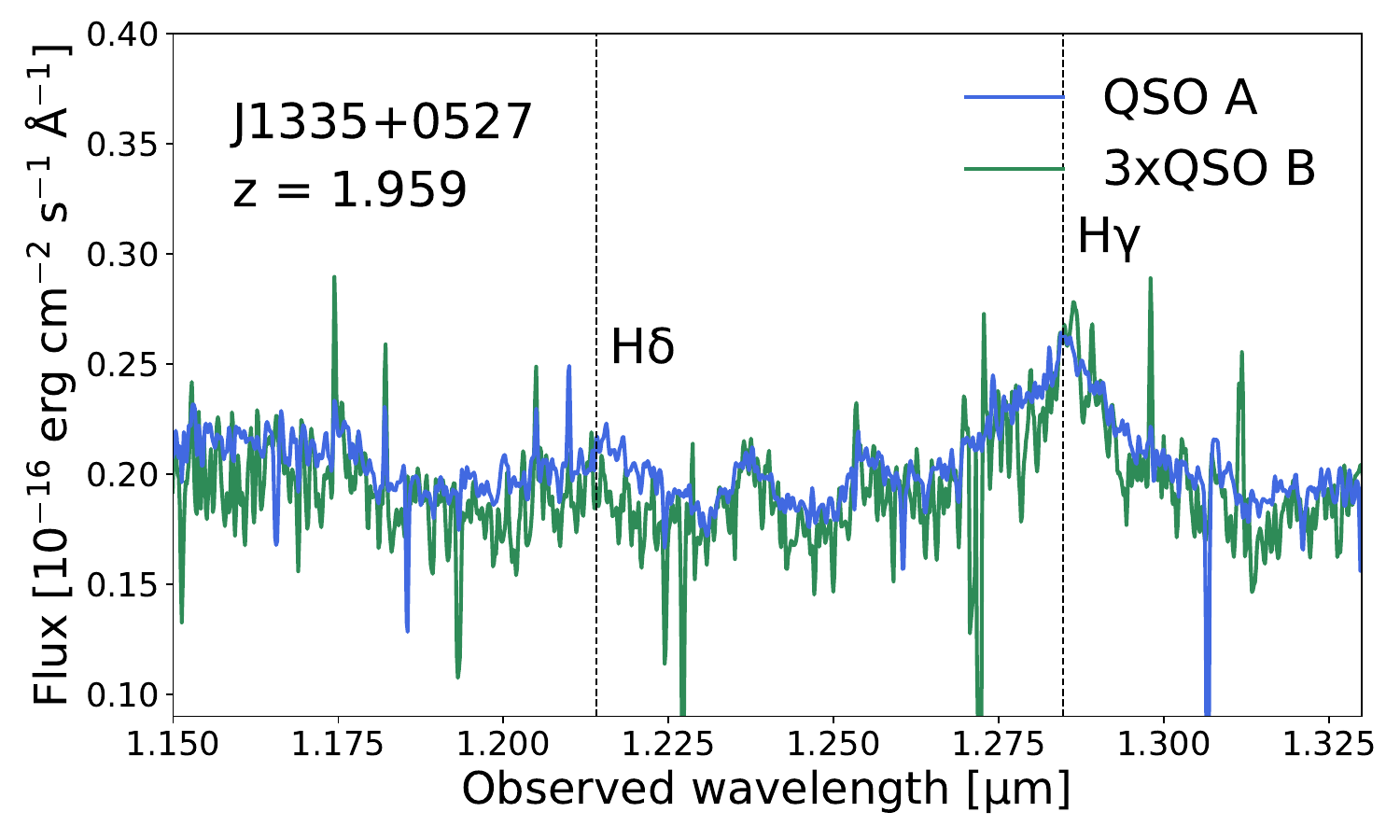}}
\resizebox{\hsize}{!}{
\includegraphics{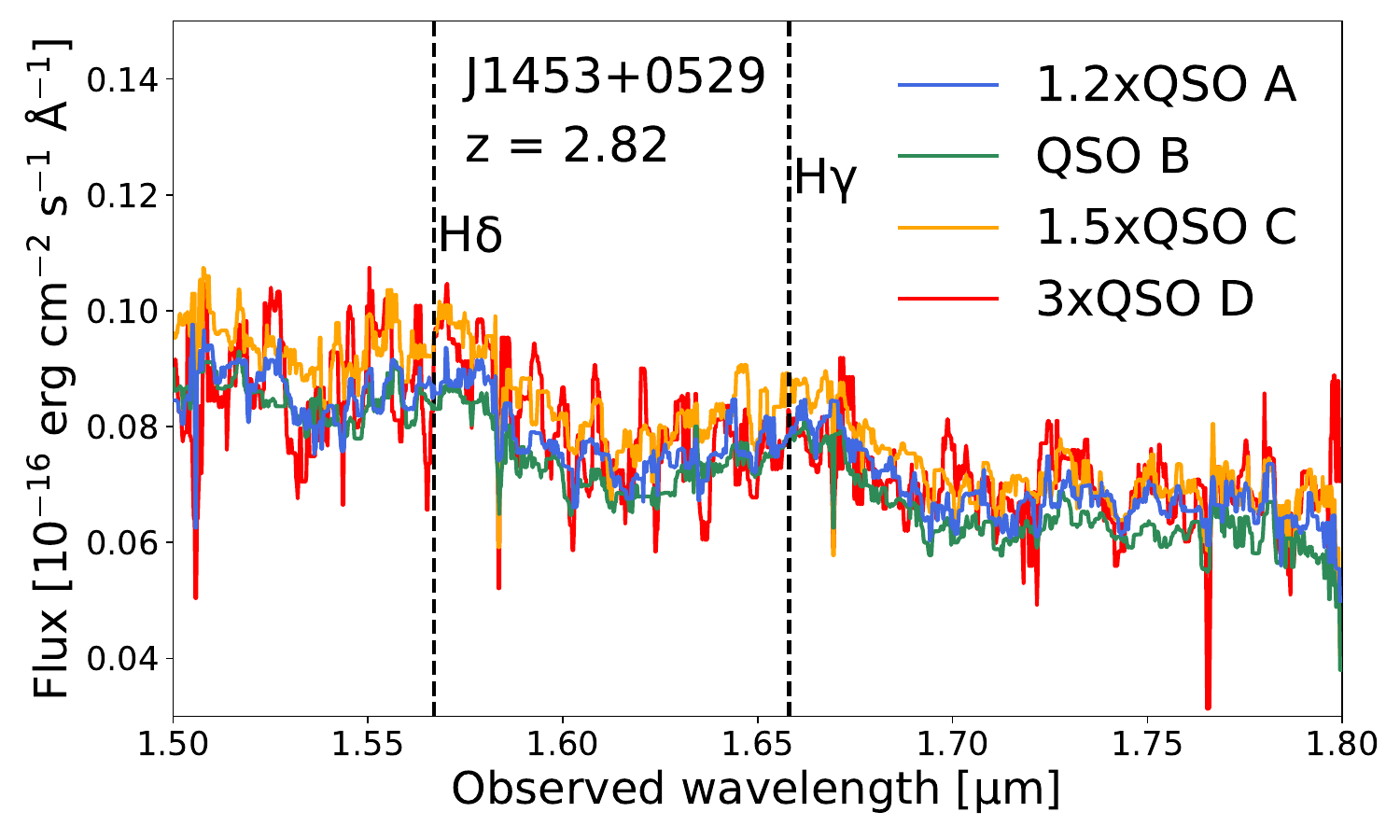}
\includegraphics{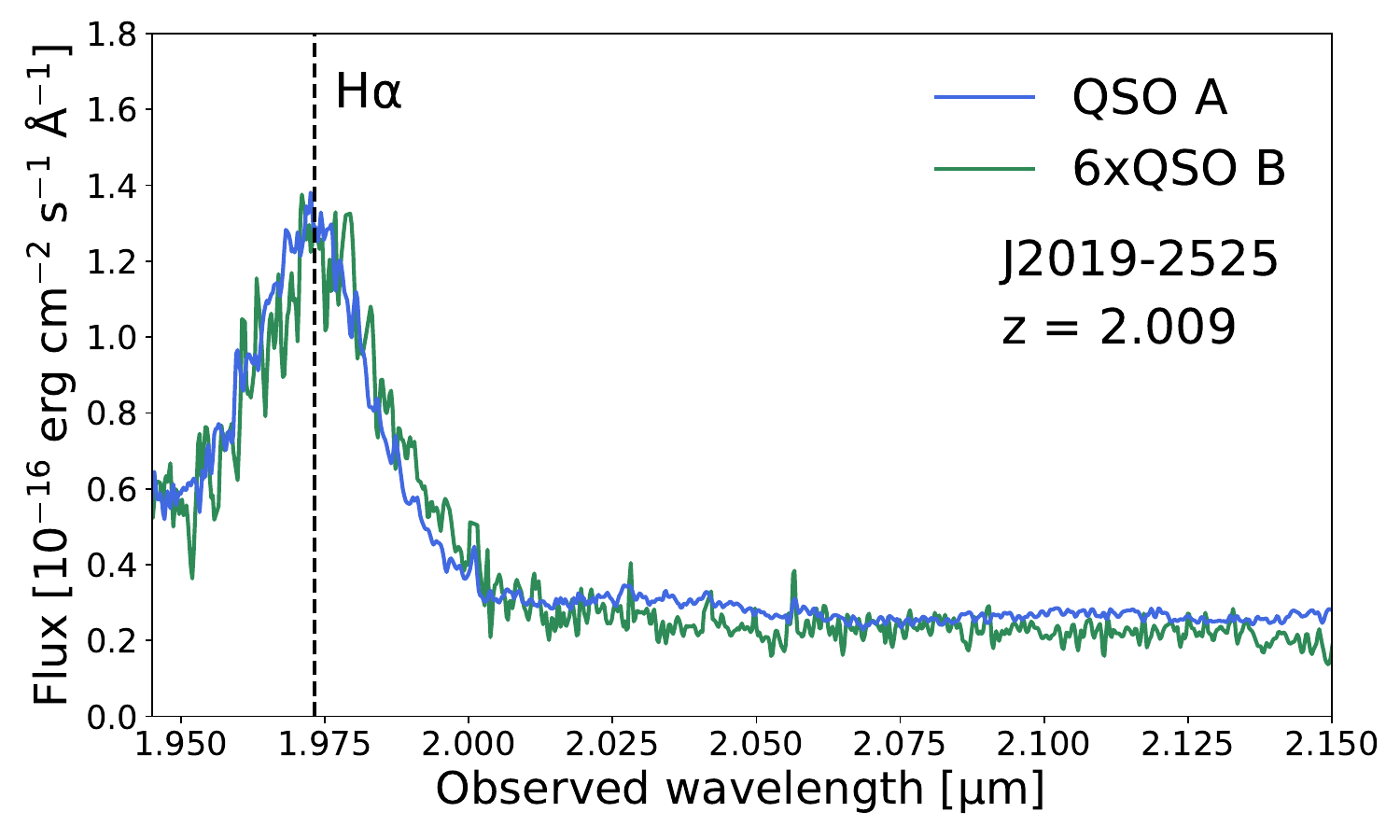}}
\begin{center}
\resizebox{0.5\hsize}{!}{\includegraphics{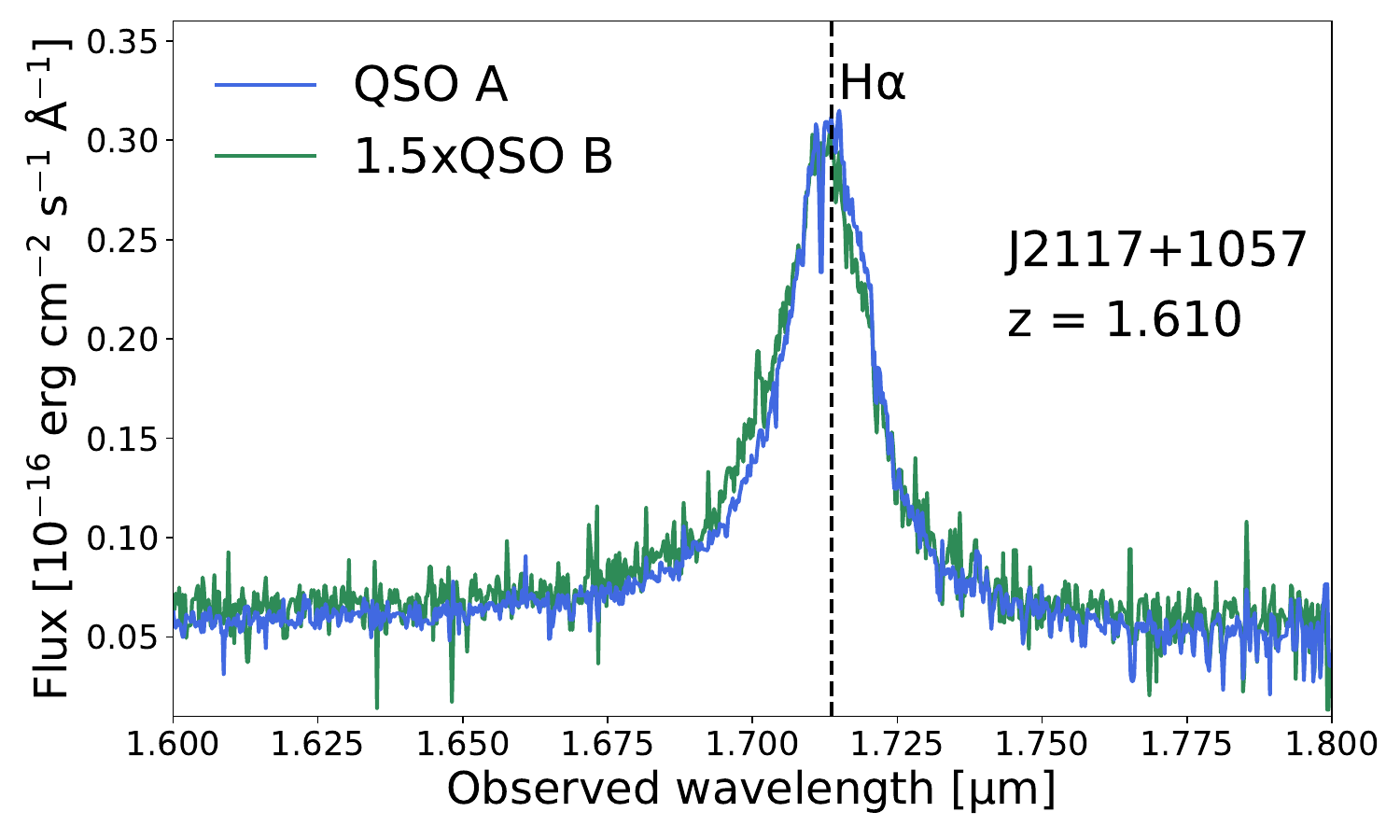}}
\end{center}
\caption{Same as Fig. \ref{fig:eris_spectra_dual} but for the systems classified as lensed AGN (both double and quadruple). The spectra have the same color-coding of the labels of Figures \ref{fig:eris_maps} and \ref{fig:eris_maps_quad}. For J1453+0529, a median filter was applied to the spectra for visualization purposes.}
\label{fig:eris_spectra_lens}
\end{figure*}

\section{Classification of the multiple systems}\label{sec:classification}

The classification is mainly based on the emission-line properties of the two components, including line-profile morphology, line widths, the presence of broad and narrow emission features, and their relative velocity offset. Since the spectral coverage of ERIS is limited, the continuum shape provides little diagnostic power and is therefore not used as a primary classification criterion.

\noindent The velocity offset $\Delta v$ is defined as the line of sight radial velocity difference between the two components, computed using eq. (10) in \citet{hogg1999} and using the redshift difference $|z_A - z_B|$ obtained through the template-fitting procedure described in Section~\ref{subsec:spectra}. Since the primary component, and in many cases also the secondary, hosts a type~1 AGN, the spectra are dominated by broad \ha\ emission arising from the broad-line region (BLR), which introduces uncertainties in the line-centroid determination. To assess the robustness of our $\Delta v$ measurements, we estimated the \ha\ line center independently using two additional methods: the flux-weighted centroid and the position of the line peak. The three methods yield consistent results when the emission line is symmetric, and well peaked; however, they can diverge in the presence of asymmetric profiles, as expected for type~1 AGN, where BLR kinematics, outflow components, or blended emission may shift the observed line centroid with respect to the systemic velocity. We adopt the $\Delta v$ values from the template fitting as our fiducial estimates, and the quoted uncertainties reflect the scatter among all three methods. 

Table~\ref{tab:sample} summarizes the classifications obtained for the 34 systems observed with ERIS--SPIFFIER analyzed in this work. We identify 13 systems resolved into multiple AGN-like components, either dual or lensed (see Table~\ref{tab:classified}); 18 systems classified as AGN+star, corresponding to chance alignments between an AGN and a foreground star (see Appendix~\ref{app:agn_star}); one system classified as a single AGN, with no clear evidence for a secondary component (see Appendix~\ref{app:singleAGN}); and two systems with an AGN primary and an unclassified secondary component (see Section~\ref{subsec:unclass}).

\subsection{Dual AGN}\label{subsec:dual}

We classify six systems as dual AGN; their spectra are shown in Figure~\ref{fig:eris_spectra_dual}, separations, redshift and velocity offsets are reported in Table~\ref{tab:classified}. In the following, we describe the properties of each system individually.
\paragraph{J0109--3529} The system is a blue, low-extinction QSO pair at z = 1.57. The ERIS \hlow\ data reveal two components separated by $\sim$6~kpc ($\sim$\SI{0.69}{\arcsecond}), with a velocity offset of $\Delta v = 290 \pm 160$~\kms. Component~A shows negligible dust attenuation, with $E(B-V) \simeq 0$, while component~B is moderately obscured, with $E(B-V) = 0.17^{+0.03}_{-0.01}$. 
The two components show very different luminosities, with component~A being $\sim$12 times more luminous than component~B.
The two spectra show clearly distinct line profiles of the \ha+[\ion{N}{ii}] complex. Component~A displays a broad, asymmetric line profile with a possible red wing and a FWHM of $\sim$2000~\kms.
Component~B shows a narrower and more symmetric line profile, in which both the peak of the narrow \ha\ emission and the [\ion{N}{ii}] doublet are clearly detected. In addition, the spectrum of component~B requires a broad \ha\ component with $\mathrm{FWHM}=3120\pm180$~\kms, blueshifted by $\sim$500~\kms\ with respect to the narrow \ha\ emission. This broad and velocity-shifted component may originate from the broad-line region, although a contribution from outflowing gas cannot be excluded.
The different line widths, profile shapes, attenuation properties, luminosities and velocity offset of the two components support the classification of J0109--3529 as a dual AGN rather than a lensed quasar.
\paragraph{J0216+1430} DESI-DR1 spectrum of J0216+1430 shows little evidence for significant dust attenuation. Consistently, the analysis described in Appendix~\ref{app:temple} yields low attenuation values for both components, with $E(B-V)_A = 0.11 \pm 0.01$ and $E(B-V)_B \simeq 0.0$. The system lies at $z\sim1$ and is resolved into two components separated by $\sim$\SI{0.55}{\arcsecond} (4.4~kpc), with a luminosity ratio close to unity. Both components show broad \ha\ emission, with $\mathrm{FWHM}_A \sim 4000$~\kms\ and $\mathrm{FWHM}_B \sim 2600$~\kms, and a significant velocity offset of $\Delta v \sim 800$~\kms\ ($z_A \sim 0.973$, $z_B \sim 0.978$). The high velocity shift, together with the different line widths of the two components, favors the dual AGN interpretation.
\paragraph{J0300--0532} The system was firstly observed with \textit{HST}/STIS long-slit spectroscopy using the G750L grating, as part of the Cycle~31 program 17423 (PI: G.~Tozzi). 
STIS spectra cover the \hb+[\ion{O}{III}] emission lines and the underlying continuum. The two AGN components clearly show different continuum and lines shapes thus confirming the dual AGN nature of the system. The velocity offset between the two AGN measured from the [\ion{O}{III}] emission lines is $\Delta v \sim 270$~\kms . The full analysis of the STIS data will be discussed in Tozzi et al., (in prep.).
Since the signal-to-noise ratio of the \hb\ line in the STIS spectrum is low, J0300--0532 was also observed with ERIS in the \jlow\ band, targeting the \ha+[\ion{N}{II}] emission lines. The high-quality \ha\ profile provides a more reliable estimate of the black hole mass through single-epoch virial methods \citep{scialpi2026_bhmass}.
From ERIS data we measure a projected separation of $\sim$\SI{0.47}{\arcsecond} (3.4~kpc at $z = 0.679$) between the two AGN. Figure~\ref{fig:j0300_temple_mcmc} shows the joint spectral fit of J0300--0532 system SDSS integrated spectrum and of ERIS spectra of the primary and secondary AGN. According to our analysis, the primary AGN is well modeled with $E(B-V)_A = 0.24 \pm 0.01$, while the secondary AGN does not show clear evidences for dust extinction.
Both AGN exhibit broad \ha\ emission lines shifted with respect to each other of $\Delta v = 950 \pm 250$~\kms\ ($z_A \sim 0.679$, $z_B \sim 0.684$). In addition, component~B shows a double-peaked profile, which correspond to the position of the \ha\ and [\ion{N}{II}]$\lambda6583$ lines at $z_B = 0.684$ . 
Unfortunately, the \ha\ emission line of component~A, and partially that of component~B, falls near the edge of the ERIS spectral band, resulting in large uncertainties in the measurements of the FWHM, luminosity ratio, and velocity offset. We obtain $\mathrm{FWHM}_A \sim 4600$~\kms, $\mathrm{FWHM}_B \sim 2700$~\kms, and a luminosity ratio of $\sim$2.
\paragraph{J0533--2303} is a dual AGN system at $z\sim1.37$, composed of two nuclei separated by $\sim$\SI{0.32}{\arcsecond} (2.7~kpc). The ERIS \hlow\ observations reveal broad \ha\ emission from both components, with no evidence for significant dust extinction in either AGN.
The two nuclei display markedly different broad-line morphologies. Component~A shows a highly asymmetric line profile, whereas component~B exhibits a more symmetric emission line with tentative evidence for a triple-peaked structure consistent with the position of the narrow line emission from \ha+[\ion{N}{II}]. The measured line widths are $\mathrm{FWHM}_A \sim 2800$~\kms\ and $\mathrm{FWHM}_B \sim 3000$~\kms. Despite their comparable line widths, the line luminosity of component~A is substantially larger, with a luminosity ratio of $\sim$8 between the two components.
We measure a velocity offset of $\Delta v = 435 \pm 90$~\kms\ between the nuclei, corresponding to $z_A \sim 1.373$ and $z_B \sim 1.377$. The combination of distinct broad-line profiles, different line luminosities, and a significant velocity shift supports the presence of two physically associated active nuclei and therefore the classification of J0533--2303 as a dual AGN.
\paragraph{J0602--3118} ERIS \hlow\ observations of J0602--3118 resolve two AGN separated by $\sim$\SI{0.26}{\arcsecond}, corresponding to a projected distance of 2.16~kpc. The system lies at $z\sim1.30$, and both components show broad \ha\ emission. A moderate amount of dust extinction is required by the spectral modeling, with $E(B-V)_A = 0.20^{+0.01}_{-0.03}$ and $E(B-V)_B = 0.08^{+0.01}_{-0.02}$.
Although the \ha+[\ion{N}{II}] complex profiles of both AGN are relatively symmetric, they differ in both shape and width. The primary nucleus displays a broader shape, with $\mathrm{FWHM}_A \sim 4100$~\kms, while the secondary AGN has $\mathrm{FWHM}_B \sim 3000$~\kms. The luminosity is also larger in the primary component, exceeding that of the secondary by a factor of $\sim$2.7.
The two broad-line systems are offset in velocity by $\Delta v = 400 \pm 70$~\kms, corresponding to redshifts of $z_A \sim 1.305$ and $z_B \sim 1.302$. The presence of two spatially resolved broad-line AGN with distinct spectral properties supports the classification of J0602--3118 as a dual AGN system.
\paragraph{J1423+1544} At a redshift of $z\sim1.33$, J1423+1544 is resolved into two AGN separated by \SI{0.80}{\arcsecond}, corresponding to a projected distance of 6.7~kpc. The spectral decomposition indicates little to no dust extinction in either component, consistent with intrinsically unobscured broad-line AGN.
The two nuclei exhibit different line profiles. Component~A is characterized by a complex, asymmetric broad emission line, with a dominant blueshifted peak accompanied by a weaker redshifted component. Component~B, on the other hand, displays an extremely broad but fainter \ha\ line, with a peak approximately 2.5 times lower than that of component~A. 
In addition, the spectrum of component~B shows two barely detectable narrow peaks, together with a tentative third peak at a position consistent with the expected redshifted wavelengths of the \ha+[\ion{N}{II}] doublet.
The broad-line widths are $\mathrm{FWHM}_A \sim 5900$~\kms\ and $\mathrm{FWHM}_B \sim 9700$~\kms, among the largest measured in our sample.
Although the broad-line centroids imply redshifts of $z_A = 1.327$ and $z_B = 1.326$, the peculiar morphology of the emission lines prevents a robust determination of the relative velocity between the two nuclei. Adopting the centroid of the broad \ha\ profile for component~A yields $\Delta v = 60 \pm 370$~\kms, consistent with no measurable velocity offset within the uncertainties. Not blended narrow emission lines would be required to obtain a more reliable estimate of the systemic redshift of each AGN. Nevertheless, the presence of two spatially distinct broad-line sources with substantially different spectral characteristics supports the interpretation of J1423+1544 as a dual AGN.

\subsection{AGN with unclassified secondary components}\label{subsec:unclass}

\paragraph{J0150+0809} system lies close to the limits of the GMP selection, with $G=20.4$ and \texttt{FMP}~=~8. It was observed with ERIS--SPIFFIER in both the \hlow\ and \jlow\ bands, covering the \ha+[\ion{N}{II}] and \hb+[\ion{O}{III}] spectral regions, respectively.
The continuum emission in both bands is dominated by a single component associated with the primary source. The presence of broad \ha\ and \hb\ emission confirms its classification as a type~1 AGN. In addition to the broad-line emission, visual inspection of the spectra reveals evidence for narrow emission-line components. However, a proper characterization of these features requires an accurate removal of the unresolved BLR contribution.
 
\noindent We modeled and removed the unresolved BLR emission independently in the \hlow\ and \jlow\ data cubes using the spectral regions around broad \ha\ and \hb, respectively. We followed the datacube-fitting procedure of \citet{marasco2020}, also adopted by \citet{Tozzi2021}. For each band, we first constructed a BLR template from the best-fitting nuclear model, including the broad Balmer emission, the [\ion{Fe}{ii}] emission where covered by the data \citep{kovacevic2010_feiitemplate}, and a polynomial continuum component. We then fitted each spaxel of the data cube with \texttt{pPXF} \citep{Cappellari2017} using a combination of this BLR template, allowed to vary only in normalization, and multiple Gaussian components describing the spatially resolved narrow-line emission. In the fit of the narrow-line component, the flux ratios among different atomic species were allowed to vary, while the central velocities and velocity widths of the Gaussian components were tied across all lines, assuming a common large-scale gas kinematics. Finally, the best-fitting BLR component was subtracted from the original data cube, producing a BLR--subtracted cube used for the subsequent analysis.
After subtraction of the BLR contribution, the residual cubes were analyzed through multi-Gaussian fitting of the narrow emission lines. In particular, we modeled \ha, [\ion{N}{II}]$\lambda\lambda6548,6583$, and [\ion{S}{II}]$\lambda\lambda6717,6731$ in the \hlow\ cube, and \hb\ and [\ion{O}{III}]$\lambda\lambda4959,5007$ in the \jlow\ cube.
The ratios of components of the [\ion{N}{II}] and [\ion{O}{III}] doublets were fixed to their theoretical values.
This analysis revealed a region located north-west of the broad AGN emission that exhibits several narrow emission lines. In particular, the \ha, [\ion{N}{II}]$\lambda6584$, \hb, and [\ion{O}{III}]$\lambda5007$ lines are blueshifted by $\sim$200~\kms\ with respect to the AGN systemic redshift ($z = 1.589$). Based on its distinct kinematics and spatial location, we identify this feature as a secondary component, hereafter referred to as the \emph{blueshifted clump}. 

\noindent Figure~\ref{fig:bpt} shows the continuum-- and BLR--subtracted spectra of the QSO (component A, magenta) and of the \emph{blueshifted clump}, together with the velocity map of the narrow \ha\ emission.
Emission-line ratios measured from the \emph{blueshifted clump} spectrum place it in the composite/H\,{\sc ii} region of the BPT diagrams \citep{Baldwin1981}, as shown in Fig.~\ref{fig:bpt}, indicating that its ionization is not consistent with pure AGN activity. Additionally, it does not exhibit any detectable continuum counterpart in either the \hlow\ or \jlow\ data. The lack of detectable continuum emission, together with its position in the BPT diagrams, suggests that the \emph{blueshifted clump} could be associated with a low-mass companion interacting or merging with the AGN.
To assess whether this component could account for the multiple \gaia~detections, we estimated its expected flux in the \gaia~$G$ band using photoionization models representative of star-forming regions \citep{Gutkin2016}. 
For this purpose we used models with emission-line ratios best matching those observed in our target. Specifically, the measured ERIS emission-line fluxes were used to normalize the models and predict the integrated rest-frame UV emission falling within the \gaia~$G$ band. We obtain an expected magnitude of $G\sim28$ mag, substantially fainter than the \gaia~detection limit.
Since the \emph{blueshifted clump} is not detected in the continuum and its predicted UV luminosity is far below the \gaia~sensitivity limit, it is unlikely that the \gaia~multiplicity is produced by this narrow-line emitting region identified in the ERIS data, thus origin of the multiple \gaia~detections remains unclear.
Given the extended nature of the \emph{blueshifted clump}, the absence of a detectable continuum counterpart, and its location in the BPT diagrams, is not included in Table~\ref{tab:classified}.
\paragraph{J2301--1906} At a redshift of $z\sim0.81$, J2301--1906 is resolved into two components separated by $\sim$\SI{0.52}{\arcsecond} (3.93~kpc). The spectral fit of the primary component requires moderate attenuation, with $E(B-V)_A = 0.33 \pm 0.01$, while the secondary component is significantly fainter, with a luminosity ratio of $\sim$16.
The primary AGN displays a clear broad component with $\mathrm{FWHM}_A \sim 2800$~\kms. The spectrum of the secondary component is instead dominated by narrow emission lines. We detect both \ha\ and \nii\ emission, with line widths of approximately 700 and 600~\kms, respectively. Although a weak broad \ha\ component may be present, its properties cannot be reliably constrained with the available data.
The redshifts measured for the two components are $z_A \sim 0.811$ and $z_B \sim 0.814$, corresponding to a velocity offset of $\Delta v = 590 \pm 170$~\kms. However, the classification of the secondary source remains uncertain. Since the observations only cover the \ha\ and [\ion{N}{ii}] spectral region, key diagnostic lines such as \hb\ and [\ion{O}{III}] are unavailable, preventing the use of standard emission-line diagnostic diagrams \citep[BPT diagrams;][]{Baldwin1981} based on $\log_{10}([\ion{O}{III}]\lambda5007$/\hb) versus $\log_{10}([\ion{N}{II}]\lambda6583$/\ha) to discriminate between AGN ionization and star formation. Additional spectroscopic observations covering these emission lines are therefore required to clarify the nature of the secondary source and securely establish the dual-AGN nature of the system. The system is therefore classified as a dual AGN candidate in Table~\ref{tab:classified}.

Both J0150+0809 and J2301--1906 therefore consist of a primary AGN and a spatially distinct secondary component whose nature cannot be unambiguously established from the available observations. We group them observationally as AGN + unclassified secondary systems (Table~\ref{tab:sample}). However, their secondary components show different properties. In J2301--1906, the secondary component has emission-line properties that make an AGN interpretation plausible, but the limited spectral coverage prevents a definitive classification; we therefore classify the system as a dual AGN candidate. In J0150+0809, instead, the secondary component is extended, lacks a detectable continuum counterpart, and has emission-line ratios inconsistent with pure AGN ionization.

\begingroup
\setlength{\tabcolsep}{8pt}
\renewcommand{\arraystretch}{0.9}
\begin{table*}
\caption[]{Results of the analysis of systems classified as dual AGN, double lensed and quadruply lensed AGN.}\label{tab:classified}
\centering                          
\begin{tabular}{c c c c c c c c c c c}        
\hline  
\hline
Target & Emission & PA & \multicolumn{2}{c}{Separation}  & \multicolumn{2}{c}{Redshift} &  $\Delta v$ & Class. & Mag. \\
& Line & $[^\circ$] (N→E)& [\SI{}{\arcsecond}] & [kpc] & $z_A$ & $z_B$ & [\kms] & &  \\
\hline
J0109--3529 & \ha & $-$137 & 0.69 & 5.85 & 1.570 & 1.572 & 290$\pm$160 & Dual AGN & 17.90$^{**}$ \\ 
J0216+1430 & \ha & 168 & 0.55 & 4.35 & 0.973 & 0.978 & 800$\pm$150 & Dual AGN & 18.16$^{**}$ \\
J0300--0532 & \ha & $-$29 & 0.47 & 3.40 & 0.679 & 0.684 & 950$\pm$250 & Dual AGN & 18.11$^*$ \\
J0459--0413 & \ha & $-$104 & 0.52 & -- & 1.276 & -- & -- & Quadruply lensed & 16.95$^{**}$ \\
J0533--2303 & \ha & 109 & 0.32 & 2.68 & 1.373 & 1.377 & 435$\pm$90 & Dual AGN & 15.88$^{*}$ \\
J0602--3118 & \ha & $-$140 & 0.26 & 2.16 & 1.305 & 1.302 & 400$\pm$70 & Dual AGN & 17.74$^{**}$ \\ 
J0602--3852 & \hb+[\ion{O}{III}] & $-$8 & 0.69 & -- & 3.037 & -- & -- & Quadruply lensed & 17.31$^{*}$ \\
J1305--0232 & \ha & $-$2 & 0.83 & -- & 1.374 & -- & -- & Lensed AGN & 17.02$^{**}$ \\ 
J1335+0527 & \hg,~\hd & $-$90 & 0.81 & -- & 1.959 & -- & -- & Lensed AGN & 17.61$^{**}$ \\
J1423+1544 & \ha & $-$95 & 0.80 & 6.69 & 1.327 & 1.326 & $<$370 & Dual AGN & 17.45$^{**}$  \\
J1453+0529 & \hg,~\hd & $-$143 & 0.21 & -- & 2.82 & -- & -- & Quadruply lensed & 16.89$^{**}$ \\
J2019--2525 & \ha & 150 & 0.78 & -- & 2.009 & -- & - & Lensed AGN & 15.20$^*$\\
J2117+1057 & \ha & 20 & 0.43 & -- & 1.610 & -- & -- & Lensed AGN & 17.21$^{**}$ \\
J2301--1906 & \ha & 85 & 0.52 & 3.93 & 0.811 & 0.814  & 590$\pm$170 & Dual AGN cand. & 16.98$^*$ \\
\hline
\end{tabular}
\tablefoot{The position angle (PA) is the angle measured clockwise from North toward East, calculated from the primary component A to the secondary B. The separation corresponds to the distance between the two components in case of a double system (lensed or dual AGN), to the radius of the circle enclosing all the four AGN images for J0459--0413 and J0602--3852, and to the Einstein ring radius in the case of J1463+0529 \citep{damato2026}. 
$^{(*)}$: Jap3, Hap3 or Kap3 magnitudes from the fifth data release (DR5) of the VISTA Hemisphere survey \citep[VHS;][]{mcmahon2013}. The 2MASS \citep{skrutskie2006} H magnitude was adopted for J0533--2303 since the Hap3 magnitude from VHS DR5 was not available. $^{(**)}$: J, H or K magnitudes derived from the available ones, see Appendix \ref{app:mag}. For J0602--3852, we report the results from the observation taken in \klow\ band, see Section \ref{subsec:lensed} for details on the \jlow~data. J1335+0527 is classified as double lensed AGN thanks to VLT/MUSE NFM data, see Sections~\ref{app:j1335+0527} and~\ref{subsec:lensed}. J0300--0532 was first observed and classified using \textit{HST}/STIS data (Tozzi et al. in prep.). }
\end{table*}
\endgroup

\subsection{Lensed AGN}\label{subsec:lensed}

We classify seven systems as lensed AGN, including four doubly imaged and three quadruply imaged systems. Their spectra and main properties are reported in Figure~\ref{fig:eris_spectra_lens} and Table~\ref{tab:classified}, respectively. In the following, we discuss the properties of each observed system.

\subsubsection{Double lensed AGN}

\paragraph{J1305--0232} is resolved into two components separated by $\sim$\SI{0.83}{\arcsecond} at $z = 1.374$. Both components show virtually identical AGN spectra, characterized by broad \ha\ emission lines. The spectra become almost indistinguishable after applying a normalization factor of $\sim$2.5, allowing us to establish the double lensed AGN nature of the system.
\paragraph{J1335+0527} At the source redshift of $z=1.959$, the \ha\ and \hb\ emission lines fall outside the ERIS spectral coverage. The ERIS observations therefore probe only the rest-frame optical region around \hg\ and \hd\ and the underlying continuum, allowing us to resolved the two components of the system at $\sim$\SI{0.81}{\arcsecond}. 
While the lack of strong emission lines prevents a robust classification, both components, after applying a flux-normalization factor of $\sim$3, display highly consistent continuum shapes  with no apparent spectral differences. Consequently, although the ERIS data alone cannot provide a definitive confirmation, the striking similarity between the two spectra strongly points towards a gravitational lensing scenario rather than an intrinsic AGN pair, while allowing us to exclude the possibility of an AGN+star superposition. 

\noindent J1335+0527 was subsequently observed with VLT/MUSE as part of the \textit{Cosmic Duets} ESO Large Program (114.27BY.001, PI: Scialpi; \citealt{scialpi2026}), see Section~\ref{app:j1335+0527}.
MUSE spectra of the two components show identical broad \ion{C}{III}], \ion{C}{II}], and \ion{Mg}{II} emission lines. The MUSE white-light image and spectra are shown in Figure~\ref{fig:j1335+0527_muse_map_spec}. In addition, the continuum shapes are consistent after applying a flux-normalization factor of $\sim$3.5. Furthermore, the lensing galaxy is detected in the MUSE data and is highlighted in purple as component~G in Figure~\ref{fig:j1335+0527_muse_map_spec}. The combination of identical broad emission lines, consistent continuum shapes, and the direct detection of the lensing galaxy confirms that J1335+0527 is a gravitationally lensed AGN.
\paragraph{J2019--2525} The system is resolved into two AGN-like components separated by $\sim$\SI{0.78}{\arcsecond} at redshift $z = 2.009$.
In this case, the two spectra exhibit almost identical profile of the \ha+[\ion{N}{ii}] doublet complex. The spectra become consistent after applying
a normalization factor of $\sim$6 to component~B. This supports the classification of the system as a double lensed AGN.
\paragraph{J2117+1057} At a redshift of $z\sim1.61$, J2117+1057 is resolved into two components separated by $\sim$\SI{0.43}{\arcsecond}.
The spectra of the two components are remarkably similar, with nearly identical line profiles. A simple normalization by a factor of $\sim$1.5 is sufficient to match the two spectra, consistent with the double lensed AGN classification.

\subsubsection{Quadruply lensed AGN}

\paragraph{J0459--0413} is a quadruply imaged AGN at $z=1.276$ with a radius of the circle enclosing all the four AGN images of $0.52''$. The ERIS observations detect broad \ha\ emission from all four lensed images. Three images form a compact group, while the fourth image is located to the north at a larger separation. All four spectra are remarkably similar, showing consistent line profiles and continuum shapes. The tight spectral agreement among the four components allow us to classify the systems as a quadruply lensed AGN.
The lensing galaxy is also detected in the ERIS data. Although its spectrum does not allow a robust measurement of the lens redshift because of the lack of detectable stellar absorption features, its detection demonstrates the capability of ERIS to detect the lens galaxy even in compact quadruple systems
\paragraph{J0602--3852} The system is a quadruply imaged AGN at $z=3.037$, with a radius of the circle enclosing all the four AGN images of $0.69''$. In the ERIS \klow\ band, we detect strong broad \hb\ emission together with the [\ion{O}{III}]$\lambda\lambda4959,5007$ doublet in all four images. 
The emission-line profiles and relative intensities are consistent across the system, indicating that all four components originate from the same background source.
The system was also observed in the \jlow~band for a total integration time of 2700~s, targeting the \ion{Mg}{II} emission line.
In the \jlow~band, \ion{Mg}{II} emission is detected only in the two brightest images, A and B, while it is not significantly detected in components~C and~D. From the analysis of the \jlow~image, we obtain a ${\rm PA} = -12^{\circ}$ between components A and B and a radius of the circle enclosing the AGN images of $\rm r = \SI{0.68}{\arcsecond} \pm \SI{0.14}{\arcsecond}$. Since the \ion{Mg}{II} emission is barely detected at the positions of C and D, for this target we adopt the parameters derived from the analysis of the \klow~band data.
The lensing galaxy is clearly detected in the \klow\ data, where it is identified as component~G and highlighted in purple in the map shown in Figure~\ref{fig:eris_maps_quad}. However, its spectrum does not exhibit significant stellar absorption features, preventing a robust measurement of the lens redshift.  
We therefore securely classify J0602--3852 as a high-redshift gravitationally lensed AGN, the most distant system in our sample.
\paragraph{J1453+0529} \citet{damato2026} recently reported the discovery of the quadruply imaged quasar J1453+0529 based on ERIS/NIX $K_{\rm s}$-band imaging part of the INAF ERIS GTO program (P.I. F. Mannucci). With an Einstein radius of $r_{\rm E}=\SI{0.21}{\arcsecond}$, corresponding to a projected physical scale of $\sim$1.71~kpc at the source redshift of $z=2.82$, J1453+0529 is the smallest-separation quadruply imaged quasar known to date.
In this work, we present complementary ERIS--SPIFFIER observations obtained in the \hlow\ band, covering the spectral region around the \hg\ and \hd\ emission lines. Using the spectral modeling approach described in Section~\ref{subsec:spectra}, we derive a robust spectroscopic redshift of $z=2.82\pm0.01$ for the background quasar. This precise redshift measurement provides a fundamental constraint for the lens reconstruction and was adopted in the lens modeling presented by \citet{damato2026}.

\section{Discussion}\label{sec:discussion}

\begin{figure}
\resizebox{\hsize}{!}{
\includegraphics{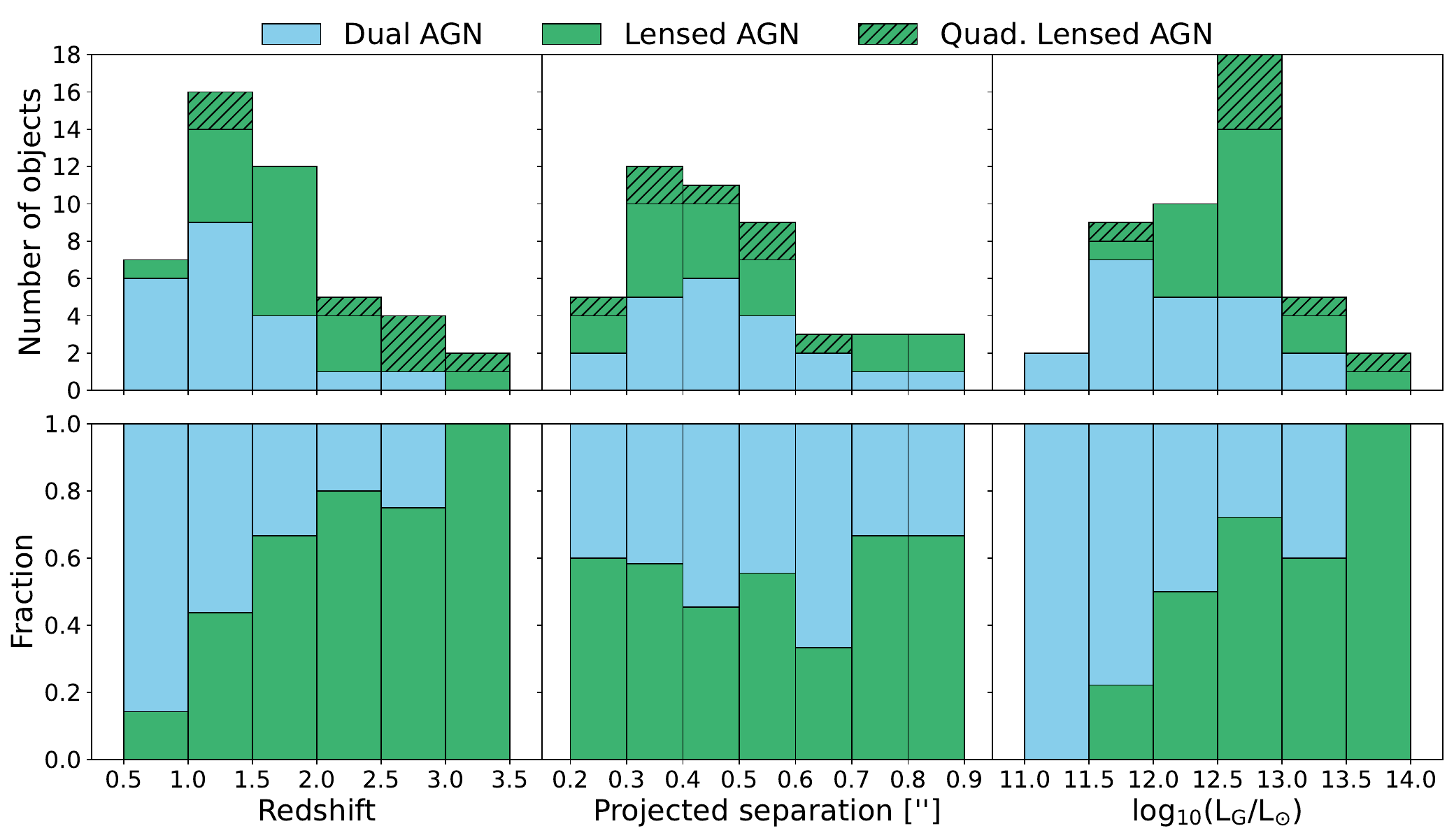}}
\caption{Number distributions and fractions of dual and lensed AGN in the sample of GMP systems classified either as dual or lensed (double or quadruple) AGN. The sample includes the systems listed in Table~\ref{tab:classified} and the GMP systems classified as dual or lensed AGN in previous works \citep{ciurlo23, mannucci2023, scialpi24, scialpi2026}. Top panels: Number distributions of dual AGN and the dual AGN candidate J2301--1906 (blue) and lensed AGN (green) as a function of redshift, projected separation, and total \gaia~$G$-band luminosity. Quadruply imaged lensed systems are shown in green with black hatching. Bottom panels: Fractions of dual AGN (blue) and lensed AGN (green) in this sample as a function of the same quantities. For quadruply imaged lensed systems, the radius of the circle enclosing all four AGN images is used as a proxy for the projected separation. In the case of J1453+0529 and J1608+2716 the Einstein ring radius calculated by \citet{damato2026} and \citet{Livodka2023} is used.}
\label{fig:histo}
\end{figure}

Figure~\ref{fig:histo} shows the number distributions (top panels) and fractions (bottom panels) of dual AGN and the dual AGN candidate J2301--1906 (blue) and lensed AGN (green) relative to the total sample of GMP multiple AGN systems (i.e. dual+dual cand.+lensed) as a function of redshift, projected separation, and total \gaia~$G$-band luminosity.
In the top panels we also distinguish between double lensed AGN (green) and quadruply lensed AGN (green with black hatching).
When multiple components are present in the \gaia~archive, the total luminosity is computed by converting the individual $G$-band magnitudes into luminosities and summing their contributions. 
The sample comprises 20 dual AGN, the dual AGN candidate J2301--1906, 18 doubly imaged lensed AGN, and 7 quadruply imaged lensed AGN, all selected through the GMP method, combining the results obtained with VLT/ERIS in this work with those from Keck/OSIRIS \citep{ciurlo23, mannucci2023} and VLT/MUSE \citep{scialpi24, scialpi2026}.
The quadruply imaged lensed systems are included by adopting the radius of the circle enclosing all four AGN images as a proxy for the image-separation scale. In the case of J1453+0529, we adopt the Einstein ring radius calculated by \citet{damato2026}.
The GMP system J1608+2716, at $z = 2.575$, was reported as a triple-quasar candidate with sub--arcsecond separations ($\sim$\SI{0.25}{\arcsecond}) based on Keck AO-assisted IFU spectroscopy \citep{ciurlo23}; deep \textit{HST} imaging subsequently revealed a fourth point-like component, indicating that J1608+2716 is a single quasar lensed into four images \citep{Livodka2023}. We therefore adopt $r_{\rm E} = \SI{0.42}{\arcsecond}$ \citep[${\sim}3.4$~kpc at $z = 2.575$;][]{Livodka2023} as proxy for the separation in this system.

\noindent Figure~\ref{fig:histo} shows that the fraction of dual AGN in our sample is higher at lower redshifts ($z < 1.5$) and at lower \gaia\ $G$-band luminosities ($\log (L_G/L_\odot) < 12.5$), while the fraction of lensed AGN increases toward higher redshifts and higher luminosities. 
To quantitatively test these trends despite the relatively small sample size, we performed a two-sample Kolmogorov-Smirnov (K-S) test. The test indicates that these differences are statistically significant, yielding $p$-values of $p_z = 0.005$ and $p_L = 0.003$ for the redshift and luminosity distributions, respectively. 
This trend is qualitatively expected from magnification bias: gravitational lensing preferentially amplifies the flux of background sources and can therefore boost lensed quasars above the flux limit of the GMP selection, making them over-represented in flux-limited samples.
However, the increase in lensed systems at higher redshift may also reflect the increasing geometric lensing probability with source redshift. The luminosity trend could also partly reflect the $L_G$--$z$ correlation in flux-limited samples (Malmquist-type bias) rather than an independent luminosity dependence. 

\noindent The distribution of dual versus lensed systems as a function of projected separation does not show a clear trend over the range probed ($0.2''$--$0.9''$). 
Indeed, a two-sample K-S test on the projected separations yields $p_{\rm sep} = 0.61$ indicating no significant difference between the two populations. While the limited sample size prevents us from drawing definitive conclusions, no distinct separation-dependent trend is observed. This comparison, restricted to projected angular separation, is intended to test for possible observational selection biases rather than to compare physical separations.

The AGN+star contamination fraction varies across the different instruments and selection sub-samples: ${\sim}40\%$ in the VLT/MUSE-based follow-up \citep{scialpi24, scialpi2026}, and ${\sim}50\%$ in the combined Keck/OSIRIS and VLT/ERIS samples \citep{ciurlo23, mannucci2023}, including this work. At wider separations, the GMP selection becomes increasingly sensitive to chance projections between an AGN and a foreground star, since the contamination probability increases monotonically with $r$ (scaling as ${\sim}r^2$ in the small-$r$ limit; \citealt{mannucci2022, scialpi2026}).
To quantify this behavior across the full combined sample, we fit the empirical contamination model given by Eq.~(3) of \citet{scialpi2026} to the binned AGN+star fraction as a function of projected separation, using all systems observed with VLT/MUSE, VLT/ERIS, and Keck/OSIRIS. This functional form describes the probability of a chance alignment between an AGN and a foreground star within a projected radius $r$, where $D$ represents the mean distance to the nearest stellar contaminant. We obtain a best-fit value of $D \sim 0.60''$. At the mean separation of our sample ($\bar{r} \simeq 0.5''$), the resulting contamination probability is $P(\bar{r}) \simeq 0.43$. This value is higher than the ${\sim}30\%$ stellar-contamination fraction assumed by \citet{mannucci2022}.
A fraction of these contaminants is removed prior to IFU follow-up by the spectral-decomposition analysis \citep[][Scialpi \& Marconcini, in prep.]{scialpi24, scialpi2026}, which was primarily used to rank candidates by their probability of stellar contamination. Nevertheless, observational constraints inherent to a GTO program, namely fixed observing nights and the consequent limited flexibility in target selection, meant that some lower-priority, higher-contamination candidates were still observed whenever higher-priority targets were unavailable. Consequently, a fraction of AGN+star contaminants remains in the final IFU sample. In addition, the faintest stellar contaminants are not detected by the spectral-decomposition analysis, and therefore cannot be excluded a priori.

Compared with classical strong-lens searches, which have built statistical samples of galaxy-scale lenses mostly at image separations larger than $\sim1''$ and source redshifts up to $z\sim2$ \citep[e.g.][]{Bolton2008_slacs, Faure2008, Inada2012, Shajib2021}, GMP-selected lensed systems extend the explored parameter space toward compact sub-arcsecond lensed quasars at higher redshifts up to $z\sim3$ \citep[][and this work]{ciurlo23, mannucci2023, scialpi24, scialpi2026}. Indeed, the GMP method targets Einstein radii smaller than the effective radius of the lens galaxy and therefore enables studies of its inner regions even at high redshift \citep[e.g.][$z_{\rm QSO}=2.82$, $r_E \sim 1.71$~kpc]{damato2026}.

The six dual AGN analyzed in this work exhibit line-of-sight velocity offsets spanning $\Delta v \sim 290$--$950$~\kms, as measured via template fitting, see Section~\ref{subsec:dual}. 
These values are broadly consistent with those expected for physically associated galaxy pairs undergoing mergers. Using interacting galaxy pairs with recent close passages ($<1$ Gyr) from the Illustris-TNG100 simulation \citep{Patton2024}, \citet{Pfeifle2025} found that most systems in the relevant projected-separation range exhibit line-of-sight velocity offsets of $|\Delta v| \lesssim 600$~\kms.
Four of our six dual AGN fall within this range, while the other two systems lie above this threshold, J0300$-$0532 ($\Delta v \sim 950$~\kms) and J0216$+$1430 ($\Delta v \sim 800$~\kms). We caution, however, that all $\Delta v$ measurements are derived from the total \ha\ line profile, dominated by the BLR component, whose centroid can be shifted with respect to the systemic velocity by BLR kinematics, outflow components, or blended emission. Therefore, these velocity offsets should be interpreted as indicative broad-line shifts rather than precise systemic velocity differences. In the absence of spectrally resolved narrow emission lines, the systemic redshift of each component cannot be determined unambiguously and the quoted $\Delta v$ values may be affected by large uncertainties.

The dust attenuation values across the dual AGN sample are modest, with $E(B-V)$ ranging from $0.0$ to ${\sim}0.3$, indicating that GMP-selected systems predominantly probe the less obscured end of the merger sequence. Indeed, \gaia-based approaches are mainly sensitive to unobscured, optically bright type~1 AGN.
Late-stage mergers are expected to be associated with increased gas column densities and dust obscuration \citep{Koss18, Blecha18, Ricci17, derosa2023, Battistini2026}, implying that a population of compact dual AGN in heavily obscured systems likely remains undetected by GMP-based searches. Complementary approaches based on radio/infrared selection and high-resolution imaging, including upcoming or ongoing searches with \textit{Euclid} and dedicated obscured-AGN selections, are therefore essential to extend the census to fainter or dustier merger phases \citep[e.g.][]{Ulivi2025, damato2026_skao}.

Finally, black-hole masses, bolometric luminosities, and Eddington ratios for the dual AGN will be presented in \citet{scialpi2026_bhmass}, together with the GMP dual AGN reported in \citet{mannucci2023, ciurlo23, scialpi24,scialpi2026}.

\section{Conclusions}\label{sec:concl}

We have presented the analysis of 34 sub-arcsecond dual AGN candidates selected through the \gaia\ Multi-Peak (GMP) method and observed with VLT/ERIS--SPIFFIER.

\noindent Among the 34 targets, 13 (${\sim}40\%$) are resolved into two or more AGN or AGN-like components, 18 (${\sim}53\%$) are identified as AGN+star superpositions, one (${\sim}3\%$) is classified as a single AGN, and two as systems with an AGN primary and an unclassified secondary component.
The fraction of single AGN systems is consistent with the false-positive rate predicted for GMP-selected sources \citep{mannucci2022, mannucci2023}.
Of the 13 multiple AGN systems, we classify six as dual AGN, and seven as gravitationally lensed quasars, including four doubly imaged and three quadruply imaged systems. The dual AGN span the redshift range $0.68 \leq z \leq 1.57$, projected separations of $0.26''$--$0.80''$ ($2.2$--$6.7$~kpc), and broad-line velocity offsets of $\Delta v \sim 290$--$950$~\kms. Dust attenuation is generally modest, with $E(B-V) \lesssim 0.3$, indicating that GMP-selected dual AGN predominantly probe unobscured systems.
The lensed sample includes J1453$+$0529, currently the smallest-separation quadruply imaged quasar known ($r_{\rm E}=0.21''\pm0.02''$; \citealt{damato2026}), and J0602$-$3852, the most distant system in our sample ($z=3.04$). Combined with previous GMP follow-up campaigns \citep{ciurlo23, mannucci2023, scialpi24, scialpi2026}, these results further demonstrate that the GMP method efficiently identifies compact lensed quasars that occupy a region of parameter space largely unexplored by traditional lens searches, namely sub-arcsecond systems at $z \gtrsim 2$.

\noindent By combining the systems presented here with previously confirmed GMP-selected sources, the current sample comprises 20 dual AGN, the dual AGN candidate J2301--1906, 18 doubly imaged lensed quasars, and 7 quadruply imaged lensed quasars. Within this population, dual AGN are preferentially found at lower redshifts and lower \gaia\ luminosities, whereas lensed quasars dominate at higher redshifts and luminosities.

Overall, these observations confirm six new dual AGN and six new gravitationally lensed quasars and significantly expand the census of spectroscopically confirmed compact multiple AGN systems at $z>0.5$. They demonstrate that the combination of the GMP selection and high-angular-resolution IFU spectroscopy with VLT/ERIS provides a powerful and efficient way to identify both dual AGN and compact lensed quasars at sub-arcsecond separations. 
In a forthcoming series of papers (\citealt{scialpi2026_bhmass}, Marconcini \& Scialpi, Tozzi et al. in prep.) these targets will be used to test models of galaxy/SMBH formation and co-evolution and obtain new estimates of the GW background in PTA experiments and the GW event rate in LISA  (Mannucci et al in prep.).

\begin{acknowledgements}
We acknowledge financial support 
from the INAF project ``VLT-MOONS'' CRAM 1.05.03.07,
from the ASI-INAF agreements n. 2024-36-HH.1-2025 and 2025-29-HH.0,
from INAF Bando di Ricerca fondamentale 
(2022) "Dual and binary supermassive black holes in the multi-messenger era: from galaxy mergers to gravitational waves",
(2023) "The MOONS Extragalactic Survey" 1.05.23.01.11, 
(2024) "The Quest for dual and binary massive black holes in the gravitational wave era " 1.05.24.01.02,
and 
(2024) "A JWST/MIRI MIRACLE: Mid-IR Activity of Circumnuclear Line Emission" and MiniGrant RSN1 1.05.24.07.01.
IL is supported by the European Research Council (ERC) under the European Union’s Horizon 2020 research and innovation program (DistantDust, Grant agreement No. 101117541).
GV acknowledges financial support the INAF under the IAF - Astrophysics Fellowships in Italy grant CUP C59J21034720001 - ``AD MAJORA''.
MP acknowledges support through the grants PID2021-127718NB-I00, PID2024-159902NA-I00, and RYC2023-044853-I, funded by the Spain Ministry of Science and Innovation/State Agency of Research MCIN/AEI/10.13039/501100011033 and El Fondo Social Europeo Plus FSE+.
Based on observations collected at the European Southern Observatory (ESO) under programs 111.24QJ, 112.25GT, 113.26BW, 114.26ZT, 115.27WA, and 116.28MD with VLT/ERIS--SPIFFIER, and run ID 112.25GT with ERIS/NIX.
We are grateful to M.~Brazzini, D.~Fedele, F.~Maio, D.~Mattolini, B.~Trefoloni, L.~Scholz-Díaz, S.~Piranomonte, E.~Parlanti for carrying out part of the ERIS visitor-mode observations.
This work also makes use of observations collected at ESO under programmes 109.22W4, 112.25CT, 113.26TB, 114.27FC, 115.28CQ, and 116.28PX with NTT/EFOSC2 and VLT/FORS2 and 114.27BY.001 with VLT/MUSE. 
The authors sincerely thank the ESO staff in Garching and at Paranal for their availability, patience, and support during both the preparation and execution of the observations. 
Based on observations made with the Italian Telescopio Nazionale Galileo (TNG), operated on the island of La Palma by the Fundaci\'{o}n Galileo Galilei of INAF (Istituto Nazionale di Astrofisica) at the Spanish Observatorio del Roque de los Muchachos of the Instituto de Astrof\'{i}sica de Canarias, under programmes \texttt{A47TAC\_28}, \texttt{A48TAC\_67}, and \texttt{AOT51\_24}.
Based on observations made with the NASA/ESA Hubble Space Telescope, obtained from the Data Archive at the Space Telescope Science Institute (STScI), which is operated by the Association of Universities for Research in Astronomy, Inc., under NASA contract NAS 5-26555. These observations are associated with program 17423 under Cycle~31.
This work has made use of data from the European Space Agency (ESA) mission
{\it Gaia} (\url{https://www.cosmos.esa.int/gaia}), processed by the {\it Gaia}
Data Processing and Analysis Consortium (DPAC,
\url{https://www.cosmos.esa.int/web/gaia/dpac/consortium}). Funding for the DPAC
has been provided by national institutions, in particular the institutions
participating in the {\it Gaia} Multilateral Agreement.
This work has made use of data from the Sloan Digital Sky Survey (SDSS, \url{www.sdss.org}).
This research used data obtained with the Dark Energy Spectroscopic Instrument (DESI, \url{www.data.desi.lbl.gov}).
This work made use of data from the VISTA Hemisphere survey. The VISTA Hemisphere Survey data products served at Astro Data Lab are based on observations collected at the European Organisation for Astronomical Research in the Southern Hemisphere under ESO program 179.A-2010, and/or data products created thereof.
This publication makes use of data products from the Two Micron All Sky Survey, which is a joint project of the University of Massachusetts and the Infrared Processing and Analysis Center/California Institute of Technology, funded by the National Aeronautics and Space Administration and the National Science Foundation.
This work made use of the following Python packages:
Astropy \citep{astropy2013,astropy2018,astropy2022},
NumPy \citep{harris2020},
SciPy \citep{virtanen2020},
Matplotlib \citep{hunter2007},
Photutils \citep{bradley2022},
and emcee \citep{foremanmackey2013}.

\end{acknowledgements}

\bibliographystyle{aa}
\bibliography{bibliografia}

@ARTICLE{yan2019,
       author = {{Yan}, Renbin and {Chen}, Yanping and {Lazarz}, Daniel and {Bizyaev}, Dmitry and {Maraston}, Claudia and {Stringfellow}, Guy S. and {McCarthy}, Kyle and {Meneses-Goytia}, Sofia and {Law}, David R. and {Thomas}, Daniel and {Falcon Barroso}, Jesus and {S{\'a}nchez-Gallego}, Jos{\'e} R. and {Schlafly}, Edward and {Zheng}, Zheng and {Argudo-Fern{\'a}ndez}, Maria and {Beaton}, Rachael L. and {Beers}, Timothy C. and {Bershady}, Matthew and {Blanton}, Michael R. and {Brownstein}, Joel and {Bundy}, Kevin and {Chambers}, Kenneth C. and {Cherinka}, Brian and {De Lee}, Nathan and {Drory}, Niv and {Galbany}, Llu{\'\i}s and {Holtzman}, Jon and {Imig}, Julie and {Kaiser}, Nick and {Kinemuchi}, Karen and {Liu}, Chao and {Luo}, A.-Li and {Magnier}, Eugene and {Majewski}, Steven and {Nair}, Preethi and {Oravetz}, Audrey and {Oravetz}, Daniel and {Pan}, Kaike and {Sobeck}, Jennifer and {Stassun}, Keivan and {Talbot}, Michael and {Tremonti}, Christy and {Waters}, Christopher and {Weijmans}, Anne-Marie and {Wilhelm}, Ronald and {Zasowski}, Gail and {Zhao}, Gang and {Zhao}, Yong-Heng},
        title = "{SDSS-IV MaStar: A Large and Comprehensive Empirical Stellar Spectral Library{\textemdash}First Release}",
      journal = {\apj},
         year = 2019,
        month = oct,
       volume = {883},
       number = {2},
          eid = {175},
        pages = {175},
          doi = {10.3847/1538-4357/ab3ebc},
archivePrefix = {arXiv},
       eprint = {1812.02745},
 primaryClass = {astro-ph.IM},
       adsurl = {https://ui.adsabs.harvard.edu/abs/2019ApJ...883..175Y}
}

@ARTICLE{derosa2023,
       author = {{De Rosa}, Alessandra and {Vignali}, Cristian and {Severgnini}, Paola and {Bianchi}, Stefano and {Bogdanovi{\'c}}, Tamara and {Charisi}, Maria and {Guainazzi}, Matteo and {Haiman}, Zoltan and {Komossa}, S. and {Paragi}, Zsolt and {Perez-Torres}, Miguel and {Piconcelli}, Enrico and {Ducci}, Lorenzo and {Parvatikar}, Manali and {Serafinelli}, Roberto},
        title = "{The X-ray view of optically selected dual AGN}",
      journal = {\mnras},
         year = 2023,
        month = mar,
       volume = {519},
       number = {4},
        pages = {5149-5160},
          doi = {10.1093/mnras/stac3664},
archivePrefix = {arXiv},
       eprint = {2212.05837},
 primaryClass = {astro-ph.GA},
       adsurl = {https://ui.adsabs.harvard.edu/abs/2023MNRAS.519.5149D}
}

@ARTICLE{foremanmackey2013,
       author = {Foreman-Mackey, Daniel and Hogg, David W. and Lang, Dustin and Goodman, Jonathan},
        title = "{emcee: The MCMC Hammer}",
      journal = {PASP},
         year = 2013,
       volume = {125},
        pages = {306--312},
          doi = {10.1086/670067}
}

@ARTICLE{astropy2013,
       author = {{Astropy Collaboration} and others},
        title = "{Astropy: A community Python package for astronomy}",
      journal = {A\&A},
         year = 2013,
       volume = {558},
          eid = {A33}
}

@ARTICLE{astropy2018,
       author = {{Astropy Collaboration} and others},
        title = "{The Astropy Project: Building an Open-science Project and Status of the v2.0 Core Package}",
      journal = {AJ},
         year = 2018,
       volume = {156},
          eid = {123}
}

@ARTICLE{astropy2022,
       author = {{Astropy Collaboration} and others},
        title = "{The Astropy Project: Sustaining and Growing a Community-oriented Open-source Project and the Latest Major Release (v5.0) of the Core Package}",
      journal = {ApJ},
         year = 2022,
       volume = {935},
          eid = {167}
}

@ARTICLE{harris2020,
       author = {Harris, C. R. and others},
        title = "{Array programming with NumPy}",
      journal = {Nature},
         year = 2020,
       volume = {585},
        pages = {357--362}
}

@ARTICLE{virtanen2020,
       author = {Virtanen, P. and others},
        title = "{SciPy 1.0: Fundamental Algorithms for Scientific Computing in Python}",
      journal = {Nature Methods},
         year = 2020,
       volume = {17},
        pages = {261--272}
}

@ARTICLE{hunter2007,
       author = {Hunter, J. D.},
        title = "{Matplotlib: A 2D Graphics Environment}",
      journal = {Computing in Science \& Engineering},
         year = 2007,
       volume = {9},
        pages = {90--95}
}

@ARTICLE{bradley2022,
       author = {Bradley, L. and others},
        title = "{Astropy Photutils}",
         year = 2022,
          doi = {10.5281/zenodo.6825092}
}

@ARTICLE{damato2026_skao,
       author = {{D'Amato}, Q. and {Cui}, L. and {Deane}, R. and {Komossa}, S. and {Pillay}, C. and {Tripathi}, A. and {Kharb}, P. and {Guo}, H. and {Nandi}, S. and {Rubinur}, K. and {Anton}, S. and {An}, T. and {Bonoli}, S. and {Chang}, N. and {Dave}, R. and {De Rosa}, A. and {Habouzit}, M. and {Mannucci}, F. and {Prandoni}, I. and {Severgnini}, P. and {Scialpi}, M. and {Spingola}, C. and {Vignali}, C. and {Xu}, W. and {Yan}, X. and {Zhang}, Y.},
        title = "{Dual AGN and Multiple SMBH Systems in the Era of SKAO}",
      journal = {arXiv e-prints},
         year = 2026,
        month = jun,
          eid = {arXiv:2606.28512},
        pages = {arXiv:2606.28512},
          doi = {10.48550/arXiv.2606.28512},
archivePrefix = {arXiv},
       eprint = {2606.28512},
 primaryClass = {astro-ph.HE},
       adsurl = {https://ui.adsabs.harvard.edu/abs/2026arXiv260628512D}
}

@ARTICLE{zamora2025,
       author = {{Zamora}, Sandra and {Venturi}, Giacomo and {Carniani}, Stefano and {Bertola}, Elena and {Parlanti}, Eleonora and {Perna}, Michele and {Arribas}, Santiago and {B{\"o}ker}, Torsten and {Bunker}, Andrew J. and {Charlot}, St{\'e}phane and {D'Eugenio}, Francesco and {Maiolino}, Roberto and {Del Pino}, Bruno Rodr{\'\i}guez and {{\"U}bler}, Hannah and {Cresci}, Giovanni and {Jones}, Gareth C. and {Lamperti}, Isabella},
        title = "{GA-NIFS: The highly overdense system BR1202-0725 at z {\ensuremath{\sim}} 4.7: A double active galactic nucleus with fast outflows plus eight companion galaxies}",
      journal = {\aap},
         year = 2025,
        month = oct,
       volume = {702},
          eid = {A102},
        pages = {A102},
          doi = {10.1051/0004-6361/202453236},
archivePrefix = {arXiv},
       eprint = {2412.02751},
 primaryClass = {astro-ph.GA},
       adsurl = {https://ui.adsabs.harvard.edu/abs/2025A&A...702A.102Z}
}

@ARTICLE{lemon_2020,
       author = {{Lemon}, C. and {Auger}, M.~W. and {McMahon}, R. and {Anguita}, T. and {Apostolovski}, Y. and {Chen}, G.~C.-F. and {Fassnacht}, C.~D. and {Melo}, A.~D. and {Motta}, V. and {Shajib}, A. and {Treu}, T. and {Agnello}, A. and {Buckley-Geer}, E. and {Schechter}, P.~L. and {Birrer}, S. and {Collett}, T. and {Courbin}, F. and {Rusu}, C.~E. and {Abbott}, T.~M.~C. and {Allam}, S. and {Annis}, J. and {Avila}, S. and {Bertin}, E. and {Brooks}, D. and {Burke}, D.~L. and {Carnero Rosell}, A. and {Carrasco Kind}, M. and {Carretero}, J. and {Costanzi}, M. and {da Costa}, L.~N. and {De Vicente}, J. and {Desai}, S. and {Eifler}, T.~F. and {Flaugher}, B. and {Frieman}, J. and {Garc{\'\i}a-Bellido}, J. and {Gaztanaga}, E. and {Gerdes}, D.~W. and {Gruen}, D. and {Gruendl}, R.~A. and {Gschwend}, J. and {Gutierrez}, G. and {Honscheid}, K. and {James}, D.~J. and {Kim}, A. and {Krause}, E. and {Kuehn}, K. and {Kuropatkin}, N. and {Lahav}, O. and {Lima}, M. and {Lin}, H. and {Maia}, M.~A.~G. and {March}, M. and {Marshall}, J.~L. and {Menanteau}, F. and {Miquel}, R. and {Palmese}, A. and {Paz-Chinch{\'o}n}, F. and {Plazas}, A.~A. and {Roodman}, A. and {Sanchez}, E. and {Schubnell}, M. and {Serrano}, S. and {Smith}, M. and {Soares-Santos}, M. and {Suchyta}, E. and {Tarle}, G. and {Walker}, A.~R.},
        title = "{The STRong lensing Insights into the Dark Energy Survey (STRIDES) 2017/2018 follow-up campaign: discovery of 10 lensed quasars and 10 quasar pairs}",
      journal = {\mnras},
         year = 2020,
        month = may,
       volume = {494},
       number = {3},
        pages = {3491-3511},
          doi = {10.1093/mnras/staa652},
archivePrefix = {arXiv},
       eprint = {1912.09133},
 primaryClass = {astro-ph.GA},
       adsurl = {https://ui.adsabs.harvard.edu/abs/2020MNRAS.494.3491L}
}

@ARTICLE{kovacevic2010_feiitemplate,
       author = {{Kova{\v{c}}evi{\'c}}, Jelena and {Popovi{\'c}}, Luka {\v{C}}. and {Dimitrijevi{\'c}}, Milan S.},
        title = "{Analysis of Optical Fe II Emission in a Sample of Active Galactic Nucleus Spectra}",
      journal = {\apjs},
         year = 2010,
        month = jul,
       volume = {189},
       number = {1},
        pages = {15-36},
          doi = {10.1088/0067-0049/189/1/15},
archivePrefix = {arXiv},
       eprint = {1004.2212},
 primaryClass = {astro-ph.CO},
       adsurl = {https://ui.adsabs.harvard.edu/abs/2010ApJS..189...15K}
}

@ARTICLE{Lamperti2026,
       author = {{Lamperti}, I. and {Mannucci}, F. and {Bertola}, E. and {Marconi}, A. and {Cresci}, G. and {Nardini}, E. and {D'Amato}, Q. and {Perna}, M. and {Rojas-Lilay{\'u}}, A. and {Bracci}, C. and {Braito}, V. and {Cataldi}, E. and {Ceci}, M. and {Chakraborty}, A. and {Cicone}, C. and {De Rosa}, A. and {Feltre}, A. and {Ginolfi}, M. and {Lusso}, E. and {Marconcini}, C. and {Moreschini}, B. and {Portaluri}, E. and {Rubinur}, K. and {Scialpi}, M. and {Severgnini}, P. and {Tozzi}, G. and {Trindade Falc{\~a}o}, A. and {Ulivi}, L. and {Venturi}, G. and {Vignali}, C. and {Zanchettin}, M.~V.},
        title = "{Fast, dust-poor outflows in the local candidate dual AGN MCG-03-34-64 observed with VLT/ERIS}",
      journal = {\aap},
         year = 2026,
        month = jun,
       volume = {710},
          eid = {A193},
        pages = {A193},
          doi = {10.1051/0004-6361/202659161},
archivePrefix = {arXiv},
       eprint = {2601.22254},
 primaryClass = {astro-ph.GA},
       adsurl = {https://ui.adsabs.harvard.edu/abs/2026A&A...710A.193L}
}

@ARTICLE{Barrows2023,
       author = {{Barrows}, R. Scott and {Comerford}, Julia M. and {Stern}, Daniel and {Assef}, Roberto J.},
        title = "{A Census of WISE-selected Dual and Offset AGNs Across the Sky: New Constraints on Merger-driven Triggering of Obscured AGNs}",
      journal = {\apj},
         year = 2023,
        month = jul,
       volume = {951},
       number = {2},
          eid = {92},
        pages = {92},
          doi = {10.3847/1538-4357/acd2d3},
archivePrefix = {arXiv},
       eprint = {2305.03760},
 primaryClass = {astro-ph.GA},
       adsurl = {https://ui.adsabs.harvard.edu/abs/2023ApJ...951...92B}
}

@article{hogg1999,
  author        = {Hogg, David W.},
  title         = {Distance Measures in Cosmology},
  year          = {1999},
  eprint        = {astro-ph/9905116},
  archivePrefix = {arXiv},
  primaryClass  = {astro-ph}
}

@ARTICLE{Jing2026_desi,
       author = {{Jing}, Liang and {Chen}, Qihang and {Zhu}, Xingyu and {Deng}, Zhuojun and {Zou}, Hu and {Xia}, Jun-Qing and {Wu}, Jianghua},
        title = "{A Quasar Pair Sample Compiled from DESI Data Release 1}",
      journal = {\apj},
         year = 2026,
        month = apr,
       volume = {1000},
       number = {2},
          eid = {311},
        pages = {311},
          doi = {10.3847/1538-4357/ae517a},
archivePrefix = {arXiv},
       eprint = {2505.03103},
 primaryClass = {astro-ph.GA},
       adsurl = {https://ui.adsabs.harvard.edu/abs/2026ApJ..1000..311J}
}

@ARTICLE{ubler2025,
       author = {{{\"U}bler}, Hannah and {Mazzolari}, Giovanni and {Maiolino}, Roberto and {D'Eugenio}, Francesco and {Davari}, Nazanin and {Juod{\v{z}}balis}, Ignas and {Perna}, Michele and {Matthee}, Jorryt and {Schneider}, Raffaella and {Valiante}, Rosa and {Arribas}, Santiago and {Bertola}, Elena and {Bunker}, Andrew J. and {Bromm}, Volker and {Carniani}, Stefano and {Charlot}, St{\'e}phane and {Cresci}, Giovanni and {Curti}, Mirko and {Davies}, Richard and {Eisenhauer}, Frank and {Fabian}, Andrew and {F{\"o}rster Schreiber}, Natascha M. and {Genzel}, Reinhard and {Inayoshi}, Kohei and {Ivey}, Lucy R. and {Jones}, Gareth C. and {Liu}, Boyuan and {Lutz}, Dieter and {Kashino}, Daichi and {Mackenzie}, Ruari and {Parlanti}, Eleonora and {Robertson}, Brant and {Rodr{\'\i}guez del Pino}, Bruno and {Shimizu}, T. Taro and {Sijacki}, Debora and {Sturm}, Eckhard and {Tacchella}, Sandro and {Tacconi}, Linda and {Tozzi}, Giulia and {Trinca}, Alessandro and {Venturi}, Giacomo and {Volonteri}, Marta and {Willott}, Chris and {Zhang}, Saiyang},
        title = "{BlackTHUNDER: Evidence of three massive black holes in a z\raisebox{-0.5ex}\textasciitilde5 galaxy}",
      journal = {arXiv e-prints},
         year = 2025,
        month = sep,
          eid = {arXiv:2509.21575},
        pages = {arXiv:2509.21575},
          doi = {10.48550/arXiv.2509.21575},
archivePrefix = {arXiv},
       eprint = {2509.21575},
 primaryClass = {astro-ph.GA},
       adsurl = {https://ui.adsabs.harvard.edu/abs/2025arXiv250921575U}
}

@ARTICLE{Faure2008,
       author = {{Faure}, Cecile and {Kneib}, Jean-Paul and {Covone}, Giovanni and {Tasca}, Lidia and {Leauthaud}, Alexie and {Capak}, Peter and {Jahnke}, Knud and {Smolcic}, Vernesa and {de la Torre}, Sylvain and {Ellis}, Richard and {Finoguenov}, Alexis and {Koekemoer}, Anton and {Le Fevre}, Oliver and {Massey}, Richard and {Mellier}, Yannick and {Refregier}, Alexandre and {Rhodes}, Jason and {Scoville}, Nick and {Schinnerer}, Eva and {Taylor}, James and {Van Waerbeke}, Ludovic and {Walcher}, Jakob},
        title = "{First Catalog of Strong Lens Candidates in the COSMOS Field}",
      journal = {\apjs},
         year = 2008,
        month = may,
       volume = {176},
       number = {1},
        pages = {19-38},
          doi = {10.1086/526426},
archivePrefix = {arXiv},
       eprint = {0802.2174},
 primaryClass = {astro-ph},
       adsurl = {https://ui.adsabs.harvard.edu/abs/2008ApJS..176...19F}
}

@ARTICLE{Bolton2008_slacs,
       author = {{Bolton}, Adam S. and {Burles}, Scott and {Koopmans}, L{\'e}on V.~E. and {Treu}, Tommaso and {Gavazzi}, Rapha{\"e}l and {Moustakas}, Leonidas A. and {Wayth}, Randall and {Schlegel}, David J.},
        title = "{The Sloan Lens ACS Survey. V. The Full ACS Strong-Lens Sample}",
      journal = {\apj},
         year = 2008,
        month = aug,
       volume = {682},
       number = {2},
        pages = {964-984},
          doi = {10.1086/589327},
archivePrefix = {arXiv},
       eprint = {0805.1931},
 primaryClass = {astro-ph},
       adsurl = {https://ui.adsabs.harvard.edu/abs/2008ApJ...682..964B}
}

@ARTICLE{Inada2012,
       author = {{Inada}, Naohisa and {Oguri}, Masamune and {Shin}, Min-Su and {Kayo}, Issha and {Strauss}, Michael A. and {Morokuma}, Tomoki and {Rusu}, Cristian E. and {Fukugita}, Masataka and {Kochanek}, Christopher S. and {Richards}, Gordon T. and {Schneider}, Donald P. and {York}, Donald G. and {Bahcall}, Neta A. and {Frieman}, Joshua A. and {Hall}, Patrick B. and {White}, Richard L.},
        title = "{The Sloan Digital Sky Survey Quasar Lens Search. V. Final Catalog from the Seventh Data Release}",
      journal = {\aj},
         year = 2012,
        month = may,
       volume = {143},
       number = {5},
          eid = {119},
        pages = {119},
          doi = {10.1088/0004-6256/143/5/119},
archivePrefix = {arXiv},
       eprint = {1203.1087},
 primaryClass = {astro-ph.CO},
       adsurl = {https://ui.adsabs.harvard.edu/abs/2012AJ....143..119I}
}

@ARTICLE{Sonnenfeld2019,
       author = {{Sonnenfeld}, Alessandro and {Jaelani}, Anton T. and {Chan}, James and {More}, Anupreeta and {Suyu}, Sherry H. and {Wong}, Kenneth C. and {Oguri}, Masamune and {Lee}, Chien-Hsiu},
        title = "{Survey of gravitationally-lensed objects in HSC imaging (SuGOHI). III. Statistical strong lensing constraints on the stellar IMF of CMASS galaxies}",
      journal = {\aap},
         year = 2019,
        month = oct,
       volume = {630},
          eid = {A71},
        pages = {A71},
          doi = {10.1051/0004-6361/201935743},
archivePrefix = {arXiv},
       eprint = {1904.10465},
 primaryClass = {astro-ph.GA},
       adsurl = {https://ui.adsabs.harvard.edu/abs/2019A&A...630A..71S}
}

@ARTICLE{Shajib2021,
       author = {{Shajib}, Anowar J. and {Treu}, Tommaso and {Birrer}, Simon and {Sonnenfeld}, Alessandro},
        title = "{Dark matter haloes of massive elliptical galaxies at z {\ensuremath{\sim}} 0.2 are well described by the Navarro-Frenk-White profile}",
      journal = {\mnras},
         year = 2021,
        month = may,
       volume = {503},
       number = {2},
        pages = {2380-2405},
          doi = {10.1093/mnras/stab536},
archivePrefix = {arXiv},
       eprint = {2008.11724},
 primaryClass = {astro-ph.GA},
       adsurl = {https://ui.adsabs.harvard.edu/abs/2021MNRAS.503.2380S}
}

@ARTICLE{Sonnenfeld2015,
       author = {{Sonnenfeld}, Alessandro and {Treu}, Tommaso and {Marshall}, Philip J. and {Suyu}, Sherry H. and {Gavazzi}, Rapha{\"e}l and {Auger}, Matthew W. and {Nipoti}, Carlo},
        title = "{The SL2S Galaxy-scale Lens Sample. V. Dark Matter Halos and Stellar IMF of Massive Early-type Galaxies Out to Redshift 0.8}",
      journal = {\apj},
         year = 2015,
        month = feb,
       volume = {800},
       number = {2},
          eid = {94},
        pages = {94},
          doi = {10.1088/0004-637X/800/2/94},
archivePrefix = {arXiv},
       eprint = {1410.1881},
 primaryClass = {astro-ph.GA},
       adsurl = {https://ui.adsabs.harvard.edu/abs/2015ApJ...800...94S}
}

@ARTICLE{Newman2017,
       author = {{Newman}, Andrew B. and {Smith}, Russell J. and {Conroy}, Charlie and {Villaume}, Alexa and {van Dokkum}, Pieter},
        title = "{The Initial Mass Function in the Nearest Strong Lenses from SNELLS: Assessing the Consistency of Lensing, Dynamical, and Spectroscopic Constraints}",
      journal = {\apj},
         year = 2017,
        month = aug,
       volume = {845},
       number = {2},
          eid = {157},
        pages = {157},
          doi = {10.3847/1538-4357/aa816d},
archivePrefix = {arXiv},
       eprint = {1612.00065},
 primaryClass = {astro-ph.GA},
       adsurl = {https://ui.adsabs.harvard.edu/abs/2017ApJ...845..157N}
}

@ARTICLE{Massey2010,
       author = {{Massey}, Richard and {Kitching}, Thomas and {Richard}, Johan},
        title = "{The dark matter of gravitational lensing}",
      journal = {Reports on Progress in Physics},
         year = 2010,
        month = aug,
       volume = {73},
       number = {8},
          eid = {086901},
        pages = {086901},
          doi = {10.1088/0034-4885/73/8/086901},
archivePrefix = {arXiv},
       eprint = {1001.1739},
 primaryClass = {astro-ph.CO},
       adsurl = {https://ui.adsabs.harvard.edu/abs/2010RPPh...73h6901M}
}

@ARTICLE{Treu2010,
       author = {{Treu}, Tommaso and {Auger}, Matthew W. and {Koopmans}, L{\'e}on V.~E. and {Gavazzi}, Rapha{\"e}l and {Marshall}, Philip J. and {Bolton}, Adam S.},
        title = "{The Initial Mass Function of Early-Type Galaxies}",
      journal = {\apj},
         year = 2010,
        month = feb,
       volume = {709},
       number = {2},
        pages = {1195-1202},
          doi = {10.1088/0004-637X/709/2/1195},
archivePrefix = {arXiv},
       eprint = {0911.3392},
 primaryClass = {astro-ph.CO},
       adsurl = {https://ui.adsabs.harvard.edu/abs/2010ApJ...709.1195T}
}

@ARTICLE{Wong2018A,
       author = {{Wong}, Kenneth C. and {Sonnenfeld}, Alessandro and {Chan}, James H.~H. and {Rusu}, Cristian E. and {Tanaka}, Masayuki and {Jaelani}, Anton T. and {Lee}, Chien-Hsiu and {More}, Anupreeta and {Oguri}, Masamune and {Suyu}, Sherry H. and {Komiyama}, Yutaka},
        title = "{Survey of Gravitationally Lensed Objects in HSC Imaging (SuGOHI). II. Environments and Line-of-Sight Structure of Strong Gravitational Lens Galaxies to z {\ensuremath{\sim}} 0.8}",
      journal = {\apj},
         year = 2018,
        month = nov,
       volume = {867},
       number = {2},
          eid = {107},
        pages = {107},
          doi = {10.3847/1538-4357/aae381},
archivePrefix = {arXiv},
       eprint = {1809.07341},
 primaryClass = {astro-ph.GA},
       adsurl = {https://ui.adsabs.harvard.edu/abs/2018ApJ...867..107W}
}

@ARTICLE{Kauffmann2003,
       author = {{Kauffmann}, Guinevere and {Heckman}, Timothy M. and {Tremonti}, Christy and {Brinchmann}, Jarle and {Charlot}, St{\'e}phane and {White}, Simon D.~M. and {Ridgway}, Susan E. and {Brinkmann}, Jon and {Fukugita}, Masataka and {Hall}, Patrick B. and {Ivezi{\'c}}, {\v{Z}}eljko and {Richards}, Gordon T. and {Schneider}, Donald P.},
        title = "{The host galaxies of active galactic nuclei}",
      journal = {\mnras},
         year = 2003,
        month = dec,
       volume = {346},
       number = {4},
        pages = {1055-1077},
          doi = {10.1111/j.1365-2966.2003.07154.x},
archivePrefix = {arXiv},
       eprint = {astro-ph/0304239},
 primaryClass = {astro-ph},
       adsurl = {https://ui.adsabs.harvard.edu/abs/2003MNRAS.346.1055K}
}

@ARTICLE{Kewley2006,
       author = {{Kewley}, Lisa J. and {Groves}, Brent and {Kauffmann}, Guinevere and {Heckman}, Tim},
        title = "{The host galaxies and classification of active galactic nuclei}",
      journal = {\mnras},
         year = 2006,
        month = nov,
       volume = {372},
       number = {3},
        pages = {961-976},
          doi = {10.1111/j.1365-2966.2006.10859.x},
archivePrefix = {arXiv},
       eprint = {astro-ph/0605681},
 primaryClass = {astro-ph},
       adsurl = {https://ui.adsabs.harvard.edu/abs/2006MNRAS.372..961K}
}

@ARTICLE{Kewley2001,
       author = {{Kewley}, L.~J. and {Dopita}, M.~A. and {Sutherland}, R.~S. and {Heisler}, C.~A. and {Trevena}, J.},
        title = "{Theoretical Modeling of Starburst Galaxies}",
      journal = {\apj},
         year = 2001,
        month = jul,
       volume = {556},
       number = {1},
        pages = {121-140},
          doi = {10.1086/321545},
archivePrefix = {arXiv},
       eprint = {astro-ph/0106324},
 primaryClass = {astro-ph},
       adsurl = {https://ui.adsabs.harvard.edu/abs/2001ApJ...556..121K}
}

@ARTICLE{Gutkin2016,
       author = {{Gutkin}, Julia and {Charlot}, St{\'e}phane and {Bruzual}, Gustavo},
        title = "{Modelling the nebular emission from primeval to present-day star-forming galaxies}",
      journal = {\mnras},
         year = 2016,
        month = oct,
       volume = {462},
       number = {2},
        pages = {1757-1774},
          doi = {10.1093/mnras/stw1716},
archivePrefix = {arXiv},
       eprint = {1607.06086},
 primaryClass = {astro-ph.GA},
       adsurl = {https://ui.adsabs.harvard.edu/abs/2016MNRAS.462.1757G}
}

@ARTICLE{Patton2024,
       author = {{Patton}, David R. and {Faria}, Lawrence and {Hani}, Maan H. and {Torrey}, Paul and {Ellison}, Sara L. and {Thakur}, Shivani D. and {Westlake}, Raven I.},
        title = "{Interacting galaxies in the IllustrisTNG simulations - VI: Reconstructed orbits, close encounters, and mergers}",
      journal = {\mnras},
         year = 2024,
        month = apr,
       volume = {529},
       number = {2},
        pages = {1493-1506},
          doi = {10.1093/mnras/stae608},
archivePrefix = {arXiv},
       eprint = {2402.17889},
 primaryClass = {astro-ph.GA},
       adsurl = {https://ui.adsabs.harvard.edu/abs/2024MNRAS.529.1493P}
}

@ARTICLE{Pfeifle2025,
       author = {{Pfeifle}, Ryan W. and {Weaver}, Kimberly A. and {Secrest}, Nathan J. and {Rothberg}, Barry and {Patton}, David R.},
        title = "{Super-size Me: The Big Multi-AGN Catalog (The Big MAC) Data Release 1: The Source Catalog}",
      journal = {\apjs},
         year = 2025,
        month = nov,
       volume = {281},
       number = {1},
          eid = {25},
        pages = {25},
          doi = {10.3847/1538-4365/adf845},
archivePrefix = {arXiv},
       eprint = {2411.12799},
 primaryClass = {astro-ph.GA},
       adsurl = {https://ui.adsabs.harvard.edu/abs/2025ApJS..281...25P}
}

@ARTICLE{Tozzi2021,
       author = {{Tozzi}, G. and {Cresci}, G. and {Marasco}, A. and {Nardini}, E. and {Marconi}, A. and {Mannucci}, F. and {Chartas}, G. and {Rizzo}, F. and {Amiri}, A. and {Brusa}, M. and {Comastri}, A. and {Dadina}, M. and {Lanzuisi}, G. and {Mainieri}, V. and {Mingozzi}, M. and {Perna}, M. and {Venturi}, G. and {Vignali}, C.},
        title = "{Connecting X-ray nuclear winds with galaxy-scale ionised outflows in two z {\ensuremath{\sim}} 1.5 lensed quasars}",
      journal = {\aap},
         year = 2021,
        month = apr,
       volume = {648},
          eid = {A99},
        pages = {A99},
          doi = {10.1051/0004-6361/202040190},
archivePrefix = {arXiv},
       eprint = {2102.07789},
 primaryClass = {astro-ph.GA},
       adsurl = {https://ui.adsabs.harvard.edu/abs/2021A&A...648A..99T}
}

@ARTICLE{Baldwin1981,
       author = {{Baldwin}, J.~A. and {Phillips}, M.~M. and {Terlevich}, R.},
        title = "{Classification parameters for the emission-line spectra of extragalactic objects.}",
      journal = {\pasp},
         year = 1981,
        month = feb,
       volume = {93},
        pages = {5-19},
          doi = {10.1086/130766},
       adsurl = {https://ui.adsabs.harvard.edu/abs/1981PASP...93....5B}
}

@ARTICLE{Planck2020,
       author = {{Planck Collaboration} and {Aghanim}, N. and {Akrami}, Y. and {Ashdown}, M. and {Aumont}, J. and {Baccigalupi}, C. and {Ballardini}, M. and {Banday}, A.~J. and {Barreiro}, R.~B. and {Bartolo}, N. and {Basak}, S. and {Battye}, R. and {Benabed}, K. and {Bernard}, J.-P. and {Bersanelli}, M. and {Bielewicz}, P. and {Bock}, J.~J. and {Bond}, J.~R. and {Borrill}, J. and {Bouchet}, F.~R. and {Boulanger}, F. and {Bucher}, M. and {Burigana}, C. and {Butler}, R.~C. and {Calabrese}, E. and {Cardoso}, J.-F. and {Carron}, J. and {Challinor}, A. and {Chiang}, H.~C. and {Chluba}, J. and {Colombo}, L.~P.~L. and {Combet}, C. and {Contreras}, D. and {Crill}, B.~P. and {Cuttaia}, F. and {de Bernardis}, P. and {de Zotti}, G. and {Delabrouille}, J. and {Delouis}, J.-M. and {Di Valentino}, E. and {Diego}, J.~M. and {Dor{\'e}}, O. and {Douspis}, M. and {Ducout}, A. and {Dupac}, X. and {Dusini}, S. and {Efstathiou}, G. and {Elsner}, F. and {En{\ss}lin}, T.~A. and {Eriksen}, H.~K. and {Fantaye}, Y. and {Farhang}, M. and {Fergusson}, J. and {Fernandez-Cobos}, R. and {Finelli}, F. and {Forastieri}, F. and {Frailis}, M. and {Fraisse}, A.~A. and {Franceschi}, E. and {Frolov}, A. and {Galeotta}, S. and {Galli}, S. and {Ganga}, K. and {G{\'e}nova-Santos}, R.~T. and {Gerbino}, M. and {Ghosh}, T. and {Gonz{\'a}lez-Nuevo}, J. and {G{\'o}rski}, K.~M. and {Gratton}, S. and {Gruppuso}, A. and {Gudmundsson}, J.~E. and {Hamann}, J. and {Handley}, W. and {Hansen}, F.~K. and {Herranz}, D. and {Hildebrandt}, S.~R. and {Hivon}, E. and {Huang}, Z. and {Jaffe}, A.~H. and {Jones}, W.~C. and {Karakci}, A. and {Keih{\"a}nen}, E. and {Keskitalo}, R. and {Kiiveri}, K. and {Kim}, J. and {Kisner}, T.~S. and {Knox}, L. and {Krachmalnicoff}, N. and {Kunz}, M. and {Kurki-Suonio}, H. and {Lagache}, G. and {Lamarre}, J.-M. and {Lasenby}, A. and {Lattanzi}, M. and {Lawrence}, C.~R. and {Le Jeune}, M. and {Lemos}, P. and {Lesgourgues}, J. and {Levrier}, F. and {Lewis}, A. and {Liguori}, M. and {Lilje}, P.~B. and {Lilley}, M. and {Lindholm}, V. and {L{\'o}pez-Caniego}, M. and {Lubin}, P.~M. and {Ma}, Y.-Z. and {Mac{\'\i}as-P{\'e}rez}, J.~F. and {Maggio}, G. and {Maino}, D. and {Mandolesi}, N. and {Mangilli}, A. and {Marcos-Caballero}, A. and {Maris}, M. and {Martin}, P.~G. and {Martinelli}, M. and {Mart{\'\i}nez-Gonz{\'a}lez}, E. and {Matarrese}, S. and {Mauri}, N. and {McEwen}, J.~D. and {Meinhold}, P.~R. and {Melchiorri}, A. and {Mennella}, A. and {Migliaccio}, M. and {Millea}, M. and {Mitra}, S. and {Miville-Desch{\^e}nes}, M.-A. and {Molinari}, D. and {Montier}, L. and {Morgante}, G. and {Moss}, A. and {Natoli}, P. and {N{\o}rgaard-Nielsen}, H.~U. and {Pagano}, L. and {Paoletti}, D. and {Partridge}, B. and {Patanchon}, G. and {Peiris}, H.~V. and {Perrotta}, F. and {Pettorino}, V. and {Piacentini}, F. and {Polastri}, L. and {Polenta}, G. and {Puget}, J.-L. and {Rachen}, J.~P. and {Reinecke}, M. and {Remazeilles}, M. and {Renzi}, A. and {Rocha}, G. and {Rosset}, C. and {Roudier}, G. and {Rubi{\~n}o-Mart{\'\i}n}, J.~A. and {Ruiz-Granados}, B. and {Salvati}, L. and {Sandri}, M. and {Savelainen}, M. and {Scott}, D. and {Shellard}, E.~P.~S. and {Sirignano}, C. and {Sirri}, G. and {Spencer}, L.~D. and {Sunyaev}, R. and {Suur-Uski}, A.-S. and {Tauber}, J.~A. and {Tavagnacco}, D. and {Tenti}, M. and {Toffolatti}, L. and {Tomasi}, M. and {Trombetti}, T. and {Valenziano}, L. and {Valiviita}, J. and {Van Tent}, B. and {Vibert}, L. and {Vielva}, P. and {Villa}, F. and {Vittorio}, N. and {Wandelt}, B.~D. and {Wehus}, I.~K. and {White}, M. and {White}, S.~D.~M. and {Zacchei}, A. and {Zonca}, A.},
        title = "{Planck 2018 results. VI. Cosmological parameters}",
      journal = {\aap},
         year = 2020,
        month = sep,
       volume = {641},
          eid = {A6},
        pages = {A6},
          doi = {10.1051/0004-6361/201833910},
archivePrefix = {arXiv},
       eprint = {1807.06209},
 primaryClass = {astro-ph.CO},
       adsurl = {https://ui.adsabs.harvard.edu/abs/2020A&A...641A...6P}
}

@INPROCEEDINGS{Larkin2006,
       author = {{Larkin}, James and {Barczys}, Matthew and {Krabbe}, Alfred and {Adkins}, Sean and {Aliado}, Ted and {Amico}, Paola and {Brims}, George and {Campbell}, Randy and {Canfield}, John and {Gasaway}, Thomas and {Honey}, Allan and {Iserlohe}, Christof and {Johnson}, Chris and {Kress}, Evan and {LaFreniere}, David and {Lyke}, James and {Magnone}, Ken and {Magnone}, Nick and {McElwain}, Michael and {Moon}, Juleen and {Quirrenbach}, Andreas and {Skulason}, Gunnar and {Song}, Inseok and {Spencer}, Michael and {Weiss}, Jason and {Wright}, Shelley},
        title = "{OSIRIS: a diffraction limited integral field spectrograph for Keck}",
    booktitle = {Ground-based and Airborne Instrumentation for Astronomy},
         year = 2006,
       editor = {{McLean}, Ian S. and {Iye}, Masanori},
       series = {Society of Photo-Optical Instrumentation Engineers (SPIE) Conference Series},
       volume = {6269},
        month = jun,
          eid = {62691A},
        pages = {62691A},
          doi = {10.1117/12.672061},
       adsurl = {https://ui.adsabs.harvard.edu/abs/2006SPIE.6269E..1AL}
}

@ARTICLE{Ulivi2025,
       author = {{Ulivi}, L. and {Mannucci}, F. and {Scialpi}, M. and {Marconcini}, C. and {Cresci}, G. and {Marconi}, A. and {Feltre}, A. and {Ginolfi}, M. and {Ricci}, F. and {Sluse}, D. and {Belfiore}, F. and {Bertola}, E. and {Bracci}, C. and {Cataldi}, E. and {Ceci}, M. and {D'Amato}, Q. and {Lamperti}, I. and {Metcalf}, R.~B. and {Moreschini}, B. and {Perna}, M. and {Tozzi}, G. and {Venturi}, G. and {Zanchettin}, M.~V. and {Fu}, Y. and {Huertas-Company}, M. and {Lines}, N.~E.~P. and {Mezcua}, M. and {P{\"o}ntinen}, M. and {Rojas}, K. and {Scottez}, V. and {Siudek}, M. and {Teimoorinia}, H. and {Andika}, I.~T. and {Acevedo Barroso}, J.~A. and {Cl{\'e}ment}, B. and {Courbin}, F. and {Gavazzi}, R. and {Ecker}, L.~R. and {Nagam}, B.~C. and {Pearce-Casey}, R. and {Schuldt}, S. and {Vincken}, S.~H. and {Stern}, D. and {Chakraborty}, A. and {Andreon}, S. and {Auricchio}, N. and {Baccigalupi}, C. and {Baldi}, M. and {Balestra}, A. and {Bardelli}, S. and {Biviano}, A. and {Branchini}, E. and {Brescia}, M. and {Camera}, S. and {Ca{\~n}as-Herrera}, G. and {Capobianco}, V. and {Carbone}, C. and {Carretero}, J. and {Castellano}, M. and {Castignani}, G. and {Cavuoti}, S. and {Cimatti}, A. and {Colodro-Conde}, C. and {Congedo}, G. and {Conselice}, C.~J. and {Conversi}, L. and {Copin}, Y. and {Courtois}, H.~M. and {Cropper}, M. and {Da Silva}, A. and {Degaudenzi}, H. and {De Lucia}, G. and {Di Giorgio}, A.~M. and {Dolding}, C. and {Dole}, H. and {Dubath}, F. and {Duncan}, C.~A.~J. and {Dupac}, X. and {Dusini}, S. and {Escoffier}, S. and {Fabricius}, M. and {Farina}, M. and {Farinelli}, R. and {Faustini}, F. and {Ferriol}, S. and {Finelli}, F. and {Frailis}, M. and {Franceschi}, E. and {Fumana}, M. and {Galeotta}, S. and {George}, K. and {Gillis}, B. and {Giocoli}, C. and {Gracia-Carpio}, J. and {Grazian}, A. and {Grupp}, F. and {Haugan}, S.~V.~H. and {Holmes}, W. and {Hook}, I.~M. and {Hormuth}, F. and {Hornstrup}, A. and {Jahnke}, K. and {Jhabvala}, M. and {Joachimi}, B. and {Keih{\"a}nen}, E. and {Kermiche}, S. and {Kiessling}, A. and {Kubik}, B. and {K{\"u}mmel}, M. and {Kunz}, M. and {Kurki-Suonio}, H. and {Le Brun}, A.~M.~C. and {Ligori}, S. and {Lilje}, P.~B. and {Lindholm}, V. and {Lloro}, I. and {Mainetti}, G. and {Maino}, D. and {Maiorano}, E. and {Mansutti}, O. and {Marggraf}, O. and {Martinelli}, M. and {Martinet}, N. and {Marulli}, F. and {Massey}, R.~J. and {Medinaceli}, E. and {Mei}, S. and {Melchior}, M. and {Mellier}, Y. and {Meneghetti}, M. and {Merlin}, E. and {Meylan}, G. and {Mora}, A. and {Moresco}, M. and {Moscardini}, L. and {Nakajima}, R. and {Neissner}, C. and {Nichol}, R.~C. and {Niemi}, S.-M. and {Padilla}, C. and {Paltani}, S. and {Pasian}, F. and {Pedersen}, K. and {Percival}, W.~J. and {Pettorino}, V. and {Pires}, S. and {Polenta}, G. and {Poncet}, M. and {Popa}, L.~A. and {Pozzetti}, L. and {Raison}, F. and {Renzi}, A. and {Rhodes}, J. and {Riccio}, G. and {Romelli}, E. and {Roncarelli}, M. and {Saglia}, R. and {Sakr}, Z. and {Sapone}, D. and {Sartoris}, B. and {Schirmer}, M. and {Schneider}, P. and {Schrabback}, T. and {Secroun}, A. and {Seidel}, G. and {Serrano}, S. and {Simon}, P. and {Sirignano}, C. and {Sirri}, G. and {Stanco}, L. and {Steinwagner}, J. and {Tallada-Cresp{\'\i}}, P. and {Tavagnacco}, D. and {Taylor}, A.~N. and {Tereno}, I. and {Tessore}, N. and {Toft}, S. and {Toledo-Moreo}, R. and {Torradeflot}, F. and {Tutusaus}, I. and {Valentijn}, E.~A. and {Valenziano}, L. and {Valiviita}, J. and {Vassallo}, T. and {Verdoes Kleijn}, G. and {Veropalumbo}, A. and {Wang}, Y. and {Weller}, J. and {Zamorani}, G. and {Zerbi}, F.~M. and {Zucca}, E. and {Nucita}, A.~A.},
        title = "{Euclid: A machine-learning search for dual and lensed AGN at sub-arcsec separations}",
      journal = {arXiv e-prints},
         year = 2025,
        month = aug,
          eid = {arXiv:2508.19494},
        pages = {arXiv:2508.19494},
          doi = {10.48550/arXiv.2508.19494},
archivePrefix = {arXiv},
       eprint = {2508.19494},
 primaryClass = {astro-ph.GA},
       adsurl = {https://ui.adsabs.harvard.edu/abs/2025arXiv250819494U}
}

@ARTICLE{Gross2025,
       author = {{Gross}, Arran C. and {Chen}, Yu-Ching and {Oguri}, Masamune and {Nolan}, Liam and {Liu}, Xin and {Shen}, Yue and {Zhuang}, Ming-Yang and {Li}, Junyao and {Zakamska}, Nadia L. and {Hwang}, Hsiang-Chih and {Ishikawa}, Yuzo},
        title = "{Varstrometry for Off-nucleus and Dual Sub-Kpc AGN (VODKA): A Mix of Singles, Lenses, and True Duals at Cosmic Noon}",
      journal = {\apj},
         year = 2025,
        month = aug,
       volume = {989},
       number = {1},
          eid = {112},
        pages = {112},
          doi = {10.3847/1538-4357/ade671},
archivePrefix = {arXiv},
       eprint = {2409.16363},
 primaryClass = {astro-ph.GA},
       adsurl = {https://ui.adsabs.harvard.edu/abs/2025ApJ...989..112G}
}

@ARTICLE{Schwartzman2025,
       author = {{Schwartzman}, Emma and {Fudolig}, Paula and {Clarke}, Tracy E. and {Nyland}, Kristina and {Secrest}, Nathan J. and {Pfeifle}, Ryan W. and {Schmitt}, Henrique and {Satyapal}, Shobita and {Rothberg}, Barry},
        title = "{Varstrometry for Dual Active Galactic Nuclei Using Radio Interferometry: VaDAR with the VLBA}",
      journal = {\apj},
         year = 2025,
        month = jul,
       volume = {987},
       number = {2},
          eid = {200},
        pages = {200},
          doi = {10.3847/1538-4357/add47c},
archivePrefix = {arXiv},
       eprint = {2503.02668},
 primaryClass = {astro-ph.GA},
       adsurl = {https://ui.adsabs.harvard.edu/abs/2025ApJ...987..200S}
}

@ARTICLE{Schwartzman2024,
       author = {{Schwartzman}, Emma and {Clarke}, Tracy E. and {Nyland}, Kristina and {Secrest}, Nathan J. and {Pfeifle}, Ryan W. and {Schmitt}, Henrique and {Satyapal}, Shobita and {Rothberg}, Barry},
        title = "{VaDAR: Varstrometry for Dual AGN Using Radio Interferometry}",
      journal = {\apj},
         year = 2024,
        month = feb,
       volume = {961},
       number = {2},
          eid = {233},
        pages = {233},
          doi = {10.3847/1538-4357/ad0ed0},
archivePrefix = {arXiv},
       eprint = {2306.13219},
 primaryClass = {astro-ph.GA},
       adsurl = {https://ui.adsabs.harvard.edu/abs/2024ApJ...961..233S}
}

@ARTICLE{Hwang2020,
       author = {{Hwang}, Hsiang-Chih and {Shen}, Yue and {Zakamska}, Nadia and {Liu}, Xin},
        title = "{Varstrometry for Off-nucleus and Dual Subkiloparsec AGN (VODKA): Methodology and Initial Results with Gaia DR2}",
      journal = {\apj},
         year = 2020,
        month = jan,
       volume = {888},
       number = {2},
          eid = {73},
        pages = {73},
          doi = {10.3847/1538-4357/ab5c1a},
archivePrefix = {arXiv},
       eprint = {1908.02292},
 primaryClass = {astro-ph.GA},
       adsurl = {https://ui.adsabs.harvard.edu/abs/2020ApJ...888...73H}
}

@ARTICLE{Shen2019,
       author = {{Shen}, Yue and {Hwang}, Hsiang-Chih and {Zakamska}, Nadia and {Liu}, Xin},
        title = "{Varstrometry for Off-nucleus and Dual Sub-Kpc AGN (VODKA): How Well Centered Are Low-z AGN?}",
      journal = {\apjl},
         year = 2019,
        month = nov,
       volume = {885},
       number = {1},
          eid = {L4},
        pages = {L4},
          doi = {10.3847/2041-8213/ab4b54},
archivePrefix = {arXiv},
       eprint = {1910.02969},
 primaryClass = {astro-ph.GA},
       adsurl = {https://ui.adsabs.harvard.edu/abs/2019ApJ...885L...4S}
}

@ARTICLE{Krone-Martins2019,
       author = {{Krone-Martins}, A. and {Graham}, M.~J. and {Stern}, D. and {Djorgovski}, S.~G. and {Delchambre}, L. and {Ducourant}, C. and {Teixeira}, R. and {Drake}, A.~J. and {Scarano}, Jr., S. and {Surdej}, J. and {Galluccio}, L. and {Jalan}, P. and {Wertz}, O. and {Kl{\"u}ter}, J. and {Mignard}, F. and {Spindola-Duarte}, C. and {Dobie}, D. and {Slezak}, E. and {Sluse}, D. and {Murphy}, T. and {Boehm}, C. and {Nierenberg}, A.~M. and {Bastian}, U. and {Wambsganss}, J. and {LeCampion}, J.-F.},
        title = "{Gaia GraL: Gaia DR2 Gravitational Lens Systems. V. Doubly-imaged QSOs discovered from entropy and wavelets}",
      journal = {arXiv e-prints},
         year = 2019,
        month = dec,
          eid = {arXiv:1912.08977},
        pages = {arXiv:1912.08977},
          doi = {10.48550/arXiv.1912.08977},
archivePrefix = {arXiv},
       eprint = {1912.08977},
 primaryClass = {astro-ph.GA},
       adsurl = {https://ui.adsabs.harvard.edu/abs/2019arXiv191208977K}
}

@ARTICLE{Kochanek2006,
       author = {{Kochanek}, C.~S. and {Mochejska}, B. and {Morgan}, N.~D. and {Stanek}, K.~Z.},
        title = "{A Simple Method to Find All Lensed Quasars}",
      journal = {\apjl},
         year = 2006,
        month = feb,
       volume = {637},
       number = {2},
        pages = {L73-L76},
          doi = {10.1086/500559},
archivePrefix = {arXiv},
       eprint = {astro-ph/0512031},
 primaryClass = {astro-ph},
       adsurl = {https://ui.adsabs.harvard.edu/abs/2006ApJ...637L..73K}
}

@ARTICLE{Lemon2017,
       author = {{Lemon}, Cameron A. and {Auger}, Matthew W. and {McMahon}, Richard G. and {Koposov}, Sergey E.},
        title = "{Gravitationally lensed quasars in Gaia: I. Resolving small-separation lenses}",
      journal = {\mnras},
         year = 2017,
        month = dec,
       volume = {472},
       number = {4},
        pages = {5023-5032},
          doi = {10.1093/mnras/stx2094},
archivePrefix = {arXiv},
       eprint = {1709.08976},
 primaryClass = {astro-ph.GA},
       adsurl = {https://ui.adsabs.harvard.edu/abs/2017MNRAS.472.5023L}
}

@ARTICLE{Shen2023,
       author = {{Shen}, Yue and {Casey-Clyde}, J. Andrew and {Chen}, Yu-Ching and {Gross}, Arran and {Habouzit}, Melanie and {Hwang}, Hsiang-Chih and {Ishikawa}, Yuzo and {Li}, Jun-Yao and {Liu}, Xin and {Mingarelli}, Chiara M.~F. and {Porquet}, D. and {Stemo}, Aaron and {Zhuang}, Ming-Yang},
        title = "{Discovery and Characterization of Galactic-scale Dual Supermassive Black Holes Across Cosmic Time}",
      journal = {arXiv e-prints},
         year = 2023,
        month = jun,
          eid = {arXiv:2306.15527},
        pages = {arXiv:2306.15527},
          doi = {10.48550/arXiv.2306.15527},
archivePrefix = {arXiv},
       eprint = {2306.15527},
 primaryClass = {astro-ph.GA},
       adsurl = {https://ui.adsabs.harvard.edu/abs/2023arXiv230615527S}
}

@ARTICLE{Perna2025,
       author = {{Perna}, Michele and {Arribas}, Santiago and {Lamperti}, Isabella and {Circosta}, Chiara and {Bertola}, Elena and {P{\'e}rez-Gonz{\'a}lez}, Pablo G. and {D'Eugenio}, Francesco and {{\"U}bler}, Hannah and {Cresci}, Giovanni and {Volonteri}, Marta and {Mannucci}, Filippo and {Maiolino}, Roberto and {Rodr{\'\i}guez Del Pino}, Bruno and {B{\"o}ker}, Torsten and {Bunker}, Andrew J. and {Charlot}, St{\'e}phane and {Willott}, Chris J. and {Carniani}, Stefano and {Curti}, Mirko and {Jones}, Gareth C. and {Kumari}, Nimisha and {Marshall}, Madeline A. and {Venturi}, Giacomo and {Saxena}, Aayush and {Scholtz}, Jan and {Witstok}, Joris},
        title = "{GA-NIFS: High number of dual active galactic nuclei at z {\ensuremath{\sim}} 3}",
      journal = {\aap},
         year = 2025,
        month = apr,
       volume = {696},
          eid = {A59},
        pages = {A59},
          doi = {10.1051/0004-6361/202453430},
archivePrefix = {arXiv},
       eprint = {2310.03067},
 primaryClass = {astro-ph.GA},
       adsurl = {https://ui.adsabs.harvard.edu/abs/2025A&A...696A..59P}
}

@ARTICLE{ubler2024,
       author = {{{\"U}bler}, Hannah and {Maiolino}, Roberto and {P{\'e}rez-Gonz{\'a}lez}, Pablo G. and {D'Eugenio}, Francesco and {Perna}, Michele and {Curti}, Mirko and {Arribas}, Santiago and {Bunker}, Andrew and {Carniani}, Stefano and {Charlot}, St{\'e}phane and {Rodr{\'\i}guez Del Pino}, Bruno and {Baker}, William and {B{\"o}ker}, Torsten and {Cresci}, Giovanni and {Dunlop}, James and {Grogin}, Norman A. and {Jones}, Gareth C. and {Kumari}, Nimisha and {Lamperti}, Isabella and {Laporte}, Nicolas and {Marshall}, Madeline A. and {Mazzolari}, Giovanni and {Parlanti}, Eleonora and {Rawle}, Tim and {Scholtz}, Jan and {Venturi}, Giacomo and {Witstok}, Joris},
        title = "{GA-NIFS: JWST discovers an offset AGN 740 million years after the big bang}",
      journal = {\mnras},
         year = 2024,
        month = jun,
       volume = {531},
       number = {1},
        pages = {355-365},
          doi = {10.1093/mnras/stae943},
archivePrefix = {arXiv},
       eprint = {2312.03589},
 primaryClass = {astro-ph.GA},
       adsurl = {https://ui.adsabs.harvard.edu/abs/2024MNRAS.531..355U}
}

@ARTICLE{Glikman2023,
       author = {{Glikman}, Eilat and {Langgin}, Rachel and {Johnstone}, Makoto A. and {Yoon}, Ilsang and {Comerford}, Julia M. and {Simmons}, Brooke D. and {Stacey}, Hannah and {Lacy}, Mark and {O'Meara}, John M.},
        title = "{A Candidate Dual QSO at Cosmic Noon}",
      journal = {\apjl},
         year = 2023,
        month = jul,
       volume = {951},
       number = {1},
          eid = {L18},
        pages = {L18},
          doi = {10.3847/2041-8213/acda2f},
archivePrefix = {arXiv},
       eprint = {2306.00068},
 primaryClass = {astro-ph.GA},
       adsurl = {https://ui.adsabs.harvard.edu/abs/2023ApJ...951L..18G}
}

@ARTICLE{Inada2008,
       author = {{Inada}, Naohisa and {Oguri}, Masamune and {Becker}, Robert H. and {Shin}, Min-Su and {Richards}, Gordon T. and {Hennawi}, Joseph F. and {White}, Richard L. and {Pindor}, Bartosz and {Strauss}, Michael A. and {Kochanek}, Christopher S. and {Johnston}, David E. and {Gregg}, Michael D. and {Kayo}, Issha and {Eisenstein}, Daniel and {Hall}, Patrick B. and {Castander}, Francisco J. and {Clocchiatti}, Alejandro and {Anderson}, Scott F. and {Schneider}, Donald P. and {York}, Donald G. and {Lupton}, Robert and {Chiu}, Kuenley and {Kawano}, Yozo and {Scranton}, Ryan and {Frieman}, Joshua A. and {Keeton}, Charles R. and {Morokuma}, Tomoki and {Rix}, Hans-Walter and {Turner}, Edwin L. and {Burles}, Scott and {Brunner}, Robert J. and {Sheldon}, Erin Scott and {Bahcall}, Neta A. and {Masataka}, Fukugita},
        title = "{The Sloan Digital Sky Survey Quasar Lens Search. II. Statistical Lens Sample from the Third Data Release}",
      journal = {\aj},
         year = 2008,
        month = feb,
       volume = {135},
       number = {2},
        pages = {496-511},
          doi = {10.1088/0004-6256/135/2/496},
archivePrefix = {arXiv},
       eprint = {0708.0828},
 primaryClass = {astro-ph},
       adsurl = {https://ui.adsabs.harvard.edu/abs/2008AJ....135..496I}
}

@ARTICLE{Hennawi2010,
       author = {{Hennawi}, Joseph F. and {Myers}, Adam D. and {Shen}, Yue and {Strauss}, Michael A. and {Djorgovski}, S.~G. and {Fan}, Xiaohui and {Glikman}, Eilat and {Mahabal}, Ashish and {Martin}, Crystal L. and {Richards}, Gordon T. and {Schneider}, Donald P. and {Shankar}, Francesco},
        title = "{Binary Quasars at High Redshift. I. 24 New Quasar Pairs at z \raisebox{-0.5ex}\textasciitilde 3-4}",
      journal = {\apj},
         year = 2010,
        month = aug,
       volume = {719},
       number = {2},
        pages = {1672-1692},
          doi = {10.1088/0004-637X/719/2/1672},
archivePrefix = {arXiv},
       eprint = {0908.3907},
 primaryClass = {astro-ph.CO},
       adsurl = {https://ui.adsabs.harvard.edu/abs/2010ApJ...719.1672H}
}

@ARTICLE{Hennawi2006,
       author = {{Hennawi}, Joseph F. and {Strauss}, Michael A. and {Oguri}, Masamune and {Inada}, Naohisa and {Richards}, Gordon T. and {Pindor}, Bartosz and {Schneider}, Donald P. and {Becker}, Robert H. and {Gregg}, Michael D. and {Hall}, Patrick B. and {Johnston}, David E. and {Fan}, Xiaohui and {Burles}, Scott and {Schlegel}, David J. and {Gunn}, James E. and {Lupton}, Robert H. and {Bahcall}, Neta A. and {Brunner}, Robert J. and {Brinkmann}, Jon},
        title = "{Binary Quasars in the Sloan Digital Sky Survey: Evidence for Excess Clustering on Small Scales}",
      journal = {\aj},
         year = 2006,
        month = jan,
       volume = {131},
       number = {1},
        pages = {1-23},
          doi = {10.1086/498235},
archivePrefix = {arXiv},
       eprint = {astro-ph/0504535},
 primaryClass = {astro-ph},
       adsurl = {https://ui.adsabs.harvard.edu/abs/2006AJ....131....1H}
}

@ARTICLE{Pfeifle2019b,
       author = {{Pfeifle}, Ryan W. and {Satyapal}, Shobita and {Manzano-King}, Christina and {Cann}, Jenna and {Sexton}, Remington O. and {Rothberg}, Barry and {Canalizo}, Gabriela and {Ricci}, Claudio and {Blecha}, Laura and {Ellison}, Sara L. and {Gliozzi}, Mario and {Secrest}, Nathan J. and {Constantin}, Anca and {Harvey}, Jenna B.},
        title = "{A Triple AGN in a Mid-infrared Selected Late-stage Galaxy Merger}",
      journal = {\apj},
         year = 2019,
        month = oct,
       volume = {883},
       number = {2},
          eid = {167},
        pages = {167},
          doi = {10.3847/1538-4357/ab3a9b},
archivePrefix = {arXiv},
       eprint = {1908.01732},
 primaryClass = {astro-ph.GA},
       adsurl = {https://ui.adsabs.harvard.edu/abs/2019ApJ...883..167P}
}

@ARTICLE{Satyapal2017,
       author = {{Satyapal}, Shobita and {Secrest}, Nathan J. and {Ricci}, Claudio and {Ellison}, Sara L. and {Rothberg}, Barry and {Blecha}, Laura and {Constantin}, Anca and {Gliozzi}, Mario and {McNulty}, Paul and {Ferguson}, Jason},
        title = "{Buried AGNs in Advanced Mergers: Mid-infrared Color Selection as a Dual AGN Candidate Finder}",
      journal = {\apj},
         year = 2017,
        month = oct,
       volume = {848},
       number = {2},
          eid = {126},
        pages = {126},
          doi = {10.3847/1538-4357/aa88ca},
archivePrefix = {arXiv},
       eprint = {1707.03921},
 primaryClass = {astro-ph.GA},
       adsurl = {https://ui.adsabs.harvard.edu/abs/2017ApJ...848..126S}
}

@ARTICLE{Fu2015,
       author = {{Fu}, Hai and {Myers}, A.~D. and {Djorgovski}, S.~G. and {Yan}, Lin and {Wrobel}, J.~M. and {Stockton}, A.},
        title = "{Radio-selected Binary Active Galactic Nuclei from the Very Large Array Stripe 82 Survey}",
      journal = {\apj},
         year = 2015,
        month = jan,
       volume = {799},
       number = {1},
          eid = {72},
        pages = {72},
          doi = {10.1088/0004-637X/799/1/72},
archivePrefix = {arXiv},
       eprint = {1411.0685},
 primaryClass = {astro-ph.GA},
       adsurl = {https://ui.adsabs.harvard.edu/abs/2015ApJ...799...72F}
}

@ARTICLE{Zhou2004,
       author = {{Zhou}, Hongyan and {Wang}, Tinggui and {Zhang}, Xueguang and {Dong}, Xiaobo and {Li}, Cheng},
        title = "{Obscured Binary Quasar Cores in SDSS J104807.74+005543.5?}",
      journal = {\apjl},
         year = 2004,
        month = mar,
       volume = {604},
       number = {1},
        pages = {L33-L36},
          doi = {10.1086/383310},
archivePrefix = {arXiv},
       eprint = {astro-ph/0411167},
 primaryClass = {astro-ph},
       adsurl = {https://ui.adsabs.harvard.edu/abs/2004ApJ...604L..33Z}
}

@ARTICLE{Comerford2009,
       author = {{Comerford}, Julia M. and {Gerke}, Brian F. and {Newman}, Jeffrey A. and {Davis}, Marc and {Yan}, Renbin and {Cooper}, Michael C. and {Faber}, S.~M. and {Koo}, David C. and {Coil}, Alison L. and {Rosario}, D.~J. and {Dutton}, Aaron A.},
        title = "{Inspiralling Supermassive Black Holes: A New Signpost for Galaxy Mergers}",
      journal = {\apj},
         year = 2009,
        month = jun,
       volume = {698},
       number = {1},
        pages = {956-965},
          doi = {10.1088/0004-637X/698/1/956},
archivePrefix = {arXiv},
       eprint = {0810.3235},
 primaryClass = {astro-ph},
       adsurl = {https://ui.adsabs.harvard.edu/abs/2009ApJ...698..956C}
}

@ARTICLE{Blecha18,
       author = {{Blecha}, Laura and {Snyder}, Gregory F. and {Satyapal}, Shobita and {Ellison}, Sara L.},
        title = "{The power of infrared AGN selection in mergers: a theoretical study}",
      journal = {\mnras},
         year = 2018,
        month = aug,
       volume = {478},
       number = {3},
        pages = {3056-3071},
          doi = {10.1093/mnras/sty1274},
archivePrefix = {arXiv},
       eprint = {1711.02094},
 primaryClass = {astro-ph.GA},
       adsurl = {https://ui.adsabs.harvard.edu/abs/2018MNRAS.478.3056B}
}

@ARTICLE{Ricci17,
       author = {{Ricci}, C. and {Bauer}, F.~E. and {Treister}, E. and {Schawinski}, K. and {Privon}, G.~C. and {Blecha}, L. and {Arevalo}, P. and {Armus}, L. and {Harrison}, F. and {Ho}, L.~C. and {Iwasawa}, K. and {Sanders}, D.~B. and {Stern}, D.},
        title = "{Growing supermassive black holes in the late stages of galaxy mergers are heavily obscured}",
      journal = {\mnras},
         year = 2017,
        month = jun,
       volume = {468},
       number = {2},
        pages = {1273-1299},
          doi = {10.1093/mnras/stx173},
archivePrefix = {arXiv},
       eprint = {1701.04825},
 primaryClass = {astro-ph.HE},
       adsurl = {https://ui.adsabs.harvard.edu/abs/2017MNRAS.468.1273R}
}

@article{koss18,
	title = {A population of luminous accreting black holes with hidden mergers},
	volume = {563},
	url = {https://ui.adsabs.harvard.edu/abs/2018Natur.563..214K},
	doi = {10.1038/s41586-018-0652-7},
	journal = {{\textbackslash}nat},
	author = {Koss, M. J. and Blecha, L. and Bernhard, P. and Hung, C.-L. and Lu, J. R. and Trakhtenbrot, B. and Treister, E. and Weigel, A. and Sartori, L. F. and Mushotzky, R. and Schawinski, K. and Ricci, C. and Veilleux, S. and Sanders, D. B.},
	month = nov,
	year = {2018},
	note = {\_eprint: 1811.03641},
	pages = {214--216},
}

@ARTICLE{Rosas-Guevara19,
       author = {{Rosas-Guevara}, Yetli M. and {Bower}, Richard G. and {McAlpine}, Stuart and {Bonoli}, Silvia and {Tissera}, Patricia B.},
        title = "{The abundances and properties of Dual AGN and their host galaxies in the EAGLE simulations}",
      journal = {\mnras},
         year = 2019,
        month = feb,
       volume = {483},
       number = {2},
        pages = {2712-2720},
          doi = {10.1093/mnras/sty3251},
archivePrefix = {arXiv},
       eprint = {1805.01479},
 primaryClass = {astro-ph.GA},
       adsurl = {https://ui.adsabs.harvard.edu/abs/2019MNRAS.483.2712R}
}

@ARTICLE{Steinborn16,
       author = {{Steinborn}, Lisa K. and {Dolag}, Klaus and {Comerford}, Julia M. and {Hirschmann}, Michaela and {Remus}, Rhea-Silvia and {Teklu}, Adelheid F.},
        title = "{Origin and properties of dual and offset active galactic nuclei in a cosmological simulation at z=2}",
      journal = {\mnras},
         year = 2016,
        month = may,
       volume = {458},
       number = {1},
        pages = {1013-1028},
          doi = {10.1093/mnras/stw316},
archivePrefix = {arXiv},
       eprint = {1510.08465},
 primaryClass = {astro-ph.GA},
       adsurl = {https://ui.adsabs.harvard.edu/abs/2016MNRAS.458.1013S}
}

@ARTICLE{Agazie2023,
       author = {{Agazie}, Gabriella and {Anumarlapudi}, Akash and {Archibald}, Anne M. and {Arzoumanian}, Zaven and {Baker}, Paul T. and {B{\'e}csy}, Bence and {Blecha}, Laura and {Brazier}, Adam and {Brook}, Paul R. and {Burke-Spolaor}, Sarah and {Burnette}, Rand and {Case}, Robin and {Charisi}, Maria and {Chatterjee}, Shami and {Chatziioannou}, Katerina and {Cheeseboro}, Belinda D. and {Chen}, Siyuan and {Cohen}, Tyler and {Cordes}, James M. and {Cornish}, Neil J. and {Crawford}, Fronefield and {Cromartie}, H. Thankful and {Crowter}, Kathryn and {Cutler}, Curt J. and {Decesar}, Megan E. and {Degan}, Dallas and {Demorest}, Paul B. and {Deng}, Heling and {Dolch}, Timothy and {Drachler}, Brendan and {Ellis}, Justin A. and {Ferrara}, Elizabeth C. and {Fiore}, William and {Fonseca}, Emmanuel and {Freedman}, Gabriel E. and {Garver-Daniels}, Nate and {Gentile}, Peter A. and {Gersbach}, Kyle A. and {Glaser}, Joseph and {Good}, Deborah C. and {G{\"u}ltekin}, Kayhan and {Hazboun}, Jeffrey S. and {Hourihane}, Sophie and {Islo}, Kristina and {Jennings}, Ross J. and {Johnson}, Aaron D. and {Jones}, Megan L. and {Kaiser}, Andrew R. and {Kaplan}, David L. and {Kelley}, Luke Zoltan and {Kerr}, Matthew and {Key}, Joey S. and {Klein}, Tonia C. and {Laal}, Nima and {Lam}, Michael T. and {Lamb}, William G. and {Lazio}, T. Joseph W. and {Lewandowska}, Natalia and {Littenberg}, Tyson B. and {Liu}, Tingting and {Lommen}, Andrea and {Lorimer}, Duncan R. and {Luo}, Jing and {Lynch}, Ryan S. and {Ma}, Chung-Pei and {Madison}, Dustin R. and {Mattson}, Margaret A. and {McEwen}, Alexander and {McKee}, James W. and {McLaughlin}, Maura A. and {McMann}, Natasha and {Meyers}, Bradley W. and {Meyers}, Patrick M. and {Mingarelli}, Chiara M.~F. and {Mitridate}, Andrea and {Natarajan}, Priyamvada and {Ng}, Cherry and {Nice}, David J. and {Ocker}, Stella Koch and {Olum}, Ken D. and {Pennucci}, Timothy T. and {Perera}, Benetge B.~P. and {Petrov}, Polina and {Pol}, Nihan S. and {Radovan}, Henri A. and {Ransom}, Scott M. and {Ray}, Paul S. and {Romano}, Joseph D. and {Sardesai}, Shashwat C. and {Schmiedekamp}, Ann and {Schmiedekamp}, Carl and {Schmitz}, Kai and {Schult}, Levi and {Shapiro-Albert}, Brent J. and {Siemens}, Xavier and {Simon}, Joseph and {Siwek}, Magdalena S. and {Stairs}, Ingrid H. and {Stinebring}, Daniel R. and {Stovall}, Kevin and {Sun}, Jerry P. and {Susobhanan}, Abhimanyu and {Swiggum}, Joseph K. and {Taylor}, Jacob and {Taylor}, Stephen R. and {Turner}, Jacob E. and {Unal}, Caner and {Vallisneri}, Michele and {van Haasteren}, Rutger and {Vigeland}, Sarah J. and {Wahl}, Haley M. and {Wang}, Qiaohong and {Witt}, Caitlin A. and {Young}, Olivia and {Nanograv Collaboration}},
        title = "{The NANOGrav 15 yr Data Set: Evidence for a Gravitational-wave Background}",
      journal = {\apjl},
         year = 2023,
        month = jul,
       volume = {951},
       number = {1},
          eid = {L8},
        pages = {L8},
          doi = {10.3847/2041-8213/acdac6},
archivePrefix = {arXiv},
       eprint = {2306.16213},
 primaryClass = {astro-ph.HE},
       adsurl = {https://ui.adsabs.harvard.edu/abs/2023ApJ...951L...8A}
}

@ARTICLE{Colpi2024,
       author = {{Colpi}, Monica and {Danzmann}, Karsten and {Hewitson}, Martin and {Holley-Bockelmann}, Kelly and {Jetzer}, Philippe and {Nelemans}, Gijs and {Petiteau}, Antoine and {Shoemaker}, David and {Sopuerta}, Carlos and {Stebbins}, Robin and {Tanvir}, Nial and {Ward}, Henry and {Weber}, William Joseph and {Thorpe}, Ira and {Daurskikh}, Anna and {Deep}, Atul and {Fern{\'a}ndez N{\'u}{\~n}ez}, Ignacio and {Garc{\'\i}a Marirrodriga}, C{\'e}sar and {Gehler}, Martin and {Halain}, Jean-Philippe and {Jennrich}, Oliver and {Lammers}, Uwe and {Larra{\~n}aga}, Jonan and {Lieser}, Maike and {L{\"u}tzgendorf}, Nora and {Martens}, Waldemar and {Mondin}, Linda and {Piris Ni{\~n}o}, Ana and {Amaro-Seoane}, Pau and {Arca Sedda}, Manuel and {Auclair}, Pierre and {Babak}, Stanislav and {Baghi}, Quentin and {Baibhav}, Vishal and {Baker}, Tessa and {Bayle}, Jean-Baptiste and {Berry}, Christopher and {Berti}, Emanuele and {Boileau}, Guillaume and {Bonetti}, Matteo and {Brito}, Richard and {Buscicchio}, Riccardo and {Calcagni}, Gianluca and {Capelo}, Pedro R. and {Caprini}, Chiara and {Caputo}, Andrea and {Castelli}, Eleonora and {Chen}, Hsin-Yu and {Chen}, Xian and {Chua}, Alvin and {Davies}, Gareth and {Derdzinski}, Andrea and {Domcke}, Valerie Fiona and {Doneva}, Daniela and {Dvorkin}, Irna and {Mar{\'\i}a Ezquiaga}, Jose and {Gair}, Jonathan and {Haiman}, Zoltan and {Harry}, Ian and {Hartwig}, Olaf and {Hees}, Aurelien and {Heffernan}, Anna and {Husa}, Sascha and {Izquierdo-Villalba}, David and {Karnesis}, Nikolaos and {Klein}, Antoine and {Korol}, Valeriya and {Korsakova}, Natalia and {Kupfer}, Thomas and {Laghi}, Danny and {Lamberts}, Astrid and {Larson}, Shane and {Le Jeune}, Maude and {Lewicki}, Marek and {Littenberg}, Tyson and {Madge}, Eric and {Mangiagli}, Alberto and {Marsat}, Sylvain and {Vilchez}, Ivan Martin and {Maselli}, Andrea and {Mathews}, Josh and {van de Meent}, Maarten and {Muratore}, Martina and {Nardini}, Germano and {Pani}, Paolo and {Peloso}, Marco and {Pieroni}, Mauro and {Pound}, Adam and {Quelquejay-Leclere}, Hippolyte and {Ricciardone}, Angelo and {Rossi}, Elena Maria and {Sartirana}, Andrea and {Savalle}, Etienne and {Sberna}, Laura and {Sesana}, Alberto and {Shoemaker}, Deirdre and {Slutsky}, Jacob and {Sotiriou}, Thomas and {Speri}, Lorenzo and {Staab}, Martin and {Steer}, Dani{\`e}le and {Tamanini}, Nicola and {Tasinato}, Gianmassimo and {Torrado}, Jesus and {Torres-Orjuela}, Alejandro and {Toubiana}, Alexandre and {Vallisneri}, Michele and {Vecchio}, Alberto and {Volonteri}, Marta and {Yagi}, Kent and {Zwick}, Lorenz},
        title = "{LISA Definition Study Report}",
      journal = {arXiv e-prints},
         year = 2024,
        month = feb,
          eid = {arXiv:2402.07571},
        pages = {arXiv:2402.07571},
          doi = {10.48550/arXiv.2402.07571},
archivePrefix = {arXiv},
       eprint = {2402.07571},
 primaryClass = {astro-ph.CO},
       adsurl = {https://ui.adsabs.harvard.edu/abs/2024arXiv240207571C}
}

@ARTICLE{Arzoumanian2018,
       author = {{Arzoumanian}, Z. and {Baker}, P.~T. and {Brazier}, A. and {Burke-Spolaor}, S. and {Chamberlin}, S.~J. and {Chatterjee}, S. and {Christy}, B. and {Cordes}, J.~M. and {Cornish}, N.~J. and {Crawford}, F. and {Thankful Cromartie}, H. and {Crowter}, K. and {DeCesar}, M. and {Demorest}, P.~B. and {Dolch}, T. and {Ellis}, J.~A. and {Ferdman}, R.~D. and {Ferrara}, E. and {Folkner}, W.~M. and {Fonseca}, E. and {Garver-Daniels}, N. and {Gentile}, P.~A. and {Haas}, R. and {Hazboun}, J.~S. and {Huerta}, E.~A. and {Islo}, K. and {Jones}, G. and {Jones}, M.~L. and {Kaplan}, D.~L. and {Kaspi}, V.~M. and {Lam}, M.~T. and {Lazio}, T.~J.~W. and {Levin}, L. and {Lommen}, A.~N. and {Lorimer}, D.~R. and {Luo}, J. and {Lynch}, R.~S. and {Madison}, D.~R. and {McLaughlin}, M.~A. and {McWilliams}, S.~T. and {Mingarelli}, C.~M.~F. and {Ng}, C. and {Nice}, D.~J. and {Park}, R.~S. and {Pennucci}, T.~T. and {Pol}, N.~S. and {Ransom}, S.~M. and {Ray}, P.~S. and {Rasskazov}, A. and {Siemens}, X. and {Simon}, J. and {Spiewak}, R. and {Stairs}, I.~H. and {Stinebring}, D.~R. and {Stovall}, K. and {Swiggum}, J. and {Taylor}, S.~R. and {Vallisneri}, M. and {van Haasteren}, R. and {Vigeland}, S. and {Zhu}, W.~W. and {NANOGrav Collaboration}},
        title = "{The NANOGrav 11 Year Data Set: Pulsar-timing Constraints on the Stochastic Gravitational-wave Background}",
      journal = {\apj},
         year = 2018,
        month = may,
       volume = {859},
       number = {1},
          eid = {47},
        pages = {47},
          doi = {10.3847/1538-4357/aabd3b},
archivePrefix = {arXiv},
       eprint = {1801.02617},
 primaryClass = {astro-ph.HE},
       adsurl = {https://ui.adsabs.harvard.edu/abs/2018ApJ...859...47A}
}

@ARTICLE{AmaroSeoane2023,
       author = {{Amaro-Seoane}, Pau and {Andrews}, Jeff and {Arca Sedda}, Manuel and {Askar}, Abbas and {Baghi}, Quentin and {Balasov}, Razvan and {Bartos}, Imre and {Bavera}, Simone S. and {Bellovary}, Jillian and {Berry}, Christopher P.~L. and {Berti}, Emanuele and {Bianchi}, Stefano and {Blecha}, Laura and {Blondin}, St{\'e}phane and {Bogdanovi{\'c}}, Tamara and {Boissier}, Samuel and {Bonetti}, Matteo and {Bonoli}, Silvia and {Bortolas}, Elisa and {Breivik}, Katelyn and {Capelo}, Pedro R. and {Caramete}, Laurentiu and {Cattorini}, Federico and {Charisi}, Maria and {Chaty}, Sylvain and {Chen}, Xian and {Chru{\'s}li{\'n}ska}, Martyna and {Chua}, Alvin J.~K. and {Church}, Ross and {Colpi}, Monica and {D'Orazio}, Daniel and {Danielski}, Camilla and {Davies}, Melvyn B. and {Dayal}, Pratika and {De Rosa}, Alessandra and {Derdzinski}, Andrea and {Destounis}, Kyriakos and {Dotti}, Massimo and {Du{\c{t}}an}, Ioana and {Dvorkin}, Irina and {Fabj}, Gaia and {Foglizzo}, Thierry and {Ford}, Saavik and {Fouvry}, Jean-Baptiste and {Franchini}, Alessia and {Fragos}, Tassos and {Fryer}, Chris and {Gaspari}, Massimo and {Gerosa}, Davide and {Graziani}, Luca and {Groot}, Paul and {Habouzit}, Melanie and {Haggard}, Daryl and {Haiman}, Zoltan and {Han}, Wen-Biao and {Istrate}, Alina and {Johansson}, Peter H. and {Khan}, Fazeel Mahmood and {Kimpson}, Tomas and {Kokkotas}, Kostas and {Kong}, Albert and {Korol}, Valeriya and {Kremer}, Kyle and {Kupfer}, Thomas and {Lamberts}, Astrid and {Larson}, Shane and {Lau}, Mike and {Liu}, Dongliang and {Lloyd-Ronning}, Nicole and {Lodato}, Giuseppe and {Lupi}, Alessandro and {Ma}, Chung-Pei and {Maccarone}, Tomas and {Mandel}, Ilya and {Mangiagli}, Alberto and {Mapelli}, Michela and {Mathis}, St{\'e}phane and {Mayer}, Lucio and {McGee}, Sean and {McKernan}, Berry and {Miller}, M. Coleman and {Mota}, David F. and {Mumpower}, Matthew and {Nasim}, Syeda S. and {Nelemans}, Gijs and {Noble}, Scott and {Pacucci}, Fabio and {Panessa}, Francesca and {Paschalidis}, Vasileios and {Pfister}, Hugo and {Porquet}, Delphine and {Quenby}, John and {Ricarte}, Angelo and {R{\"o}pke}, Friedrich K. and {Regan}, John and {Rosswog}, Stephan and {Ruiter}, Ashley and {Ruiz}, Milton and {Runnoe}, Jessie and {Schneider}, Raffaella and {Schnittman}, Jeremy and {Secunda}, Amy and {Sesana}, Alberto and {Seto}, Naoki and {Shao}, Lijing and {Shapiro}, Stuart and {Sopuerta}, Carlos and {Stone}, Nicholas C. and {Suvorov}, Arthur and {Tamanini}, Nicola and {Tamfal}, Tomas and {Tauris}, Thomas and {Temmink}, Karel and {Tomsick}, John and {Toonen}, Silvia and {Torres-Orjuela}, Alejandro and {Toscani}, Martina and {Tsokaros}, Antonios and {Unal}, Caner and {V{\'a}zquez-Aceves}, Ver{\'o}nica and {Valiante}, Rosa and {van Putten}, Maurice and {van Roestel}, Jan and {Vignali}, Christian and {Volonteri}, Marta and {Wu}, Kinwah and {Younsi}, Ziri and {Yu}, Shenghua and {Zane}, Silvia and {Zwick}, Lorenz and {Antonini}, Fabio and {Baibhav}, Vishal and {Barausse}, Enrico and {Bonilla Rivera}, Alexander and {Branchesi}, Marica and {Branduardi-Raymont}, Graziella and {Burdge}, Kevin and {Chakraborty}, Srija and {Cuadra}, Jorge and {Dage}, Kristen and {Davis}, Benjamin and {de Mink}, Selma E. and {Decarli}, Roberto and {Doneva}, Daniela and {Escoffier}, Stephanie and {Gandhi}, Poshak and {Haardt}, Francesco and {Lousto}, Carlos O. and {Nissanke}, Samaya and {Nordhaus}, Jason and {O'Shaughnessy}, Richard and {Portegies Zwart}, Simon and {Pound}, Adam and {Schussler}, Fabian and {Sergijenko}, Olga and {Spallicci}, Alessandro and {Vernieri}, Daniele and {Vigna-G{\'o}mez}, Alejandro},
        title = "{Astrophysics with the Laser Interferometer Space Antenna}",
      journal = {Living Reviews in Relativity},
         year = 2023,
        month = dec,
       volume = {26},
       number = {1},
          eid = {2},
        pages = {2},
          doi = {10.1007/s41114-022-00041-y},
archivePrefix = {arXiv},
       eprint = {2203.06016},
 primaryClass = {gr-qc},
       adsurl = {https://ui.adsabs.harvard.edu/abs/2023LRR....26....2A}
}

@ARTICLE{White78,
       author = {{White}, S.~D.~M. and {Rees}, M.~J.},
        title = "{Core condensation in heavy halos: a two-stage theory for galaxy formation and clustering.}",
      journal = {\mnras},
         year = 1978,
        month = may,
       volume = {183},
        pages = {341-358},
          doi = {10.1093/mnras/183.3.341},
       adsurl = {https://ui.adsabs.harvard.edu/abs/1978MNRAS.183..341W}
}

@ARTICLE{Begelman80,
       author = {{Begelman}, M.~C. and {Blandford}, R.~D. and {Rees}, M.~J.},
        title = "{Massive black hole binaries in active galactic nuclei}",
      journal = {\nat},
         year = 1980,
        month = sep,
       volume = {287},
       number = {5780},
        pages = {307-309},
          doi = {10.1038/287307a0},
       adsurl = {https://ui.adsabs.harvard.edu/abs/1980Natur.287..307B}
}

@ARTICLE{Colpi14,
       author = {{Colpi}, Monica},
        title = "{Massive Binary Black Holes in Galactic Nuclei and Their Path to Coalescence}",
      journal = {\ssr},
         year = 2014,
        month = sep,
       volume = {183},
       number = {1-4},
        pages = {189-221},
          doi = {10.1007/s11214-014-0067-1},
archivePrefix = {arXiv},
       eprint = {1407.3102},
 primaryClass = {astro-ph.GA},
       adsurl = {https://ui.adsabs.harvard.edu/abs/2014SSRv..183..189C}
}

@ARTICLE{Volonteri22,
       author = {{Volonteri}, Marta and {Pfister}, Hugo and {Beckmann}, Ricarda and {Dotti}, Massimo and {Dubois}, Yohan and {Massonneau}, Warren and {Musoke}, Gibwa and {Tremmel}, Michael},
        title = "{Dual AGN in the Horizon-AGN simulation and their link to galaxy and massive black hole mergers, with an excursus on multiple AGN}",
      journal = {\mnras},
         year = 2022,
        month = jul,
       volume = {514},
       number = {1},
        pages = {640-656},
          doi = {10.1093/mnras/stac1217},
archivePrefix = {arXiv},
       eprint = {2112.07193},
 primaryClass = {astro-ph.GA},
       adsurl = {https://ui.adsabs.harvard.edu/abs/2022MNRAS.514..640V}
}

@ARTICLE{efosc2_buzzoni84,
       author = {{Buzzoni}, B. and {Delabre}, B. and {Dekker}, H. and {Dodorico}, S. and {Enard}, D. and {Focardi}, P. and {Gustafsson}, B. and {Nees}, W. and {Paureau}, J. and {Reiss}, R.},
        title = "{The ESO Faint Object Spectrograph and Camera / EFOSC}",
      journal = {The Messenger},
         year = 1984,
        month = dec,
       volume = {38},
        pages = {9},
       adsurl = {https://ui.adsabs.harvard.edu/abs/1984Msngr..38....9B}
}

@ARTICLE{fu2024,
       author = {{Fu}, Yuming and {Wu}, Xue-Bing and {Li}, Yifan and {Pang}, Yuxuan and {Joshi}, Ravi and {Zhang}, Shuo and {Wang}, Qiyue and {Yang}, Jing and {Ng}, FanLam and {Liu}, Xingjian and {Qiu}, Yu and {Zhu}, Rui and {Wang}, Huimei and {Wolf}, Christian and {Zhang}, Yanxia and {Huo}, Zhi-Ying and {Ai}, Y.~L. and {Ma}, Qinchun and {Feng}, Xiaotong and {Bouwens}, R.~J.},
        title = "{CatNorth: An Improved Gaia DR3 Quasar Candidate Catalog with Pan-STARRS1 and CatWISE}",
      journal = {\apjs},
         year = 2024,
        month = apr,
       volume = {271},
       number = {2},
          eid = {54},
        pages = {54},
          doi = {10.3847/1538-4365/ad2ae6},
archivePrefix = {arXiv},
       eprint = {2310.12704},
 primaryClass = {astro-ph.GA},
       adsurl = {https://ui.adsabs.harvard.edu/abs/2024ApJS..271...54F}
}

@ARTICLE{wise_ps1_2022,
       author = {{Kuns{\'a}gi-M{\'a}t{\'e}}, S{\'a}ndor and {Beck}, R{\'o}bert and {Szapudi}, Istv{\'a}n and {Csabai}, Istv{\'a}n},
        title = "{Photometric redshifts for quasars from WISE-PS1-STRM}",
      journal = {\mnras},
         year = 2022,
        month = oct,
       volume = {516},
       number = {2},
        pages = {2662-2670},
          doi = {10.1093/mnras/stac2411},
archivePrefix = {arXiv},
       eprint = {2206.01440},
 primaryClass = {astro-ph.IM},
       adsurl = {https://ui.adsabs.harvard.edu/abs/2022MNRAS.516.2662K}
}

@ARTICLE{quaia_Storey-Fisher2024,
       author = {{Storey-Fisher}, Kate and {Hogg}, David W. and {Rix}, Hans-Walter and {Eilers}, Anna-Christina and {Fabbian}, Giulio and {Blanton}, Michael R. and {Alonso}, David},
        title = "{Quaia, the Gaia-unWISE Quasar Catalog: An All-sky Spectroscopic Quasar Sample}",
      journal = {\apj},
         year = 2024,
        month = mar,
       volume = {964},
       number = {1},
          eid = {69},
        pages = {69},
          doi = {10.3847/1538-4357/ad1328},
archivePrefix = {arXiv},
       eprint = {2306.17749},
 primaryClass = {astro-ph.GA},
       adsurl = {https://ui.adsabs.harvard.edu/abs/2024ApJ...964...69S}
}

@ARTICLE{salvato2025,
       author = {{Salvato}, M. and {Wolf}, J. and {Dwelly}, T. and {Starck}, H. and {Buchner}, J. and {Shirley}, R. and {Merloni}, A. and {Georgakakis}, A. and {Balzer}, F. and {Brusa}, M. and {Rau}, A. and {Freund}, S. and {Lang}, D. and {Liu}, T. and {Lamer}, G. and {Schwope}, A. and {Roster}, W. and {Waddell}, S. and {Scialpi}, M. and {Igo}, Z. and {Kluge}, M. and {Mannucci}, F. and {Tiwari}, S. and {Homan}, D. and {Krumpe}, M. and {Zenteno}, A. and {Hernandez-Lang}, D. and {Comparat}, J. and {Fabricius}, M. and {Snigula}, J. and {Schlegel}, D. and {Weaver}, B.~A. and {Zhou}, R. and {Dey}, A. and {Valdes}, F. and {Myers}, A. and {Juneau}, S. and {Winkler}, H. and {Marquez}, I. and {di Mille}, F. and {Ciroi}, S. and {Schramm}, M. and {Buckley}, D.~A.~H. and {Brink}, J. and {Gromadzki}, M. and {Robrade}, J. and {Nandra}, K.},
        title = "{Counterpart identification and classification for eRASS1 and characterisation of the active galactic nuclei content}",
      journal = {\aap},
         year = 2025,
        month = dec,
       volume = {704},
          eid = {A344},
        pages = {A344},
          doi = {10.1051/0004-6361/202556142},
archivePrefix = {arXiv},
       eprint = {2509.02842},
 primaryClass = {astro-ph.HE},
       adsurl = {https://ui.adsabs.harvard.edu/abs/2025A&A...704A.344S}
}

@ARTICLE{delchambre2022,
       author = {{Delchambre}, L. and {Bailer-Jones}, C.~A.~L. and {Bellas-Velidis}, I. and {Drimmel}, R. and {Garabato}, D. and {Carballo}, R. and {Hatzidimitriou}, D. and {Marshall}, D.~J. and {Andrae}, R. and {Dafonte}, C. and {Livanou}, E. and {Fouesneau}, M. and {Licata}, E.~L. and {Lindstr{\o}m}, H.~E.~P. and {Manteiga}, M. and {Robin}, C. and {Silvelo}, A. and {Abreu Aramburu}, A. and {{\'A}lvarez}, M.~A. and {Bakker}, J. and {Bijaoui}, A. and {Brouillet}, N. and {Brugaletta}, E. and {Burlacu}, A. and {Casamiquela}, L. and {Chaoul}, L. and {Chiavassa}, A. and {Contursi}, G. and {Cooper}, W.~J. and {Creevey}, O.~L. and {Dapergolas}, A. and {de Laverny}, P. and {Demouchy}, C. and {Dharmawardena}, T.~E. and {Edvardsson}, B. and {Fr{\'e}mat}, Y. and {Garc{\'\i}a-Lario}, P. and {Garc{\'\i}a-Torres}, M. and {Gavel}, A. and {Gomez}, A. and {Gonz{\'a}lez-Santamar{\'\i}a}, I. and {Heiter}, U. and {Jean-Antoine Piccolo}, A. and {Kontizas}, M. and {Kordopatis}, G. and {Korn}, A.~J. and {Lanzafame}, A.~C. and {Lebreton}, Y. and {Lobel}, A. and {Lorca}, A. and {Magdaleno Romeo}, A. and {Marocco}, F. and {Mary}, N. and {Nicolas}, C. and {Ordenovic}, C. and {Pailler}, F. and {Palicio}, P.~A. and {Pallas-Quintela}, L. and {Panem}, C. and {Pichon}, B. and {Poggio}, E. and {Recio-Blanco}, A. and {Riclet}, F. and {Rybizki}, J. and {Santove{\~n}a}, R. and {Sarro}, L.~M. and {Schultheis}, M.~S. and {Segol}, M. and {Slezak}, I. and {Smart}, R.~L. and {Sordo}, R. and {Soubiran}, C. and {S{\"u}veges}, M. and {Th{\'e}venin}, F. and {Torralba Elipe}, G. and {Ulla}, A. and {Utrilla}, E. and {Vallenari}, A. and {van Dillen}, E. and {Zhao}, H. and {Zorec}, J.},
        title = "{Gaia Data Release 3. Apsis. III. Non-stellar content and source classification}",
      journal = {\aap},
         year = 2023,
        month = jun,
       volume = {674},
          eid = {A31},
        pages = {A31},
          doi = {10.1051/0004-6361/202243423},
archivePrefix = {arXiv},
       eprint = {2206.06710},
 primaryClass = {astro-ph.GA},
       adsurl = {https://ui.adsabs.harvard.edu/abs/2023A&A...674A..31D}
}

@ARTICLE{flesch2023,
       author = {{Flesch}, Eric Wim},
        title = "{The Million Quasars (Milliquas) Catalogue, v8}",
      journal = {The Open Journal of Astrophysics},
         year = 2023,
        month = dec,
       volume = {6},
          eid = {49},
        pages = {49},
          doi = {10.21105/astro.2308.01505},
archivePrefix = {arXiv},
       eprint = {2308.01505},
 primaryClass = {astro-ph.GA},
       adsurl = {https://ui.adsabs.harvard.edu/abs/2023OJAp....6E..49F}
}

@ARTICLE{flesch2021,
       author = {{Flesch}, Eric Wim},
        title = "{The Million Quasars (Milliquas) v7.2 Catalogue, now with VLASS associations. The inclusion of SDSS-DR16Q quasars is detailed}",
      journal = {arXiv e-prints},
         year = 2021,
        month = may,
          eid = {arXiv:2105.12985},
        pages = {arXiv:2105.12985},
archivePrefix = {arXiv},
       eprint = {2105.12985},
 primaryClass = {astro-ph.GA},
       adsurl = {https://ui.adsabs.harvard.edu/abs/2021arXiv210512985F}
}

@ARTICLE{DESI_Karim2026,
       author = {{DESI Collaboration} and {Abdul Karim}, M. and {Adame}, A.~G. and {Aguado}, D. and {Aguilar}, J. and {Ahlen}, S. and {Alam}, S. and {Aldering}, G. and {Alexander}, D.~M. and {Alfarsy}, R. and {Allen}, L. and {Allende Prieto}, C. and {Alves}, O. and {Anand}, A. and {Andrade}, U. and {Armengaud}, E. and {Avila}, S. and {Aviles}, A. and {Awan}, H. and {Bailey}, S. and {Baleato Lizancos}, A. and {Ballester}, O. and {Bault}, A. and {Bautista}, J. and {Bean}, R. and {Behera}, J. and {BenZvi}, S. and {Beraldo e Silva}, L. and {Bermejo-Climent}, J.~R. and {Beutler}, F. and {Bianchi}, D. and {Blake}, C. and {Blum}, R. and {Bolton}, A.~S. and {Bonici}, M. and {Brieden}, S. and {Brodzeller}, A. and {Brooks}, D. and {Buckley-Geer}, E. and {Burtin}, E. and {Bystr{\"o}m}, A. and {Canning}, R. and {Carnero Rosell}, A. and {Carr}, A. and {Carrilho}, P. and {Casas}, L. and {Castander}, F.~J. and {Cereskaite}, R. and {Cervantes-Cota}, J.~L. and {Chaussidon}, E. and {Chaves-Montero}, J. and {Chen}, S. and {Chen}, X. and {Circosta}, C. and {Claybaugh}, T. and {Cole}, S. and {Cooper}, A.~P. and {Cousinou}, M.-C. and {Cuceu}, A. and {Davis}, T.~M. and {Dawson}, K.~S. and {de Belsunce}, R. and {de la Cruz}, R. and {de la Macorra}, A. and {de Mattia}, A. and {Deiosso}, N. and {Della Costa}, J. and {Demina}, R. and {Demirbozan}, U. and {DeRose}, J. and {Dey}, A. and {Dey}, B. and {Ding}, J. and {Ding}, Z. and {Doel}, P. and {Douglass}, K. and {Dowicz}, M. and {Ebina}, H. and {Edelstein}, J. and {Eisenstein}, D.~J. and {Elbers}, W. and {Emas}, N. and {Escoffier}, S. and {Fagrelius}, P. and {Fan}, X. and {Fanning}, K. and {Favole}, G. and {Fawcett}, V.~A. and {Fern{\'a}ndez-Garc{\'\i}a}, E. and {Ferraro}, S. and {Findlay}, N. and {Font-Ribera}, A. and {Forero-Romero}, J.~E. and {Forero-S{\'a}nchez}, D. and {Frenk}, C.~S. and {G{\"a}nsicke}, B.~T. and {Galbany}, L. and {Garc{\'\i}a-Bellido}, J. and {Garcia-Quintero}, C. and {Garrison}, L.~H. and {Gazta{\~n}aga}, E. and {Gil-Mar{\'\i}n}, H. and {Gloudemans}, A. and {Gnedin}, O.~Y. and {Gontcho A Gontcho}, S. and {Gonzalez}, D. and {Gonzalez-Morales}, A.~X. and {Gonzalez-Perez}, V. and {Gordon}, C. and {Graur}, O. and {Green}, D. and {Gruen}, D. and {Gsponer}, R. and {Guandalin}, C. and {Gutierrez}, G. and {Guy}, J. and {Hahn}, C. and {Han}, J.~J. and {Han}, J. and {He}, S. and {Herrera-Alcantar}, H.~K. and {Heydenreich}, S. and {Honscheid}, K. and {Hou}, J. and {Howlett}, C. and {Huterer}, D. and {Ir{\v{s}}i{\v{c}}}, V. and {Ishak}, M. and {Jacques}, A. and {Jiang}, L. and {Jimenez}, J. and {Jing}, Y.~P. and {Joachimi}, B. and {Joudaki}, S. and {Joyce}, R. and {Jullo}, E. and {Juneau}, S. and {Kara{\c{c}}ayl{\i}}, N.~G. and {Karim}, T. and {Kehoe}, R. and {Kent}, S. and {Khederlarian}, A. and {Kirkby}, D. and {Kisner}, T. and {Kitaura}, F.-S. and {Kizhuprakkat}, N. and {Kong}, H. and {Koposov}, S.~E. and {Kremin}, A. and {Krolewski}, A. and {Lahav}, O. and {Lai}, Y. and {Lamman}, C. and {Lan}, T.-W. and {Landriau}, M. and {Lang}, D. and {Lange}, J.~U. and {Lasker}, J. and {Le Goff}, J.~M. and {Le Guillou}, L. and {Leauthaud}, A. and {Levi}, M.~E. and {Li}, S. and {Li}, T.~S. and {Liu}, W. and {Lodha}, K. and {Lokken}, M. and {Luo}, Y. and {Magneville}, C. and {Manera}, M. and {Manser}, C.~J. and {Margala}, D. and {Martini}, P. and {Maus}, M. and {McCullough}, J. and {McDonald}, P. and {Medina}, G.~E. and {Medina-Varela}, L. and {Meisner}, A. and {Mena-Fern{\'a}ndez}, J. and {Menegas}, A. and {Meneses-Rizo}, J. and {Mezcua}, M. and {Miquel}, R. and {Montero-Camacho}, P. and {Moon}, J. and {Moustakas}, J. and {Mu{\~n}oz-Guti{\'e}rrez}, A. and {Mu noz-Santos}, D. and {Myers}, A.~D. and {Myles}, J. and {Nadathur}, S. and {Najita}, J. and {Napolitano}, L. and {Newman}, J.~A. and {Nikakhtar}, F. and {Nikutta}, R. and {Niz}, G. and {Noriega}, H.~E. and {Nugent}, P.},
        title = "{Data Release 1 of the Dark Energy Spectroscopic Instrument}",
      journal = {\aj},
         year = 2026,
        month = may,
       volume = {171},
       number = {5},
          eid = {285},
        pages = {285},
          doi = {10.3847/1538-3881/ae4c43},
archivePrefix = {arXiv},
       eprint = {2503.14745},
 primaryClass = {astro-ph.CO},
       adsurl = {https://ui.adsabs.harvard.edu/abs/2026AJ....171..285D}
}

@ARTICLE{SDSS_lyke2020,
       author = {{Lyke}, Brad W. and {Higley}, Alexandra N. and {McLane}, J.~N. and {Schurhammer}, Danielle P. and {Myers}, Adam D. and {Ross}, Ashley J. and {Dawson}, Kyle and {Chabanier}, Sol{\`e}ne and {Martini}, Paul and {Busca}, Nicol{\'a}s G. and {Mas des Bourboux}, H{\'e}lion du and {Salvato}, Mara and {Streblyanska}, Alina and {Zarrouk}, Pauline and {Burtin}, Etienne and {Anderson}, Scott F. and {Bautista}, Julian and {Bizyaev}, Dmitry and {Brandt}, W.~N. and {Brinkmann}, Jonathan and {Brownstein}, Joel R. and {Comparat}, Johan and {Green}, Paul and {de la Macorra}, Axel and {Mu{\~n}oz Guti{\'e}rrez}, Andrea and {Hou}, Jiamin and {Newman}, Jeffrey A. and {Palanque-Delabrouille}, Nathalie and {P{\^a}ris}, Isabelle and {Percival}, Will J. and {Petitjean}, Patrick and {Rich}, James and {Rossi}, Graziano and {Schneider}, Donald P. and {Smith}, Alexander and {Vivek}, M. and {Weaver}, Benjamin Alan},
        title = "{The Sloan Digital Sky Survey Quasar Catalog: Sixteenth Data Release}",
      journal = {\apjs},
         year = 2020,
        month = sep,
       volume = {250},
       number = {1},
          eid = {8},
        pages = {8},
          doi = {10.3847/1538-4365/aba623},
archivePrefix = {arXiv},
       eprint = {2007.09001},
 primaryClass = {astro-ph.GA},
       adsurl = {https://ui.adsabs.harvard.edu/abs/2020ApJS..250....8L}
}

@ARTICLE{chen23b,
       author = {{Chen}, Nianyi and {Di Matteo}, Tiziana and {Ni}, Yueying and {Tremmel}, Michael and {DeGraf}, Colin and {Shen}, Yue and {Holgado}, A. Miguel and {Bird}, Simeon and {Croft}, Rupert and {Feng}, Yu},
        title = "{Properties and evolution of dual and offset AGN in the ASTRID simulation at z   2}",
      journal = {\mnras},
         year = 2023,
        month = jun,
       volume = {522},
       number = {2},
        pages = {1895-1913},
          doi = {10.1093/mnras/stad834},
archivePrefix = {arXiv},
       eprint = {2208.04970},
 primaryClass = {astro-ph.GA},
       adsurl = {https://ui.adsabs.harvard.edu/abs/2023MNRAS.522.1895C}
}

@ARTICLE{Gaia_EDR3_1,
       author = {{Gaia Collaboration} and {Brown}, A.~G.~A. and {Vallenari}, A. and {Prusti}, T. and {de Bruijne}, J.~H.~J. and {Babusiaux}, C. and {Biermann}, M. and {Creevey}, O.~L. and {Evans}, D.~W. and {Eyer}, L. and {Hutton}, A. and {Jansen}, F. and {Jordi}, C. and {Klioner}, S.~A. and {Lammers}, U. and {Lindegren}, L. and {Luri}, X. and {Mignard}, F. and {Panem}, C. and {Pourbaix}, D. and {Randich}, S. and {Sartoretti}, P. and {Soubiran}, C. and {Walton}, N.~A. and {Arenou}, F. and {Bailer-Jones}, C.~A.~L. and {Bastian}, U. and {Cropper}, M. and {Drimmel}, R. and {Katz}, D. and {Lattanzi}, M.~G. and {van Leeuwen}, F. and {Bakker}, J. and {Cacciari}, C. and {Casta{\~n}eda}, J. and {De Angeli}, F. and {Ducourant}, C. and {Fabricius}, C. and {Fouesneau}, M. and {Fr{\'e}mat}, Y. and {Guerra}, R. and {Guerrier}, A. and {Guiraud}, J. and {Jean-Antoine Piccolo}, A. and {Masana}, E. and {Messineo}, R. and {Mowlavi}, N. and {Nicolas}, C. and {Nienartowicz}, K. and {Pailler}, F. and {Panuzzo}, P. and {Riclet}, F. and {Roux}, W. and {Seabroke}, G.~M. and {Sordo}, R. and {Tanga}, P. and {Th{\'e}venin}, F. and {Gracia-Abril}, G. and {Portell}, J. and {Teyssier}, D. and {Altmann}, M. and {Andrae}, R. and {Bellas-Velidis}, I. and {Benson}, K. and {Berthier}, J. and {Blomme}, R. and {Brugaletta}, E. and {Burgess}, P.~W. and {Busso}, G. and {Carry}, B. and {Cellino}, A. and {Cheek}, N. and {Clementini}, G. and {Damerdji}, Y. and {Davidson}, M. and {Delchambre}, L. and {Dell'Oro}, A. and {Fern{\'a}ndez-Hern{\'a}ndez}, J. and {Galluccio}, L. and {Garc{\'\i}a-Lario}, P. and {Garcia-Reinaldos}, M. and {Gonz{\'a}lez-N{\'u}{\~n}ez}, J. and {Gosset}, E. and {Haigron}, R. and {Halbwachs}, J. -L. and {Hambly}, N.~C. and {Harrison}, D.~L. and {Hatzidimitriou}, D. and {Heiter}, U. and {Hern{\'a}ndez}, J. and {Hestroffer}, D. and {Hodgkin}, S.~T. and {Holl}, B. and {Jan{\ss}en}, K. and {Jevardat de Fombelle}, G. and {Jordan}, S. and {Krone-Martins}, A. and {Lanzafame}, A.~C. and {L{\"o}ffler}, W. and {Lorca}, A. and {Manteiga}, M. and {Marchal}, O. and {Marrese}, P.~M. and {Moitinho}, A. and {Mora}, A. and {Muinonen}, K. and {Osborne}, P. and {Pancino}, E. and {Pauwels}, T. and {Petit}, J. -M. and {Recio-Blanco}, A. and {Richards}, P.~J. and {Riello}, M. and {Rimoldini}, L. and {Robin}, A.~C. and {Roegiers}, T. and {Rybizki}, J. and {Sarro}, L.~M. and {Siopis}, C. and {Smith}, M. and {Sozzetti}, A. and {Ulla}, A. and {Utrilla}, E. and {van Leeuwen}, M. and {van Reeven}, W. and {Abbas}, U. and {Abreu Aramburu}, A. and {Accart}, S. and {Aerts}, C. and {Aguado}, J.~J. and {Ajaj}, M. and {Altavilla}, G. and {{\'A}lvarez}, M.~A. and {{\'A}lvarez Cid-Fuentes}, J. and {Alves}, J. and {Anderson}, R.~I. and {Anglada Varela}, E. and {Antoja}, T. and {Audard}, M. and {Baines}, D. and {Baker}, S.~G. and {Balaguer-N{\'u}{\~n}ez}, L. and {Balbinot}, E. and {Balog}, Z. and {Barache}, C. and {Barbato}, D. and {Barros}, M. and {Barstow}, M.~A. and {Bartolom{\'e}}, S. and {Bassilana}, J. -L. and {Bauchet}, N. and {Baudesson-Stella}, A. and {Becciani}, U. and {Bellazzini}, M. and {Bernet}, M. and {Bertone}, S. and {Bianchi}, L. and {Blanco-Cuaresma}, S. and {Boch}, T. and {Bombrun}, A. and {Bossini}, D. and {Bouquillon}, S. and {Bragaglia}, A. and {Bramante}, L. and {Breedt}, E. and {Bressan}, A. and {Brouillet}, N. and {Bucciarelli}, B. and {Burlacu}, A. and {Busonero}, D. and {Butkevich}, A.~G. and {Buzzi}, R. and {Caffau}, E. and {Cancelliere}, R. and {C{\'a}novas}, H. and {Cantat-Gaudin}, T. and {Carballo}, R. and {Carlucci}, T. and {Carnerero}, M.~I. and {Carrasco}, J.~M. and {Casamiquela}, L. and {Castellani}, M. and {Castro-Ginard}, A. and {Castro Sampol}, P. and {Chaoul}, L. and {Charlot}, P. and {Chemin}, L. and {Chiavassa}, A. and {Cioni}, M. -R.~L. and {Comoretto}, G. and {Cooper}, W.~J. and {Cornez}, T. and {Cowell}, S. and {Crifo}, F. and {Crosta}, M. and {Crowley}, C. and {Dafonte}, C. and {Dapergolas}, A. and {David}, M. and {David}, P. and {de Laverny}, P. and {De Luise}, F. and {De March}, R. and {De Ridder}, J. and {de Souza}, R. and {de Teodoro}, P. and {de Torres}, A. and {del Peloso}, E.~F. and {del Pozo}, E. and {Delbo}, M. and {Delgado}, A. and {Delgado}, H.~E. and {Delisle}, J. -B. and {Di Matteo}, P. and {Diakite}, S. and {Diener}, C. and {Distefano}, E. and {Dolding}, C. and {Eappachen}, D. and {Edvardsson}, B. and {Enke}, H. and {Esquej}, P. and {Fabre}, C. and {Fabrizio}, M. and {Faigler}, S. and {Fedorets}, G. and {Fernique}, P. and {Fienga}, A. and {Figueras}, F. and {Fouron}, C. and {Fragkoudi}, F. and {Fraile}, E. and {Franke}, F. and {Gai}, M. and {Garabato}, D. and {Garcia-Gutierrez}, A. and {Garc{\'\i}a-Torres}, M. and {Garofalo}, A. and {Gavras}, P. and {Gerlach}, E. and {Geyer}, R. and {Giacobbe}, P. and {Gilmore}, G. and {Girona}, S. and {Giuffrida}, G. and {Gomel}, R. and {Gomez}, A. and {Gonzalez-Santamaria}, I. and {Gonz{\'a}lez-Vidal}, J.~J. and {Granvik}, M. and {Guti{\'e}rrez-S{\'a}nchez}, R. and {Guy}, L.~P. and {Hauser}, M. and {Haywood}, M. and {Helmi}, A. and {Hidalgo}, S.~L. and {Hilger}, T. and {H{\l}adczuk}, N. and {Hobbs}, D. and {Holland}, G. and {Huckle}, H.~E. and {Jasniewicz}, G. and {Jonker}, P.~G. and {Juaristi Campillo}, J. and {Julbe}, F. and {Karbevska}, L. and {Kervella}, P. and {Khanna}, S. and {Kochoska}, A. and {Kontizas}, M. and {Kordopatis}, G. and {Korn}, A.~J. and {Kostrzewa-Rutkowska}, Z. and {Kruszy{\'n}ska}, K. and {Lambert}, S. and {Lanza}, A.~F. and {Lasne}, Y. and {Le Campion}, J. -F. and {Le Fustec}, Y. and {Lebreton}, Y. and {Lebzelter}, T. and {Leccia}, S. and {Leclerc}, N. and {Lecoeur-Taibi}, I. and {Liao}, S. and {Licata}, E. and {Lindstr{\o}m}, E.~P. and {Lister}, T.~A. and {Livanou}, E. and {Lobel}, A. and {Madrero Pardo}, P. and {Managau}, S. and {Mann}, R.~G. and {Marchant}, J.~M. and {Marconi}, M. and {Marcos Santos}, M.~M.~S. and {Marinoni}, S. and {Marocco}, F. and {Marshall}, D.~J. and {Martin Polo}, L. and {Mart{\'\i}n-Fleitas}, J.~M. and {Masip}, A. and {Massari}, D. and {Mastrobuono-Battisti}, A. and {Mazeh}, T. and {McMillan}, P.~J. and {Messina}, S. and {Michalik}, D. and {Millar}, N.~R. and {Mints}, A. and {Molina}, D. and {Molinaro}, R. and {Moln{\'a}r}, L. and {Montegriffo}, P. and {Mor}, R. and {Morbidelli}, R. and {Morel}, T. and {Morris}, D. and {Mulone}, A.~F. and {Munoz}, D. and {Muraveva}, T. and {Murphy}, C.~P. and {Musella}, I. and {Noval}, L. and {Ord{\'e}novic}, C. and {Orr{\`u}}, G. and {Osinde}, J. and {Pagani}, C. and {Pagano}, I. and {Palaversa}, L. and {Palicio}, P.~A. and {Panahi}, A. and {Pawlak}, M. and {Pe{\~n}alosa Esteller}, X. and {Penttil{\"a}}, A. and {Piersimoni}, A.~M. and {Pineau}, F. -X. and {Plachy}, E. and {Plum}, G. and {Poggio}, E. and {Poretti}, E. and {Poujoulet}, E. and {Pr{\v{s}}a}, A. and {Pulone}, L. and {Racero}, E. and {Ragaini}, S. and {Rainer}, M. and {Raiteri}, C.~M. and {Rambaux}, N. and {Ramos}, P. and {Ramos-Lerate}, M. and {Re Fiorentin}, P. and {Regibo}, S. and {Reyl{\'e}}, C. and {Ripepi}, V. and {Riva}, A. and {Rixon}, G. and {Robichon}, N. and {Robin}, C. and {Roelens}, M. and {Rohrbasser}, L. and {Romero-G{\'o}mez}, M. and {Rowell}, N. and {Royer}, F. and {Rybicki}, K.~A. and {Sadowski}, G. and {Sagrist{\`a} Sell{\'e}s}, A. and {Sahlmann}, J. and {Salgado}, J. and {Salguero}, E. and {Samaras}, N. and {Sanchez Gimenez}, V. and {Sanna}, N. and {Santove{\~n}a}, R. and {Sarasso}, M. and {Schultheis}, M. and {Sciacca}, E. and {Segol}, M. and {Segovia}, J.~C. and {S{\'e}gransan}, D. and {Semeux}, D. and {Shahaf}, S. and {Siddiqui}, H.~I. and {Siebert}, A. and {Siltala}, L. and {Slezak}, E. and {Smart}, R.~L. and {Solano}, E. and {Solitro}, F. and {Souami}, D. and {Souchay}, J. and {Spagna}, A. and {Spoto}, F. and {Steele}, I.~A. and {Steidelm{\"u}ller}, H. and {Stephenson}, C.~A. and {S{\"u}veges}, M. and {Szabados}, L. and {Szegedi-Elek}, E. and {Taris}, F. and {Tauran}, G. and {Taylor}, M.~B. and {Teixeira}, R. and {Thuillot}, W. and {Tonello}, N. and {Torra}, F. and {Torra}, J. and {Turon}, C. and {Unger}, N. and {Vaillant}, M. and {van Dillen}, E. and {Vanel}, O. and {Vecchiato}, A. and {Viala}, Y. and {Vicente}, D. and {Voutsinas}, S. and {Weiler}, M. and {Wevers}, T. and {Wyrzykowski}, {\L}. and {Yoldas}, A. and {Yvard}, P. and {Zhao}, H. and {Zorec}, J. and {Zucker}, S. and {Zurbach}, C. and {Zwitter}, T.},
        title = "{Gaia Early Data Release 3. Summary of the contents and survey properties}",
      journal = {\aap},
         year = 2021,
        month = may,
       volume = {649},
          eid = {A1},
        pages = {A1},
          doi = {10.1051/0004-6361/202039657},
archivePrefix = {arXiv},
       eprint = {2012.01533},
 primaryClass = {astro-ph.GA},
       adsurl = {https://ui.adsabs.harvard.edu/abs/2021A&A...649A...1G}
      }

@ARTICLE{Gaia_EDR3_2,
       author = {{Gaia Collaboration} and {Vallenari}, A. and {Brown}, A.~G.~A. and {Prusti}, T. and {de Bruijne}, J.~H.~J. and {Arenou}, F. and {Babusiaux}, C. and {Biermann}, M. and {Creevey}, O.~L. and {Ducourant}, C. and {Evans}, D.~W. and {Eyer}, L. and {Guerra}, R. and {Hutton}, A. and {Jordi}, C. and {Klioner}, S.~A. and {Lammers}, U.~L. and {Lindegren}, L. and {Luri}, X. and {Mignard}, F. and {Panem}, C. and {Pourbaix}, D. and {Randich}, S. and {Sartoretti}, P. and {Soubiran}, C. and {Tanga}, P. and {Walton}, N.~A. and {Bailer-Jones}, C.~A.~L. and {Bastian}, U. and {Drimmel}, R. and {Jansen}, F. and {Katz}, D. and {Lattanzi}, M.~G. and {van Leeuwen}, F. and {Bakker}, J. and {Cacciari}, C. and {Casta{\~n}eda}, J. and {De Angeli}, F. and {Fabricius}, C. and {Fouesneau}, M. and {Fr{\'e}mat}, Y. and {Galluccio}, L. and {Guerrier}, A. and {Heiter}, U. and {Masana}, E. and {Messineo}, R. and {Mowlavi}, N. and {Nicolas}, C. and {Nienartowicz}, K. and {Pailler}, F. and {Panuzzo}, P. and {Riclet}, F. and {Roux}, W. and {Seabroke}, G.~M. and {Sordo}, R. and {Th{\'e}venin}, F. and {Gracia-Abril}, G. and {Portell}, J. and {Teyssier}, D. and {Altmann}, M. and {Andrae}, R. and {Audard}, M. and {Bellas-Velidis}, I. and {Benson}, K. and {Berthier}, J. and {Blomme}, R. and {Burgess}, P.~W. and {Busonero}, D. and {Busso}, G. and {C{\'a}novas}, H. and {Carry}, B. and {Cellino}, A. and {Cheek}, N. and {Clementini}, G. and {Damerdji}, Y. and {Davidson}, M. and {de Teodoro}, P. and {Nu{\~n}ez Campos}, M. and {Delchambre}, L. and {Dell'Oro}, A. and {Esquej}, P. and {Fern{\'a}ndez-Hern{\'a}ndez}, J. and {Fraile}, E. and {Garabato}, D. and {Garc{\'\i}a-Lario}, P. and {Gosset}, E. and {Haigron}, R. and {Halbwachs}, J.-L. and {Hambly}, N.~C. and {Harrison}, D.~L. and {Hern{\'a}ndez}, J. and {Hestroffer}, D. and {Hodgkin}, S.~T. and {Holl}, B. and {Jan{\ss}en}, K. and {Jevardat de Fombelle}, G. and {Jordan}, S. and {Krone-Martins}, A. and {Lanzafame}, A.~C. and {L{\"o}ffler}, W. and {Marchal}, O. and {Marrese}, P.~M. and {Moitinho}, A. and {Muinonen}, K. and {Osborne}, P. and {Pancino}, E. and {Pauwels}, T. and {Recio-Blanco}, A. and {Reyl{\'e}}, C. and {Riello}, M. and {Rimoldini}, L. and {Roegiers}, T. and {Rybizki}, J. and {Sarro}, L.~M. and {Siopis}, C. and {Smith}, M. and {Sozzetti}, A. and {Utrilla}, E. and {van Leeuwen}, M. and {Abbas}, U. and {{\'A}brah{\'a}m}, P. and {Abreu Aramburu}, A. and {Aerts}, C. and {Aguado}, J.~J. and {Ajaj}, M. and {Aldea-Montero}, F. and {Altavilla}, G. and {{\'A}lvarez}, M.~A. and {Alves}, J. and {Anders}, F. and {Anderson}, R.~I. and {Anglada Varela}, E. and {Antoja}, T. and {Baines}, D. and {Baker}, S.~G. and {Balaguer-N{\'u}{\~n}ez}, L. and {Balbinot}, E. and {Balog}, Z. and {Barache}, C. and {Barbato}, D. and {Barros}, M. and {Barstow}, M.~A. and {Bartolom{\'e}}, S. and {Bassilana}, J.-L. and {Bauchet}, N. and {Becciani}, U. and {Bellazzini}, M. and {Berihuete}, A. and {Bernet}, M. and {Bertone}, S. and {Bianchi}, L. and {Binnenfeld}, A. and {Blanco-Cuaresma}, S. and {Blazere}, A. and {Boch}, T. and {Bombrun}, A. and {Bossini}, D. and {Bouquillon}, S. and {Bragaglia}, A. and {Bramante}, L. and {Breedt}, E. and {Bressan}, A. and {Brouillet}, N. and {Brugaletta}, E. and {Bucciarelli}, B. and {Burlacu}, A. and {Butkevich}, A.~G. and {Buzzi}, R. and {Caffau}, E. and {Cancelliere}, R. and {Cantat-Gaudin}, T. and {Carballo}, R. and {Carlucci}, T. and {Carnerero}, M.~I. and {Carrasco}, J.~M. and {Casamiquela}, L. and {Castellani}, M. and {Castro-Ginard}, A. and {Chaoul}, L. and {Charlot}, P. and {Chemin}, L. and {Chiaramida}, V. and {Chiavassa}, A. and {Chornay}, N. and {Comoretto}, G. and {Contursi}, G. and {Cooper}, W.~J. and {Cornez}, T. and {Cowell}, S. and {Crifo}, F. and {Cropper}, M. and {Crosta}, M. and {Crowley}, C. and {Dafonte}, C. and {Dapergolas}, A. and {David}, M. and {David}, P. and {de Laverny}, P. and {De Luise}, F. and {De March}, R.},
        title = "{Gaia Data Release 3. Summary of the content and survey properties}",
      journal = {\aap},
         year = 2023,
        month = jun,
       volume = {674},
          eid = {A1},
        pages = {A1},
          doi = {10.1051/0004-6361/202243940},
archivePrefix = {arXiv},
       eprint = {2208.00211},
 primaryClass = {astro-ph.GA},
       adsurl = {https://ui.adsabs.harvard.edu/abs/2023A&A...674A...1G}
}

@ARTICLE{Weilbacher2020,
       author = {{Weilbacher}, Peter M. and {Palsa}, Ralf and {Streicher}, Ole and {Bacon}, Roland and {Urrutia}, Tanya and {Wisotzki}, Lutz and {Conseil}, Simon and {Husemann}, Bernd and {Jarno}, Aur{\'e}lien and {Kelz}, Andreas and {P{\'e}contal-Rousset}, Arlette and {Richard}, Johan and {Roth}, Martin M. and {Selman}, Fernando and {Vernet}, Jo{\"e}l},
        title = "{The data processing pipeline for the MUSE instrument}",
      journal = {\aap},
         year = 2020,
        month = sep,
       volume = {641},
          eid = {A28},
        pages = {A28},
          doi = {10.1051/0004-6361/202037855},
archivePrefix = {arXiv},
       eprint = {2006.08638},
 primaryClass = {astro-ph.IM},
       adsurl = {https://ui.adsabs.harvard.edu/abs/2020A&A...641A..28W}
}

@INPROCEEDINGS{bacon2010,
       author = {{Bacon}, R. and {Accardo}, M. and {Adjali}, L. and {Anwand}, H. and {Bauer}, S. and {Biswas}, I. and {Blaizot}, J. and {Boudon}, D. and {Brau-Nogue}, S. and {Brinchmann}, J. and {Caillier}, P. and {Capoani}, L. and {Carollo}, C.~M. and {Contini}, T. and {Couderc}, P. and {Daguis{\'e}}, E. and {Deiries}, S. and {Delabre}, B. and {Dreizler}, S. and {Dubois}, J. and {Dupieux}, M. and {Dupuy}, C. and {Emsellem}, E. and {Fechner}, T. and {Fleischmann}, A. and {Fran{\c{c}}ois}, M. and {Gallou}, G. and {Gharsa}, T. and {Glindemann}, A. and {Gojak}, D. and {Guiderdoni}, B. and {Hansali}, G. and {Hahn}, T. and {Jarno}, A. and {Kelz}, A. and {Koehler}, C. and {Kosmalski}, J. and {Laurent}, F. and {Le Floch}, M. and {Lilly}, S.~J. and {Lizon}, J.-L. and {Loupias}, M. and {Manescau}, A. and {Monstein}, C. and {Nicklas}, H. and {Olaya}, J.-C. and {Pares}, L. and {Pasquini}, L. and {P{\'e}contal-Rousset}, A. and {Pell{\'o}}, R. and {Petit}, C. and {Popow}, E. and {Reiss}, R. and {Remillieux}, A. and {Renault}, E. and {Roth}, M. and {Rupprecht}, G. and {Serre}, D. and {Schaye}, J. and {Soucail}, G. and {Steinmetz}, M. and {Streicher}, O. and {Stuik}, R. and {Valentin}, H. and {Vernet}, J. and {Weilbacher}, P. and {Wisotzki}, L. and {Yerle}, N.},
        title = "{The MUSE second-generation VLT instrument}",
    booktitle = {Ground-based and Airborne Instrumentation for Astronomy III},
         year = 2010,
       editor = {{McLean}, Ian S. and {Ramsay}, Suzanne K. and {Takami}, Hideki},
       series = {Society of Photo-Optical Instrumentation Engineers (SPIE) Conference Series},
       volume = {7735},
        month = jul,
          eid = {773508},
        pages = {773508},
          doi = {10.1117/12.856027},
archivePrefix = {arXiv},
       eprint = {2211.16795},
 primaryClass = {astro-ph.IM},
       adsurl = {https://ui.adsabs.harvard.edu/abs/2010SPIE.7735E..08B}
}

@ARTICLE{temple2021,
       author = {{Temple}, Matthew J. and {Hewett}, Paul C. and {Banerji}, Manda},
        title = "{Modelling type 1 quasar colours in the era of Rubin and Euclid}",
      journal = {\mnras},
         year = 2021,
        month = nov,
       volume = {508},
       number = {1},
        pages = {737-754},
          doi = {10.1093/mnras/stab2586},
archivePrefix = {arXiv},
       eprint = {2109.04472},
 primaryClass = {astro-ph.GA},
       adsurl = {https://ui.adsabs.harvard.edu/abs/2021MNRAS.508..737T}
}

@INPROCEEDINGS{George2016,
       author = {{George}, E.~M. and {Gr{\"a}ff}, D. and {Feuchtgruber}, H. and {Hartl}, M. and {Eisenhauer}, F. and {Buron}, A. and {Davies}, R. and {Genzel}, R. and {Huber}, H. and {Rau}, C. and {Plattner}, M. and {Wiezorrek}, E. and {Weisz}, H. and {Amico}, P. and {Glindemann}, A. and {Hau}, G. and {Kuntschner}, H.},
        title = "{Making SPIFFI SPIFFIER: upgrade of the SPIFFI instrument for use in ERIS and performance analysis from re-commissioning}",
    booktitle = {Ground-based and Airborne Instrumentation for Astronomy VI},
         year = 2016,
       editor = {{Evans}, Christopher J. and {Simard}, Luc and {Takami}, Hideki},
       series = {Society of Photo-Optical Instrumentation Engineers (SPIE) Conference Series},
       volume = {9908},
        month = aug,
          eid = {99080G},
        pages = {99080G},
          doi = {10.1117/12.2231285},
archivePrefix = {arXiv},
       eprint = {1608.02457},
 primaryClass = {astro-ph.IM},
       adsurl = {https://ui.adsabs.harvard.edu/abs/2016SPIE.9908E..0GG}
}

@ARTICLE{Livodka2023,
       author = {{Li}, Junyao and {Liu}, Xin and {Shen}, Yue and {Oguri}, Masamune and {Gross}, Arran C. and {Zakamska}, Nadia L. and {Chen}, Yu-Ching and {Hwang}, Hsiang-Chih},
        title = "{Varstrometry for Off-nucleus and Dual Subkiloparsec AGN (VODKA). SDSS J1608+2716: A Subarcsecond Quadruply Lensed Quasar at z = 2.575}",
      journal = {\apjl},
         year = 2023,
        month = sep,
       volume = {955},
       number = {1},
          eid = {L16},
        pages = {L16},
          doi = {10.3847/2041-8213/acf27a},
archivePrefix = {arXiv},
       eprint = {2306.12502},
 primaryClass = {astro-ph.GA},
       adsurl = {https://ui.adsabs.harvard.edu/abs/2023ApJ...955L..16L}
}

@ARTICLE{skrutskie2006,
       author = {{Skrutskie}, M.~F. and {Cutri}, R.~M. and {Stiening}, R. and {Weinberg}, M.~D. and {Schneider}, S. and {Carpenter}, J.~M. and {Beichman}, C. and {Capps}, R. and {Chester}, T. and {Elias}, J. and {Huchra}, J. and {Liebert}, J. and {Lonsdale}, C. and {Monet}, D.~G. and {Price}, S. and {Seitzer}, P. and {Jarrett}, T. and {Kirkpatrick}, J.~D. and {Gizis}, J.~E. and {Howard}, E. and {Evans}, T. and {Fowler}, J. and {Fullmer}, L. and {Hurt}, R. and {Light}, R. and {Kopan}, E.~L. and {Marsh}, K.~A. and {McCallon}, H.~L. and {Tam}, R. and {Van Dyk}, S. and {Wheelock}, S.},
        title = "{The Two Micron All Sky Survey (2MASS)}",
      journal = {\aj},
         year = 2006,
        month = feb,
       volume = {131},
       number = {2},
        pages = {1163-1183},
          doi = {10.1086/498708},
       adsurl = {https://ui.adsabs.harvard.edu/abs/2006AJ....131.1163S}
}

@ARTICLE{mcmahon2013,
       author = {{McMahon}, R.~G. and {Banerji}, M. and {Gonzalez}, E. and {Koposov}, S.~E. and {Bejar}, V.~J. and {Lodieu}, N. and {Rebolo}, R. and {VHS Collaboration}},
        title = "{First Scientific Results from the VISTA Hemisphere Survey (VHS)}",
      journal = {The Messenger},
         year = 2013,
        month = dec,
       volume = {154},
        pages = {35-37},
       adsurl = {https://ui.adsabs.harvard.edu/abs/2013Msngr.154...35M}
}

@ARTICLE{2016A&A...595A...1G,
       author = {{Gaia Collaboration} and {Prusti}, T. and {de Bruijne}, J.~H.~J. and {Brown}, A.~G.~A. and {Vallenari}, A. and {Babusiaux}, C. and {Bailer-Jones}, C.~A.~L. and {Bastian}, U. and {Biermann}, M. and {Evans}, D.~W. and {Eyer}, L. and {Jansen}, F. and {Jordi}, C. and {Klioner}, S.~A. and {Lammers}, U. and {Lindegren}, L. and {Luri}, X. and {Mignard}, F. and {Milligan}, D.~J. and {Panem}, C. and {Poinsignon}, V. and {Pourbaix}, D. and {Randich}, S. and {Sarri}, G. and {Sartoretti}, P. and {Siddiqui}, H.~I. and {Soubiran}, C. and {Valette}, V. and {van Leeuwen}, F. and {Walton}, N.~A. and {Aerts}, C. and {Arenou}, F. and {Cropper}, M. and {Drimmel}, R. and {H{\o}g}, E. and {Katz}, D. and {Lattanzi}, M.~G. and {O'Mullane}, W. and {Grebel}, E.~K. and {Holland}, A.~D. and {Huc}, C. and {Passot}, X. and {Bramante}, L. and {Cacciari}, C. and {Casta{\~n}eda}, J. and {Chaoul}, L. and {Cheek}, N. and {De Angeli}, F. and {Fabricius}, C. and {Guerra}, R. and {Hern{\'a}ndez}, J. and {Jean-Antoine-Piccolo}, A. and {Masana}, E. and {Messineo}, R. and {Mowlavi}, N. and {Nienartowicz}, K. and {Ord{\'o}{\~n}ez-Blanco}, D. and {Panuzzo}, P. and {Portell}, J. and {Richards}, P.~J. and {Riello}, M. and {Seabroke}, G.~M. and {Tanga}, P. and {Th{\'e}venin}, F. and {Torra}, J. and {Els}, S.~G. and {Gracia-Abril}, G. and {Comoretto}, G. and {Garcia-Reinaldos}, M. and {Lock}, T. and {Mercier}, E. and {Altmann}, M. and {Andrae}, R. and {Astraatmadja}, T.~L. and {Bellas-Velidis}, I. and {Benson}, K. and {Berthier}, J. and {Blomme}, R. and {Busso}, G. and {Carry}, B. and {Cellino}, A. and {Clementini}, G. and {Cowell}, S. and {Creevey}, O. and {Cuypers}, J. and {Davidson}, M. and {De Ridder}, J. and {de Torres}, A. and {Delchambre}, L. and {Dell'Oro}, A. and {Ducourant}, C. and {Fr{\'e}mat}, Y. and {Garc{\'\i}a-Torres}, M. and {Gosset}, E. and {Halbwachs}, J. -L. and {Hambly}, N.~C. and {Harrison}, D.~L. and {Hauser}, M. and {Hestroffer}, D. and {Hodgkin}, S.~T. and {Huckle}, H.~E. and {Hutton}, A. and {Jasniewicz}, G. and {Jordan}, S. and {Kontizas}, M. and {Korn}, A.~J. and {Lanzafame}, A.~C. and {Manteiga}, M. and {Moitinho}, A. and {Muinonen}, K. and {Osinde}, J. and {Pancino}, E. and {Pauwels}, T. and {Petit}, J. -M. and {Recio-Blanco}, A. and {Robin}, A.~C. and {Sarro}, L.~M. and {Siopis}, C. and {Smith}, M. and {Smith}, K.~W. and {Sozzetti}, A. and {Thuillot}, W. and {van Reeven}, W. and {Viala}, Y. and {Abbas}, U. and {Abreu Aramburu}, A. and {Accart}, S. and {Aguado}, J.~J. and {Allan}, P.~M. and {Allasia}, W. and {Altavilla}, G. and {{\'A}lvarez}, M.~A. and {Alves}, J. and {Anderson}, R.~I. and {Andrei}, A.~H. and {Anglada Varela}, E. and {Antiche}, E. and {Antoja}, T. and {Ant{\'o}n}, S. and {Arcay}, B. and {Atzei}, A. and {Ayache}, L. and {Bach}, N. and {Baker}, S.~G. and {Balaguer-N{\'u}{\~n}ez}, L. and {Barache}, C. and {Barata}, C. and {Barbier}, A. and {Barblan}, F. and {Baroni}, M. and {Barrado y Navascu{\'e}s}, D. and {Barros}, M. and {Barstow}, M.~A. and {Becciani}, U. and {Bellazzini}, M. and {Bellei}, G. and {Bello Garc{\'\i}a}, A. and {Belokurov}, V. and {Bendjoya}, P. and {Berihuete}, A. and {Bianchi}, L. and {Bienaym{\'e}}, O. and {Billebaud}, F. and {Blagorodnova}, N. and {Blanco-Cuaresma}, S. and {Boch}, T. and {Bombrun}, A. and {Borrachero}, R. and {Bouquillon}, S. and {Bourda}, G. and {Bouy}, H. and {Bragaglia}, A. and {Breddels}, M.~A. and {Brouillet}, N. and {Br{\"u}semeister}, T. and {Bucciarelli}, B. and {Budnik}, F. and {Burgess}, P. and {Burgon}, R. and {Burlacu}, A. and {Busonero}, D. and {Buzzi}, R. and {Caffau}, E. and {Cambras}, J. and {Campbell}, H. and {Cancelliere}, R. and {Cantat-Gaudin}, T. and {Carlucci}, T. and {Carrasco}, J.~M. and {Castellani}, M. and {Charlot}, P. and {Charnas}, J. and {Charvet}, P. and {Chassat}, F. and {Chiavassa}, A. and {Clotet}, M. and {Cocozza}, G. and {Collins}, R.~S. and {Collins}, P. and {Costigan}, G.},
        title = "{The Gaia mission}",
      journal = {\aap},
         year = 2016,
        month = nov,
       volume = {595},
          eid = {A1},
        pages = {A1},
          doi = {10.1051/0004-6361/201629272},
archivePrefix = {arXiv},
       eprint = {1609.04153},
 primaryClass = {astro-ph.IM},
       adsurl = {https://ui.adsabs.harvard.edu/abs/2016A&A...595A...1G}
}

@ARTICLE{chen2022,
       author = {{Chen}, Yu-Ching and {Hwang}, Hsiang-Chih and {Shen}, Yue and {Liu}, Xin and {Zakamska}, Nadia L. and {Yang}, Qian and {Li}, Jennifer I.},
        title = "{Varstrometry for Off-nucleus and Dual Subkiloparsec AGN (VODKA): Hubble Space Telescope Discovers Double Quasars}",
      journal = {\apj},
         year = 2022,
        month = feb,
       volume = {925},
       number = {2},
          eid = {162},
        pages = {162},
          doi = {10.3847/1538-4357/ac401b},
archivePrefix = {arXiv},
       eprint = {2108.01672},
 primaryClass = {astro-ph.HE},
       adsurl = {https://ui.adsabs.harvard.edu/abs/2022ApJ...925..162C}
}

@ARTICLE{lemon2023,
       author = {{Lemon}, C. and {Anguita}, T. and {Auger-Williams}, M.~W. and {Courbin}, F. and {Galan}, A. and {McMahon}, R. and {Neira}, F. and {Oguri}, M. and {Schechter}, P. and {Shajib}, A. and {Treu}, T. and {Agnello}, A. and {Spiniello}, C.},
        title = "{Gravitationally lensed quasars in Gaia - IV. 150 new lenses, quasar pairs, and projected quasars}",
      journal = {\mnras},
         year = 2023,
        month = apr,
       volume = {520},
       number = {3},
        pages = {3305-3328},
          doi = {10.1093/mnras/stac3721},
archivePrefix = {arXiv},
       eprint = {2206.07714},
 primaryClass = {astro-ph.GA},
       adsurl = {https://ui.adsabs.harvard.edu/abs/2023MNRAS.520.3305L}
}

@ARTICLE{ciurlo23,
       author = {{Ciurlo}, A. and {Mannucci}, F. and {Yeh}, S. and {Amiri}, A. and {Carniani}, S. and {Cicone}, C. and {Cresci}, G. and {Lusso}, E. and {Marasco}, A. and {Marconcini}, C. and {Marconi}, A. and {Nardini}, E. and {Pancino}, E. and {Rosati}, P. and {Rubinur}, K. and {Severgnini}, P. and {Scialpi}, M. and {Tozzi}, G. and {Venturi}, G. and {Vignali}, C. and {Volonteri}, M.},
        title = "{New multiple AGN systems with subarcsec separation: Confirmation of candidates selected via the novel GMP method}",
      journal = {\aap},
         year = 2023,
        month = mar,
       volume = {671},
          eid = {L4},
        pages = {L4},
          doi = {10.1051/0004-6361/202345853},
archivePrefix = {arXiv},
       eprint = {2301.03091},
 primaryClass = {astro-ph.GA},
       adsurl = {https://ui.adsabs.harvard.edu/abs/2023A&A...671L...4C}
}

@ARTICLE{Davies2023,
       author = {{Davies}, R. and {Absil}, O. and {Agapito}, G. and {Agudo Berbel}, A. and {Baruffolo}, A. and {Biliotti}, V. and {Black}, M. and {Bonaglia}, M. and {Bonse}, M. and {Briguglio}, R. and {Campana}, P. and {Cao}, Y. and {Carbonaro}, L. and {Cortes}, A. and {Cresci}, G. and {Dallilar}, Y. and {Dannert}, F. and {De Rosa}, R.~J. and {Deysenroth}, M. and {Di Antonio}, I. and {Di Cianno}, A. and {Di Rico}, G. and {Doelman}, D. and {Dolci}, M. and {Dorn}, R. and {Eisenhauer}, F. and {Esposito}, S. and {Fantinel}, D. and {Ferruzzi}, D. and {Feuchtgruber}, H. and {Finger}, G. and {F{\"o}rster Schreiber}, N.~M. and {Gao}, X. and {Gemperlein}, H. and {Genzel}, R. and {Gillessen}, S. and {Ginski}, C. and {Glauser}, A.~M. and {Glindemann}, A. and {Grani}, P. and {Hartl}, M. and {Hayoz}, J. and {Heida}, M. and {Henry}, D. and {Hofmann}, R. and {Huber}, H. and {Kasper}, M. and {Keller}, C. and {Kenworthy}, M. and {Kravchenko}, K. and {Kuntschner}, H. and {Lacour}, S. and {Lightfoot}, J. and {Lunney}, D. and {Lutz}, D. and {Macintosh}, M. and {Mannucci}, F. and {Marsset}, M. and {Modigliani}, A. and {Neeser}, M. and {Orban de Xivry}, G. and {Ott}, T. and {Pallanca}, L. and {Patapis}, P. and {Pearson}, D. and {Pe{\~n}a}, E. and {Percheron}, I. and {Puglisi}, A. and {Quanz}, S.~P. and {Rabien}, S. and {Rau}, C. and {Riccardi}, A. and {Salasnich}, B. and {Schmid}, H.-M. and {Schubert}, J. and {Serra}, B. and {Shimizu}, T. and {Snik}, F. and {Sturm}, E. and {Tacconi}, L. and {Taylor}, W. and {Valentini}, A. and {Waring}, C. and {Wiezorrek}, E. and {Xompero}, M.},
        title = "{The Enhanced Resolution Imager and Spectrograph for the VLT}",
      journal = {\aap},
         year = 2023,
        month = jun,
       volume = {674},
          eid = {A207},
        pages = {A207},
          doi = {10.1051/0004-6361/202346559},
archivePrefix = {arXiv},
       eprint = {2304.02343},
 primaryClass = {astro-ph.IM},
       adsurl = {https://ui.adsabs.harvard.edu/abs/2023A&A...674A.207D}
}

@ARTICLE{Scialpi2026,
       author = {{Scialpi}, M. and {Mannucci}, F. and {D'Amato}, Q. and {Marconcini}, C. and {Cresci}, G. and {Marconi}, A. and {Ulivi}, L. and {Fumagalli}, M. and {Rosati}, P. and {Tozzi}, G. and {Zanchettin}, M.~V. and {Battistini}, L. and {Bertola}, E. and {Bracci}, C. and {Carniani}, S. and {Cataldi}, E. and {Ceci}, M. and {Chakraborty}, A. and {Cicone}, C. and {Ciurlo}, A. and {De Rosa}, A. and {Di Rosa}, G. and {Feltre}, A. and {Ginolfi}, M. and {Lamperti}, I. and {Lusso}, E. and {Moreschini}, B. and {Nardini}, E. and {Parvatikar}, M. and {Perna}, M. and {Rubinur}, K. and {Severgnini}, P. and {Singh}, J. and {Spingola}, C. and {Venturi}, G. and {Vignali}, C. and {Volonteri}, M.},
        title = "{Cosmic Duets: I. High-spatial resolution spectroscopy of dual and lensed active galactic nuclei with MUSE-NFM}",
      journal = {\aap},
         year = 2026,
        month = may,
       volume = {709},
          eid = {A236},
        pages = {A236},
          doi = {10.1051/0004-6361/202558548},
archivePrefix = {arXiv},
       eprint = {2512.11960},
 primaryClass = {astro-ph.GA},
       adsurl = {https://ui.adsabs.harvard.edu/abs/2026A&A...709A.236S}
}

@ARTICLE{scialpi24,
       author = {{Scialpi}, M. and {Mannucci}, F. and {Marconcini}, C. and {Venturi}, G. and {Pancino}, E. and {Marconi}, A. and {Cresci}, G. and {Belfiore}, F. and {Amiri}, A. and {Bertola}, E. and {Carniani}, S. and {Cicone}, C. and {Ciurlo}, A. and {D'Amato}, Q. and {Ginolfi}, M. and {Lusso}, E. and {Marasco}, A. and {Nardini}, E. and {Rubinur}, K. and {Severgnini}, P. and {Tozzi}, G. and {Ulivi}, L. and {Vignali}, C. and {Volonteri}, M.},
        title = "{MUSE adaptive-optics spectroscopy confirms dual active galactic nuclei and strongly lensed systems at sub-arcsec separation}",
      journal = {\aap},
         year = 2024,
        month = oct,
       volume = {690},
          eid = {A57},
        pages = {A57},
          doi = {10.1051/0004-6361/202347242},
archivePrefix = {arXiv},
       eprint = {2305.11850},
 primaryClass = {astro-ph.GA},
       adsurl = {https://ui.adsabs.harvard.edu/abs/2024A&A...690A..57S}
}

@ARTICLE{mannucci2023,
       author = {{Mannucci}, F. and {Scialpi}, M. and {Ciurlo}, A. and {Yeh}, S. and {Marconcini}, C. and {Tozzi}, G. and {Cresci}, G. and {Marconi}, A. and {Amiri}, A. and {Belfiore}, F. and {Carniani}, S. and {Cicone}, C. and {Nardini}, E. and {Pancino}, E. and {Rubinur}, K. and {Severgnini}, P. and {Ulivi}, L. and {Venturi}, G. and {Vignali}, C. and {Volonteri}, M. and {Pinna}, E. and {Rossi}, F. and {Puglisi}, A. and {Agapito}, G. and {Plantet}, C. and {Ghose}, E. and {Carbonaro}, L. and {Xompero}, M. and {Grani}, P. and {Esposito}, S. and {Power}, J. and {Guerra Ramon}, J.~C. and {Lefebvre}, M. and {Cavallaro}, A. and {Davies}, R. and {Riccardi}, A. and {Macintosh}, M. and {Taylor}, W. and {Dolci}, M. and {Baruffolo}, A. and {Feuchtgruber}, H. and {Kravchenko}, K. and {Rau}, C. and {Sturm}, E. and {Wiezorrek}, E. and {Dallilar}, Y. and {Kenworthy}, M.},
        title = "{GMP-selected dual and lensed AGNs: Selection function and classification based on near-IR colors and resolved spectra from VLT/ERIS, Keck/OSIRIS, and LBT/LUCI}",
      journal = {\aap},
         year = 2023,
        month = dec,
       volume = {680},
          eid = {A53},
        pages = {A53},
          doi = {10.1051/0004-6361/202346894},
archivePrefix = {arXiv},
       eprint = {2305.07396},
 primaryClass = {astro-ph.GA},
       adsurl = {https://ui.adsabs.harvard.edu/abs/2023A&A...680A..53M}
}

@ARTICLE{mannucci2022,
       author = {{Mannucci}, F. and {Pancino}, E. and {Belfiore}, F. and {Cicone}, C. and {Ciurlo}, A. and {Cresci}, G. and {Lusso}, E. and {Marasco}, A. and {Marconi}, A. and {Nardini}, E. and {Pinna}, E. and {Severgnini}, P. and {Saracco}, P. and {Tozzi}, G. and {Yeh}, S.},
        title = "{Unveiling the population of dual and lensed active galactic nuclei at sub-arcsec separations}",
      journal = {Nature Astronomy},
         year = 2022,
        month = aug,
       volume = {6},
        pages = {1185-1192},
          doi = {10.1038/s41550-022-01761-5},
archivePrefix = {arXiv},
       eprint = {2203.11234},
 primaryClass = {astro-ph.GA},
       adsurl = {https://ui.adsabs.harvard.edu/abs/2022NatAs...6.1185M}
}

@ARTICLE{derosa2019,
       author = {{De Rosa}, Alessandra and {Vignali}, Cristian and {Bogdanovi{\'c}}, Tamara and {Capelo}, Pedro R. and {Charisi}, Maria and {Dotti}, Massimo and {Husemann}, Bernd and {Lusso}, Elisabeta and {Mayer}, Lucio and {Paragi}, Zsolt and {Runnoe}, Jessie and {Sesana}, Alberto and {Steinborn}, Lisa and {Bianchi}, Stefano and {Colpi}, Monica and {del Valle}, Luciano and {Frey}, S{\'a}ndor and {Gab{\'a}nyi}, Krisztina {\'E}. and {Giustini}, Margherita and {Guainazzi}, Matteo and {Haiman}, Zoltan and {Herrera Ruiz}, Noelia and {Herrero-Illana}, Rub{\'e}n and {Iwasawa}, Kazushi and {Komossa}, S. and {Lena}, Davide and {Loiseau}, Nora and {Perez-Torres}, Miguel and {Piconcelli}, Enrico and {Volonteri}, Marta},
        title = "{The quest for dual and binary supermassive black holes: A multi-messenger view}",
      journal = {\nar},
         year = 2019,
        month = dec,
       volume = {86},
          eid = {101525},
        pages = {101525},
          doi = {10.1016/j.newar.2020.101525},
archivePrefix = {arXiv},
       eprint = {2001.06293},
 primaryClass = {astro-ph.GA},
       adsurl = {https://ui.adsabs.harvard.edu/abs/2019NewAR..8601525D}
}

@ARTICLE{2021NatRP...3..732V,
       author = {{Volonteri}, Marta and {Habouzit}, M{\'e}lanie and {Colpi}, Monica},
        title = "{The origins of massive black holes}",
      journal = {Nature Reviews Physics},
         year = 2021,
        month = sep,
       volume = {3},
       number = {11},
        pages = {732-743},
          doi = {10.1038/s42254-021-00364-9},
archivePrefix = {arXiv},
       eprint = {2110.10175},
 primaryClass = {astro-ph.GA},
       adsurl = {https://ui.adsabs.harvard.edu/abs/2021NatRP...3..732V}
}

@ARTICLE{2003ApJ...582..559V,
       author = {{Volonteri}, Marta and {Haardt}, Francesco and {Madau}, Piero},
        title = "{The Assembly and Merging History of Supermassive Black Holes in Hierarchical Models of Galaxy Formation}",
      journal = {\apj},
         year = 2003,
        month = jan,
       volume = {582},
       number = {2},
        pages = {559-573},
          doi = {10.1086/344675},
archivePrefix = {arXiv},
       eprint = {astro-ph/0207276},
 primaryClass = {astro-ph},
       adsurl = {https://ui.adsabs.harvard.edu/abs/2003ApJ...582..559V}
}

@ARTICLE{2007Sci...316.1874M,
       author = {{Mayer}, L. and {Kazantzidis}, S. and {Madau}, P. and {Colpi}, M. and {Quinn}, T. and {Wadsley}, J.},
        title = "{Rapid Formation of Supermassive Black Hole Binaries in Galaxy Mergers with Gas}",
      journal = {Science},
         year = 2007,
        month = jun,
       volume = {316},
       number = {5833},
        pages = {1874},
          doi = {10.1126/science.1141858},
archivePrefix = {arXiv},
       eprint = {0706.1562},
 primaryClass = {astro-ph},
       adsurl = {https://ui.adsabs.harvard.edu/abs/2007Sci...316.1874M}
}

@ARTICLE{1978MNRAS.183..341W,
       author = {{White}, S.~D.~M. and {Rees}, M.~J.},
        title = "{Core condensation in heavy halos: a two-stage theory for galaxy formation and clustering.}",
      journal = {\mnras},
         year = 1978,
        month = may,
       volume = {183},
        pages = {341-358},
          doi = {10.1093/mnras/183.3.341},
       adsurl = {https://ui.adsabs.harvard.edu/abs/1978MNRAS.183..341W}
}

@ARTICLE{Davies2007MNRAS.375.1099D,
       author = {{Davies}, R.~I.},
        title = "{A method to remove residual OH emission from near-infrared spectra}",
      journal = {\mnras},
         year = 2007,
        month = mar,
       volume = {375},
       number = {3},
        pages = {1099-1105},
          doi = {10.1111/j.1365-2966.2006.11383.x},
archivePrefix = {arXiv},
       eprint = {astro-ph/0612257},
 primaryClass = {astro-ph},
       adsurl = {https://ui.adsabs.harvard.edu/abs/2007MNRAS.375.1099D}
}

@INPROCEEDINGS{Eisenhauer2003,
       author = {{Eisenhauer}, Frank and {Abuter}, Roberto and {Bickert}, Klaus and {Biancat-Marchet}, Fabio and {Bonnet}, Henri and {Brynnel}, Joar and {Conzelmann}, Ralf D. and {Delabre}, Bernard and {Donaldson}, Robert and {Farinato}, Jacopo and {Fedrigo}, Enrico and {Genzel}, Reinhard and {Hubin}, Norbert N. and {Iserlohe}, Christof and {Kasper}, Markus E. and {Kissler-Patig}, Markus and {Monnet}, Guy J. and {Roehrle}, Claudia and {Schreiber}, Juergen and {Stroebele}, Stefan and {Tecza}, Matthias and {Thatte}, Niranjan A. and {Weisz}, Harald},
        title = "{SINFONI - Integral field spectroscopy at 50 milli-arcsecond resolution with the ESO VLT}",
    booktitle = {Instrument Design and Performance for Optical/Infrared Ground-based Telescopes},
         year = 2003,
       editor = {{Iye}, Masanori and {Moorwood}, Alan F.~M.},
       series = {Society of Photo-Optical Instrumentation Engineers (SPIE) Conference Series},
       volume = {4841},
        month = mar,
        pages = {1548-1561},
          doi = {10.1117/12.459468},
archivePrefix = {arXiv},
       eprint = {astro-ph/0306191},
 primaryClass = {astro-ph},
       adsurl = {https://ui.adsabs.harvard.edu/abs/2003SPIE.4841.1548E}
}

@ARTICLE{scialpi2026_bhmass,
       author = {{Scialpi}, M. and {Mannucci}, F. and {Izquierdo-Villalba}, D. and {Marconi}, A. and {Singh}, J. and {Volonteri}, M. and {Marconcini}, C. and {D'Amato}, Q. and {Moreschini}, B. and {Zanchettin}, M.~V. and {Tozzi}, G. and {Ceci}, M. and {Ulivi}, L. and {Bertola}, E. and {Bonoli}, S. and {Bracci}, C. and {Cataldi}, E. and {Cicone}, C. and {Ciurlo}, A. and {Colpi}, M. and {Cresci}, G. and {De Rosa}, A. and {Dotti}, M. and {Fumagalli}, M. and {Ginolfi}, M. and {Lamperti}, I. and {Nardini}, E. and {Parvatikar}, M. and {Perna}, M. and {Rigamonti}, F. and {Rubinur}, K. and {Sala}, B. and {Sesana}, A. and {Severgnini}, P. and {Spingola}, C. and {Spinoso}, D. and {Venturi}, G. and {Vignali}, C. and {Yeh}, S.},
        title = "{Cosmic Duets II. Growth and accretion in Gaia Multi-Peak dual AGN at Cosmic Noon}",
      journal = {arXiv e-prints},
         year = 2026,
        month = oct,
          eid = {arXiv:2610.10687},
        pages = {arXiv:2610.10687},
          doi = {10.48550/arXiv.2610.10687},
archivePrefix = {arXiv},
       eprint = {2610.10687},
 primaryClass = {astro-ph.GA},
       adsurl = {https://ui.adsabs.harvard.edu/abs/2026arXiv261010687S}
}

@ARTICLE{damato2026,
       author = {{D'Amato}, Quirino and {Mannucci}, Filippo and {Sonnenfeld}, Alessandro and {Scialpi}, Martina and {Nightingale}, James W. and {Spingola}, Cristiana and {Zibetti}, Stefano and {Marconi}, Alessandro and {Rosati}, Piero and {Marconcini}, Cosimo and {Agapito}, Guido and {Gallazzi}, Anna and {Di Teodoro}, Enrico and {Andreuzzi}, Gloria and {Belfiore}, Francesco and {Bertola}, Elena and {Bracci}, Caterina and {Carniani}, Stefano and {Cataldi}, Elisa and {Chakraborty}, Avinanda and {Ceci}, Matteo and {Cicone}, Claudia and {Ciurlo}, Anna and {Cresci}, Giovanni and {De Rosa}, Alessandra and {Di Carlo}, Elisa and {Feltre}, Anna and {Ginolfi}, Michele and {Lamperti}, Isabella and {Moreschini}, Bianca and {Nardini}, Emanuele and {Perna}, Michele and {Portaluri}, Elisa and {Rubinur}, Khatun and {Saracco}, Paolo and {Severgnini}, Paola and {Testa}, Vincenzo and {Tozzi}, Giulia and {Venturi}, Giacomo and {Ulivi}, Lorenzo and {Vignali}, Cristian and {Zanchettin}, Maria Vittoria and {Pepe}, Antonio},
        title = "{Milky-Way-like stars in a galaxy core 8 billion years ago revealed by gravitational lensing}",
      journal = {Nature Astronomy},
         year = 2026,
        month = apr,
          doi = {10.1038/s41550-026-02819-4},
archivePrefix = {arXiv},
       eprint = {2604.01828},
 primaryClass = {astro-ph.GA},
       adsurl = {https://ui.adsabs.harvard.edu/abs/2026NatAs.tmp...68D}
}

@ARTICLE{kennicutt2012,
       author = {{Kennicutt}, Robert C. and {Evans}, Neal J.},
        title = "{Star Formation in the Milky Way and Nearby Galaxies}",
      journal = {\araa},
         year = 2012,
        month = sep,
       volume = {50},
        pages = {531-608},
          doi = {10.1146/annurev-astro-081811-125610},
archivePrefix = {arXiv},
       eprint = {1204.3552},
 primaryClass = {astro-ph.GA},
       adsurl = {https://ui.adsabs.harvard.edu/abs/2012ARA&A..50..531K}
}

@ARTICLE{cidfernandes2011,
       author = {{Cid Fernandes}, R. and {Stasi{\'n}ska}, G. and {Mateus}, A. and {Vale Asari}, N.},
        title = "{A comprehensive classification of galaxies in the Sloan Digital Sky Survey: how to tell true from fake AGN?}",
      journal = {\mnras},
         year = 2011,
        month = may,
       volume = {413},
       number = {3},
        pages = {1687-1699},
          doi = {10.1111/j.1365-2966.2011.18244.x},
archivePrefix = {arXiv},
       eprint = {1012.4426},
 primaryClass = {astro-ph.CO},
       adsurl = {https://ui.adsabs.harvard.edu/abs/2011MNRAS.413.1687C}
}

@ARTICLE{madau2014,
       author = {{Madau}, Piero and {Dickinson}, Mark},
        title = "{Cosmic Star-Formation History}",
      journal = {\araa},
         year = 2014,
        month = aug,
       volume = {52},
        pages = {415-486},
          doi = {10.1146/annurev-astro-081811-125615},
archivePrefix = {arXiv},
       eprint = {1403.0007},
 primaryClass = {astro-ph.CO},
       adsurl = {https://ui.adsabs.harvard.edu/abs/2014ARA&A..52..415M}
}

\begin{appendix}

\section{Near-infrared magnitudes}\label{app:mag}

As described in Section~\ref{subsec:spectra}, we used broadband magnitudes to correct the fluxes of the extracted spectra, in order to simultaneously account for the fraction of flux lost due to the AO Strehl ratio and for the overall flux-calibration uncertainty.
When available, we adopted the Jap3, Hap3, or Kap3 magnitudes from the fifth data release (DR5) of the VISTA Hemisphere Survey \citep[VHS;][]{mcmahon2013}. For J0533$-$2303, we used the H-band magnitude from Two Micron All Sky Survey \citep[2MASS;][]{skrutskie2006}, since Hap3 magnitude from VHS DR5 was not available.
For systems lacking photometric measurements in the broadband filter corresponding to the ERIS--SPIFFIER grating used in the observations, we estimated the magnitudes using empirical relations between the observed and predicted values. 
In particular, we estimated these magnitudes using linear combinations either of the photometry in other near-infrared bands (e.g., deriving H-band magnitude from a linear combination of the J and K bands) or of the \gaia~$B_P$ and $R_P$ magnitudes. We considered all the QSOs with observed J-, H-, K-band magnitudes, from 2MASS and VHS, together with the two \gaia~magnitudes, and determined the linear combination of two bands that minimizes the difference between the observed and predicted magnitudes.
Since quasars exhibit strong and broad emission lines that can significantly affect broadband photometry, we derived these empirical relations in different redshift bins. Each bin corresponds to the redshift range in which the emission line targeted by the ERIS observations (e.g. \ha, \hb, \hg, or \hd) falls within the filter used to estimate the magnitude. This approach ensures that the inferred magnitude is appropriate for the specific spectral feature probed in each object.

The derived relations, together with their associated scatter, are reported in Table~\ref{tab:mag_relations}. For completeness, we also provide the relations computed over the full redshift range covered by the photometric data. Figure~\ref{fig:mag_relations} shows a couple of the derived relations as examples.
The derived magnitudes used along the manuscript are reported in the last column of Table~\ref{tab:classified} and labeled as $^{(**)}$.

\begin{figure}
\resizebox{\hsize}{!}{
\includegraphics{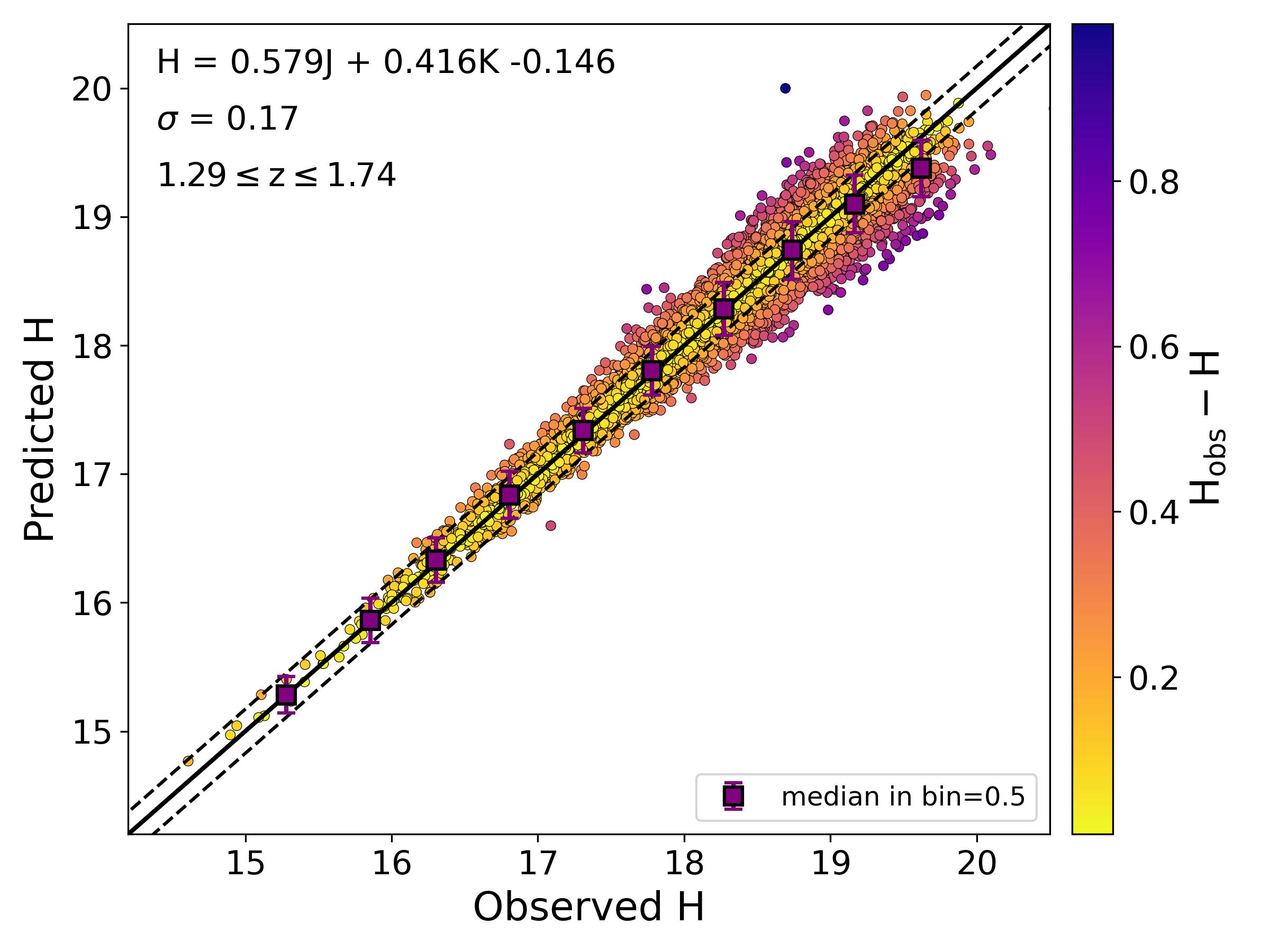}}
\resizebox{\hsize}{!}{
\includegraphics{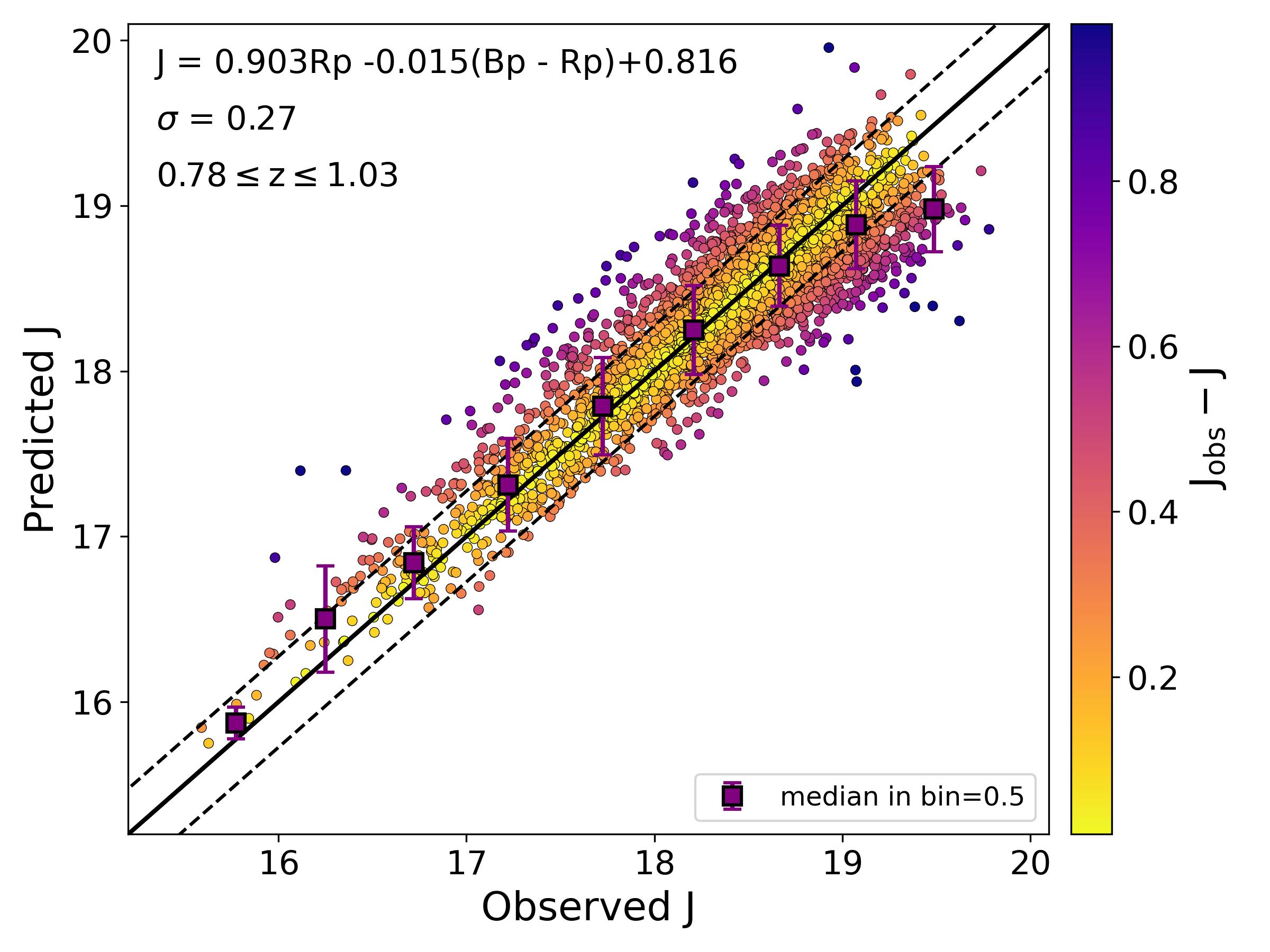}}
\caption{Comparison between the observed H- (right panel) and J- (left panel) magnitudes and the values derived from the observed near-infrared (right) and $B_P$ and $R_P$ \gaia~magnitudes (left) using the equations in table~\ref{tab:mag_relations}. Each point is color-coded as difference between the observed and predicted magnitudes. The solid line shows the one-to-one relation, the dotted lines report the observed scatter around this relation. The squares represent the median values of the points in bin of 0.5 mag. The fitted relation, the scatter and the redshift range are reported on the panels.}
\label{fig:mag_relations}
\end{figure}

\begingroup
\setlength{\tabcolsep}{6pt}
\renewcommand{\arraystretch}{1.0} 
\begin{table*}
\centering 
\caption[]{Derived relations between the observed and predicted magnitudes. }\label{tab:mag_relations}                        
\begin{tabular}{c c c c c c c c c c}        
\hline
Redshift & Emission & $a$ & $b$ & $c$ & $\sigma$ & $d$ & $e$ & $f$ & $\sigma$\\
range & line &  &  &  &  &  &  &  &  \\
\hline
 & & \multicolumn{4}{c}{$ J = a\cdot H+ b \cdot K + c$} & \multicolumn{4}{c}{$ J = d\cdot Rp + e \cdot  (Bp-Rp) + f$} \\
\hline 
all &  & 0.727 & 0.223 & 1.602 & 0.22 & 0.947 & $-$0.180 & 0.401 & 0.33\\
0.78$-$1.03 & \ha & 0.901 & 0.131 & $-$0.160 & 0.19 & 0.903 & $-$0.015 & 0.816 & 0.27 \\
1.26$-$1.88 & \hb &  0.535 & 0.83 & 0.542 & 0.20 & 0.926 & $-$0.181 & 0.937 & 0.29 \\
1.70$-$2.24 & \hg,\hd & 0.8812 & 0.183 & 0.667 & 0.18 & 0.956 & $-$0.159 & 0.356 & 0.29 \\
\hline
 & & \multicolumn{4}{c}{$ H = a\cdot J + b \cdot K + c$} & \multicolumn{4}{c}{$ H = d\cdot Rp + e \cdot  (Bp-Rp) + f$}\\
\hline
all &  & 0.545 & 0.423 & 0.535 & 0.20 & 0.928 & $-$0.275 & 0.273 & 0.36 \\
1.29$-$1.74 & \ha  & 0.579 & 0.416 & $-$0.146 & 0.17 & 0.880 & $-$0.232 & 1.159 & 0.30 \\
2.09$-$2.70 & \hb & 0.701 & 0.297 & $-$0.092 & 0.15 & 0.933 & $-$0.204 & 0.244 & 0.33\\
2.66$-$3.15 & \hg,\hd & 0.614 & 0.399 & -0.299 & 0.14 & 0.958 & $-$0.189 & $-$0.291 & 0.32 \\
\hline
& &  \multicolumn{4}{c}{$ K = a\cdot J + b \cdot H + c$} & \multicolumn{4}{c}{$ K = d\cdot Rp + e \cdot  (Bp-Rp) + f$}\\
\hline
all & & 0.298 & 0.699 & $-$0.769 & 0.25 & 0.910 & $-$0.268 & $-$0.024 & 0.44 \\
2.05$-$2.54 & \ha & 0.288 & 0.672 & $-$0.216 & 0.21 & 0.891 & $-$0.232 & 0.294 & 0.36 \\
3.11$-$3.78 & \hb & 0.163 & 0.844 & $-$0.791 & 0.19 & 0.942 & $-$0.076 & 0.701 & 0.39 \\
3.88$-$4.35 & \hg,\hd & 0.159 & 0.788 & 0.364 & 0.21 & 0.902 & $-$0.234 & 1.358 & 0.28\\
\hline
\end{tabular}
\tablefoot{The table reports the coefficient of the fitted relations together with their associated scatter for different redshift ranges.}
\end{table*}
\endgroup

\section{Template spectral decomposition}\label{app:temple}

\begin{figure}
\includegraphics[width=\hsize]{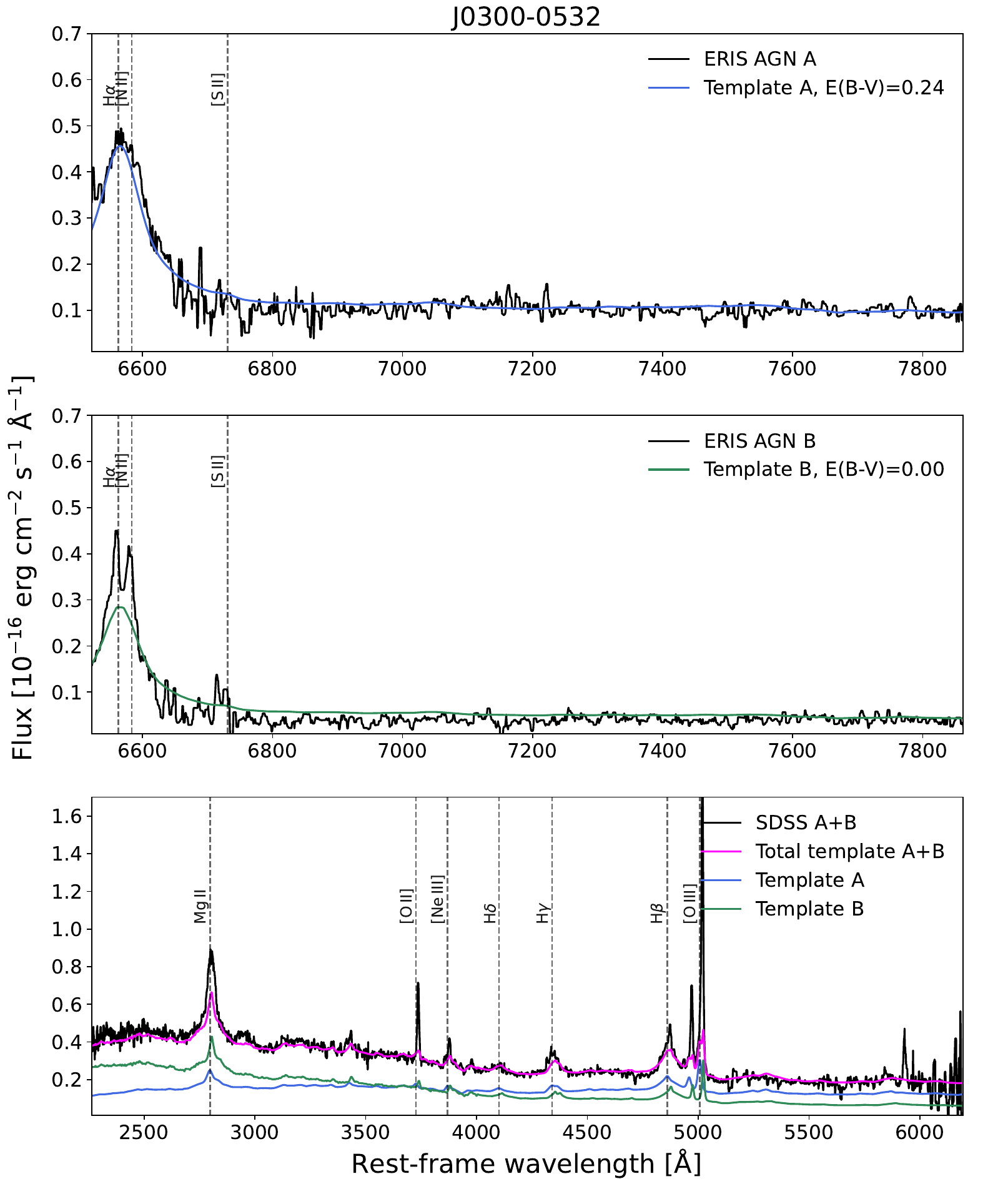}
\caption{Joint spectral fit of J0300$-$0532. From top to bottom: ERIS spectra of the primary (A) and secondary (B) AGN, and the SDSS long-slit spectrum of the system. The template for component A is shown in blue, for component B in green, while the combined model is shown in magenta. Vertical dashed black lines mark the expected positions of the most prominent emission lines at the respective redshifts of the two AGN components.}
\label{fig:j0300_temple_mcmc}
\end{figure}
In this Appendix we provide additional details on the final step of the spectral modeling, in which the optical long-slit spectrum and the ERIS spectra of the individual components are fitted simultaneously.
The redshift of each component is fixed to the values derived from the ERIS spectra alone (see Section~\ref{subsec:spectra}). Each component is modeled using empirical AGN templates from \citet{temple2021}, allowing for independent normalization and dust attenuation.
The observed ERIS spectra are modeled independently as:
\begin{equation}
F_{A,B}(\lambda) = C_{A,B} \, T_{A,B}(\lambda, z_{A,B}, E(B\!-\!V)_{A,B})
\end{equation}
where $T_{A,B}$ are the template spectra for the two components, $C_{A,B}$ are normalization factors, and $E(B\!-\!V)_{A,B}$ are the corresponding color excess values.
The long-slit spectrum is modeled as the sum of the two components:
\begin{equation}
F_{\rm tot}(\lambda) = C_A \, T_A(\lambda, z_A, E(B\!-\!V)_A) + C_B \, T_B(\lambda, z_B, E(B\!-\!V)_B).
\end{equation}
The parameter space is explored using a MCMC approach, and we adopt the median of the posterior distributions as best-fit values, with uncertainties derived from the 16th and 84th percentiles.
The long-slit spectrum is calibrated using available G-band photometry from \gaia, which also allows us to account for systematic uncertainties in the spectroscopic flux calibration. When the system is resolved by \gaia\ (i.e. individual $G_A$ and $G_B$ magnitudes are available), these provide an additional constraint on the relative normalization of the two templates.
\noindent As an illustration, in Fig.~\ref{fig:j0300_temple_mcmc} we show the result of the joint fit for J0300$-$0532. In this case, the optical spectrum is from SDSS \citep{SDSS_lyke2020}.
The best-fitting solution yields $E(B\!-\!V)_A \sim 0.2$ and $E(B\!-\!V)_B \sim 0.0$, highlighting different levels of attenuation in the two components. However, in this case the flux ratio between the two components is not independently constrained in the observed optical band. This makes it difficult to unambiguously disentangle the relative contribution of each component and to assess the robustness of the inferred attenuation values.

\section{\texttt{2D--Moffat} decomposition}\label{app:moffat_model}

To assess the reliability of the \texttt{2D--Moffat} decomposition used to extract the spectra of each source (Section~\ref{sec:method}), we computed the reduced $\chi^2_{\nu,\,\rm broadband}$ of the fit to the collapsed broad band image, as well as the median reduced  $\chi^2_{\nu,\,\rm median}$ across all fitted spectral channels, for each target in the sample. These statistics are reported in Table~\ref{tab:chi2}. In all cases the reduced $\chi^2_{\nu}$ values are close to unity, indicating that the \texttt{2D--Moffat} model provides an adequate description of the source both in the broad band image and on a channel-by-channel basis. Figure~\ref{fig:model_moffat} shows an example of the decomposition, for the broad band image of J0602--3118: the data, the best-fit \texttt{2D--Moffat} model, and the residual (data$-$model) map. The residual map shows no significant structure above the noise level, confirming that the model describes the observed flux distribution well. The specific case of J0300--0532, for which host--galaxy contamination is potentially more significant given its lower redshift, is discussed in Section~\ref{subsec:spectra}.

\begingroup
\setlength{\tabcolsep}{6pt} 
\renewcommand{\arraystretch}{1.0} 
\begin{table}[ht]
\caption[]{Reduced $\chi^2$ statistics of the \texttt{2D--Moffat} decomposition for each system.}\label{tab:chi2}
\centering                          
\begin{tabular}{l c c c }        
\hline  
\hline
Target & $\chi^2_{\nu,\,\rm broadband}$ & $\chi^2_{\nu,\,\rm median}$ \\
\hline
J0109--3529  & 0.94 & 0.98 \\
J0216$+$1430 & 0.87 & 1.01 \\
J0300--0532  & 0.93 & 1.02 \\
J0459--0413  & 1.14 & 1.05 \\
J0533--2303  & 0.98 & 1.12 \\
J0602--3118  & 0.88 & 0.97 \\
J0602--3852  & 0.83 & 0.98 \\
J1305--0232  & 0.93 & 1.01 \\ 
J1335$+$0527 & 0.96 & 0.98 \\
J1423$+$1544 & 0.84 & 0.98 \\
J1453$+$0529 & 0.97 & 1.01 \\
J2019--2525  & 0.88 & 1.15 \\
J2117$+$1057 & 0.86 & 0.98 \\
J2301--1906  & 1.01 & 1.01 \\
\hline
\end{tabular}
\tablefoot{$\chi^2_{\nu,\,\rm broadband}$ is the reduced $\chi^2$ of the fit to the collapsed broadband image; $\chi^2_{\nu,\,\rm median}$ is the median reduced $\chi^2$ across the spectral channels.}
\end{table}
\endgroup

\begin{figure}
\resizebox{\hsize}{!}{\includegraphics{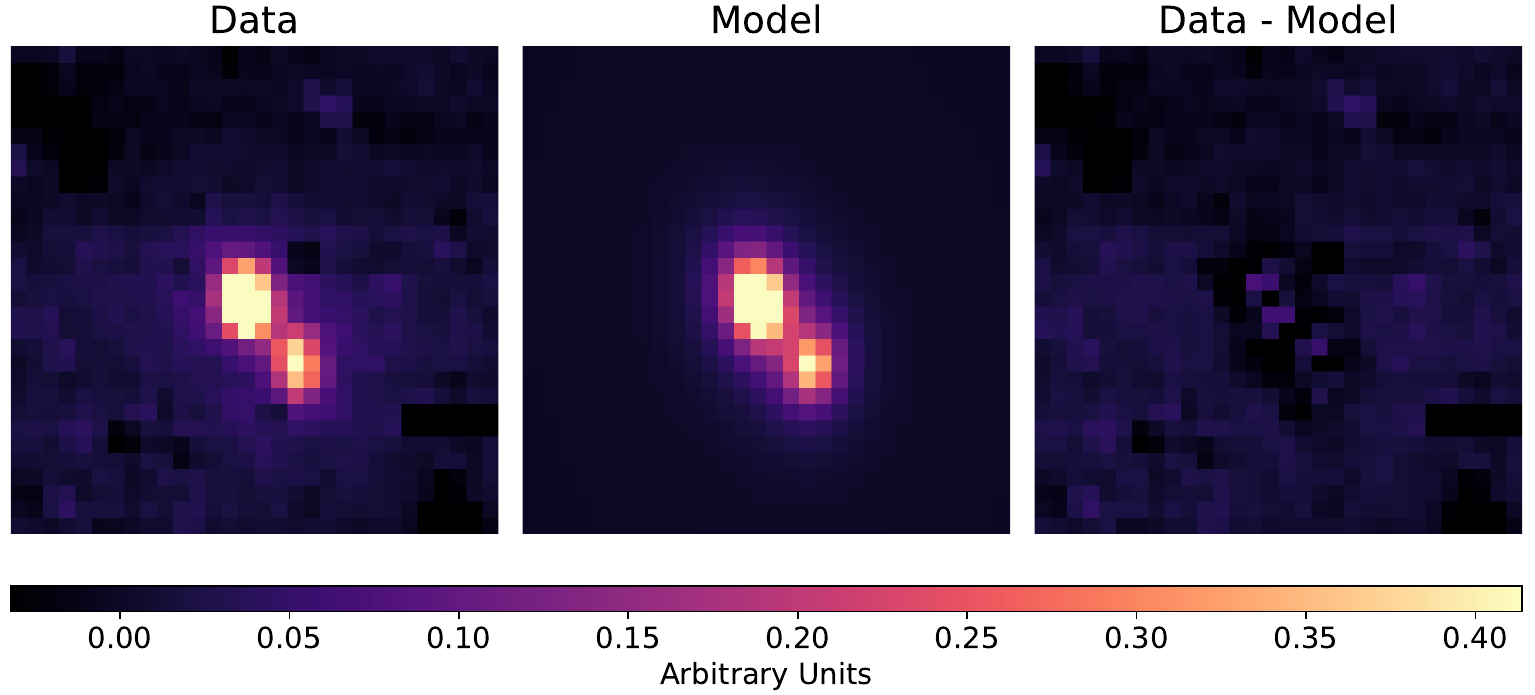}}
\caption{Example of the \texttt{2D--Moffat} decomposition for the dual AGN J0602--31118. From left to right: broad band image (data) of the system, double \texttt{2D--Moffat} model and residual (data - model) map. The maps are normalized to the maximum of the data map. }
\label{fig:model_moffat}
\end{figure}

\section{J1335$+$0527 VLT/MUSE observations}\label{app:j1335+0527}

\begin{figure}
\includegraphics[width=\hsize]{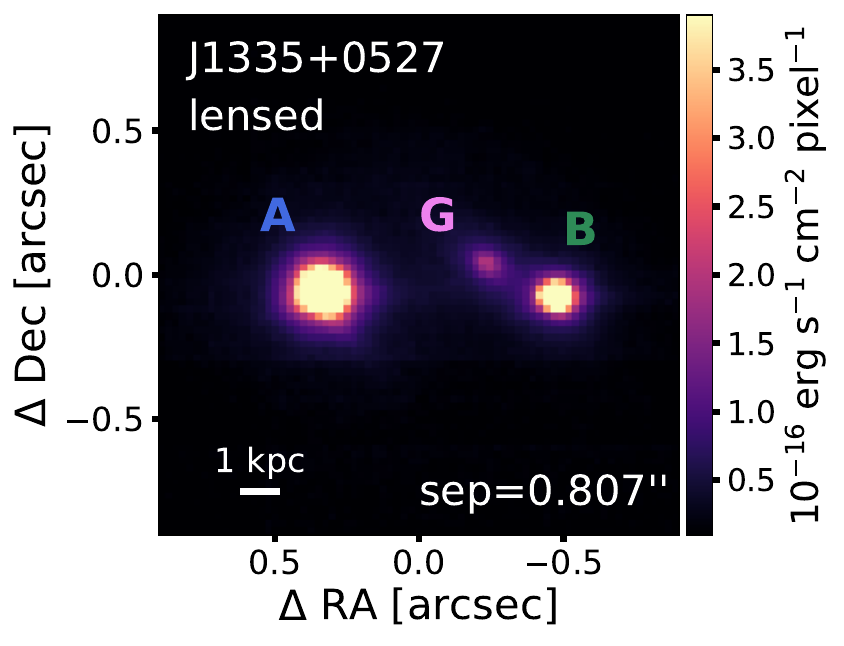}
\resizebox{\hsize}{!}{
\includegraphics{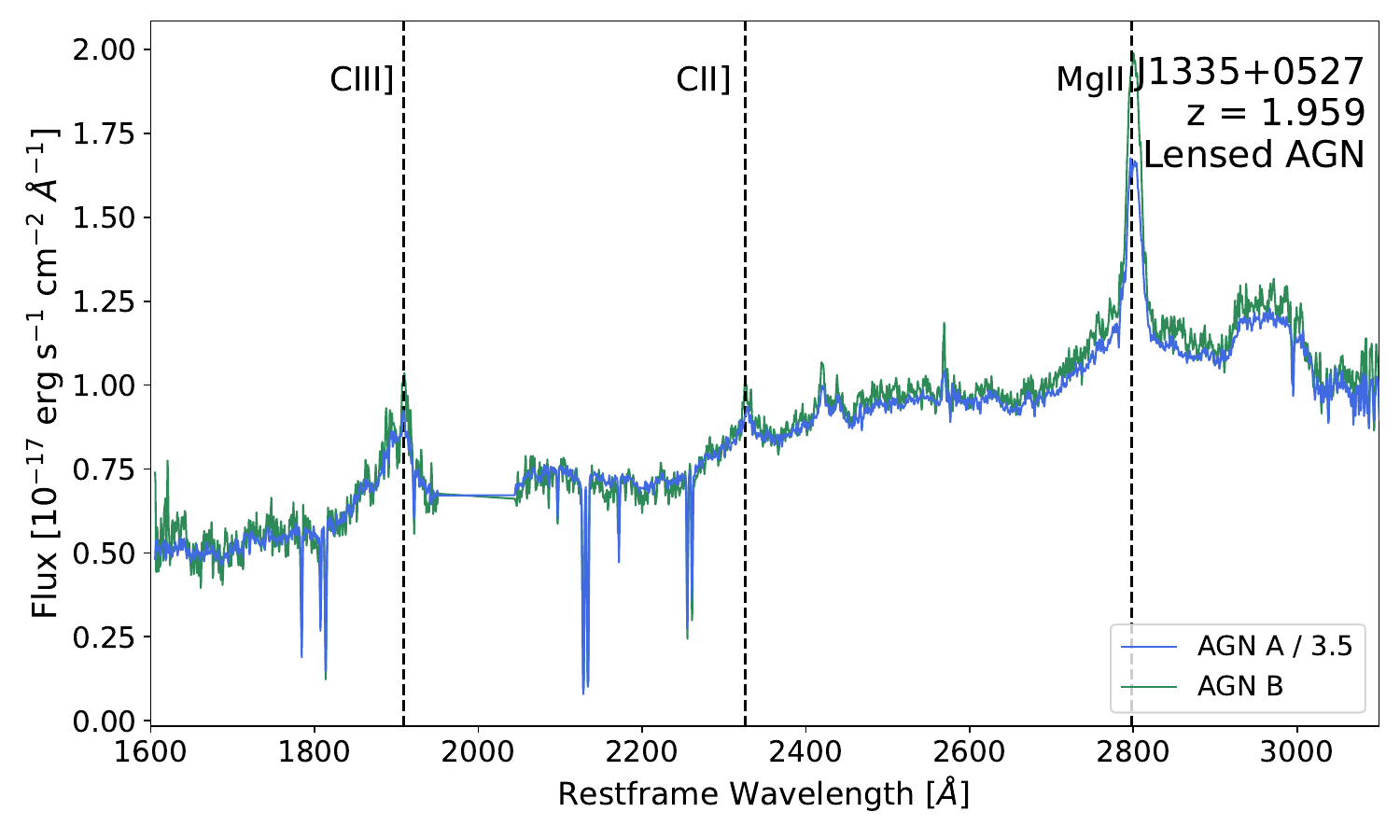}}
\caption{VLT/MUSE NFM data of the lensed AGN system J1335+0527. Top panel: White-light image of the system. The circular aperture mark the positions of the primary component (A, green), of the secondary component (B, blue) and of the lensing galaxy (G, purple). Bottom panel: VLT/MUSE spectra of the system.
The spectra were extracted from the circular apertures shown in the top panel (same color-coding), corrected for Galactic extinction, normalized to \gaia\ magnitude and shifted to the rest frame. The spectrum of the primary A AGN has been re-normalized to facilitate comparison with that of the B component; the applied renormalization factor is indicated in the legend. Vertical dashed black lines mark the expected positions of the most prominent emission lines at z = 1.959. The gap around $\lambda_{obs}$ = 6000 \AA\ (observed frame) is masked due to the presence of the NaID laser emission used for AO by New Generation Controllers (NGC).}
\label{fig:j1335+0527_muse_map_spec}
\end{figure}

One system in our sample, namely J1335+0527, could not be conclusively classified from ERIS data alone. At the redshift of this source ($z = 1.959$), \ha~and \hb~fall well outside the ERIS wavelength coverage. Only the fainter \hg~and \hd~lines are covered by the observable bands, but are only barely detected (see Fig.~\ref{fig:eris_spectra_lens}). This prevents a reliable comparison of line profiles and equivalent widths between the two components, making it impossible to discriminate between a lensed and a dual AGN configuration. We therefore obtained additional follow-up observations with the Multi Unit Spectroscopic Explorer \citep[MUSE,][]{bacon2010} in Narrow Field Mode (NFM), assisted by the GALACSI adaptive optics system at the VLT, as part of the \textit{Cosmic Duets} ESO Large Program (114.27BY.001, PI: Scialpi; \citealt{scialpi2026}).

\noindent The observations were carried out on March 10, 2026 with seeing conditions around \SI{0.4}{\arcsecond}, with a total exposure time of 2700 s split into three dithered exposures. The MUSE NFM delivers a field of view of $\sim7''\times7''$ with a pixel scale of $0.025''\times0.025''$, covering the wavelength range 4750--9350\,\AA\ at a spectral resolving power $R \sim 1700$--$3400$, corresponding to a velocity resolution of $\Delta v \sim 90$--$180$ \kms. The sodium notch filter blocks the 5780--6050\,\AA\ range to avoid contamination from the laser guide stars. The AO system delivered a PSF FWHM of $\sim 0.16 ''$ at 7000\,\AA, sufficient to spatially resolve the two components separated by $\sim$\SI{0.8}{\arcsecond} (see Tab.~\ref{tab:classified}).

\noindent Data reduction followed the procedures described in \citet{scialpi2026}, using the MUSE pipeline \citep{Weilbacher2020} within the \textsc{EsoRex} environment, including bias and dark subtraction, flat-fielding, wavelength calibration, sky subtraction, and flux calibration via a standard star observed the same night. Spectra were extracted from circular apertures of 5-pixel radius ($0.125''$), chosen to match the typical PSF FWHM, and normalized to \gaia~magnitudes following the approach of \citet{scialpi2026}.

\noindent Top panel of Figure~\ref{fig:j1335+0527_muse_map_spec} shows the white-light image obtained by summing up all spectral channels. The circular apertures mark the position of the primary AGN (A, green), of the secondary AGN (B, blue), and of the lensing galaxy (G, purple). Bottom panel of Figure~\ref{fig:j1335+0527_muse_map_spec} shows the spectra of the two components extracted using a 5-pixel radius aperture. The spectra were corrected for Galactic extinction, normalized to \gaia~magnitudes, and shifted to the rest frame.

\noindent The extracted spectra of each component were modeled independently using AGN templates \citet{temple2021}, as described in \citet{scialpi2026}. This fitting aims to derive the dust attenuation $E(B-V)$ and redshift of each AGN individually.

\section{AGN+star}\label{app:agn_star}

In our sample of 34 observed systems, 18 are revealed to be AGN+star systems.
This relatively large fraction is partly due to the difficulty of identifying very faint stellar companions in the spectral-decomposition analysis. In particular, when the stellar component is much fainter than the primary AGN, its contribution to the integrated spectrum can be too weak to produce detectable features, making the stellar contamination difficult to identify from the spectral decomposition alone.
The efficiency of the spectral decomposition also depends on the stellar type. M-type stars are generally more easily identified due to their prominent absorption features, whereas hotter stars, such as G-type stars, can be more difficult to detect because of their narrower absorption lines.
Furthermore, the spectral-decomposition analysis was primarily used to rank candidates according to their probability of stellar contamination, rather than as a strict rejection criterion. The final selection of targets for the ERIS observations was also constrained by the observing strategy of the INAF GTO program. In particular, the fixed number of allocated observing nights and the visibility of the targets, together with the prevailing observing conditions, could prevent the highest-priority targets from being observed at a given time. In such cases, lower-priority candidates were observed to make efficient use of the allocated observing time. Consequently, a fraction of AGN+star contaminants remained in the final IFU sample.

For the AGN+star systems, component~A corresponds to the AGN, showing detectable emission lines, with redshifts consistent with the expected values.
Component~B shows no emission lines, but only a few absorption features, indicating that it is a foreground star consistent with a stellar origin. The restricted wavelength coverage of the ERIS--SPIFFIER data does not allow a robust determination of the stellar spectral type, particularly because several diagnostic absorption features are affected by residual sky-line contamination. 
Table~\ref{tab:classified_app} reports the detected emission lines in the data cube, the position angle and the projected separation between the primary (the AGN) and the secondary (the star) components. The separations are calculated following the procedure described in Sect.~\ref{subsec:image}.
For completeness, we include in the sample J1318--0136, which was already presented and classified as an AGN+star system by \citet{mannucci2023}, as it was observed during the first INAF GTO run with ERIS (run ID 111.24QJ).

\begingroup
\setlength{\tabcolsep}{6pt} 
\renewcommand{\arraystretch}{0.8}
\begin{table}[ht]
\caption[]{Results of ERIS analysis of systems classified as AGN+star.}\label{tab:classified_app}
\centering                          
\begin{tabular}{l c c c c c c c c }        
\hline  
\hline
Target & Emission & PA & Separation \\
& Line & $^\circ$ (N→E)& \SI{}{\arcsecond} \\
\hline
J0008$-$3655 & \ha & $-$104 & 0.48 \\
J0018$-$4815 & \ha & $-$9 & 0.84\\
J0025$+$0750 & \ion{Mg}{II} & $-$52 & 0.28 \\
J0434$+$0846 & \ha & $-$136 & 0.43 \\
J0917$-$0223 & \ha & 151 & 0.44 \\
J1245$-$0908 & \ha & $-$78 & 0.46 \\
J1251$-$2857 & \ha & $-$77 & 0.67 \\
J1318$-$0136 & \ha & $-$82 & 0.30 \\
J1502$+$0257 & \ha & 139 & 0.58 \\
J2013$-$1306 & \hb & $-$102 & 0.61 \\
J2018$-$0351 & \ha & $-$170 & 0.73 \\
J2054$-$2900 & \ha & 126 & 0.71 \\
J2056$-$1656 & \ha & 144 & 0.44 \\ 
J2104$-$0215 & \ha & 80 & 0.93\\
J2118$-$1640 & \hb & $-$29 & 0.54 \\
J2135$-$2025 & \ha & $-$40 & 0.56 \\
J2208$-$3529 & \ha & $-$10 & 0.29 \\
J2211$+$1310 & \ha & 154 & 0.63 \\
\hline
\end{tabular} 
\tablefoot{The position angle (PA) is the angle measured clockwise from North toward East, calculated from the primary component A to the secondary B. The separation corresponds to the projected distance between the primary component A to the secondary B. In this case, the separation is reported only in \SI{}{\arcsecond}.~A refers to the AGN while B refers to the star.}
\end{table}
\endgroup

\section{Single AGN}\label{app:singleAGN}

\paragraph{J1518+0605} By visual inspection of the data cubes and of the broad-band images, we identify one system, namely J1518+0605, which does not show any clear evidence for multiple point-like components. This system is therefore classified as a single AGN and labeled as ``AGN'' in Table~\ref{tab:sample}. 
We note that the target has \texttt{FMP}~=~8, the minimum value adopted for the GMP selection, and a \gaia~G-band magnitude of 20.42, placing the system at the boundaries of the GMP selection criteria.  
J1518+0605 was first observed in March 2025 for a total on-source integration time of 900\,s, split into two individual exposures of 450\,s each. The observations were affected by sub-optimal atmospheric conditions, with poor seeing and strong winds, which limited the overall data quality. Therefore, the target was re-observed in April 2026 for an on-source integration time of 4050\,s, split into nine individual exposures of 450\,s.
From the analysis of the combined frames, only a single point-like component is clearly detected, showing both continuum and \ha~emission. No evidence for additional components is found in either the individual exposures or their combination. We can therefore exclude the presence of a secondary component above the detection limit of the data.
Overall, this level of contamination is in broad agreement with the predictions of \citet{mannucci2023}, who estimated a false positive rate below 2\% for sources with \texttt{FMP}=8 and $G<20.5$. Among the 46 GMP--selected systems with IFU follow-up observations \citep[this work;][]{mannucci2023, ciurlo23, scialpi24, scialpi2026}, only this source is classified as a single AGN, corresponding to a contamination fraction of 2.2\%, consistent with the expectations of \citet{mannucci2023}.

\paragraph{J1319$+$080} J1319$+$0807 (RA = 13:19:52.37, Dec = $+$08:07:21.76) is an AGN at redshift $z = 0.854$ with a \texttt{FMP}~=~8 and a \gaia\ G-band magnitude of 20.70. The target was observed on 14 April 2023 with the \jlow\ grating, for a total exposure time of $8 \times 300$ s.
Despite being included in the observing program, its \gaia~G-band magnitude places the source outside the nominal GMP selection criteria \citep{mannucci2023}, for this reason is not included in Table~\ref{tab:sample}. To further investigate its morphology, we inspected additional ERIS/NIX imaging obtained in the $J$ and $K_{\rm s}$ bands as part of the INAF GTO program (run ID: 112.25GT.001, PI: F.~Mannucci). The near-infrared images reveal a single unresolved nucleus, with no evidence for multiple components or close companions. These observations confirm the classification of J1319+0807 as a single AGN.

\section{J0150+0809}\label{sec:J0150+0809}

\begin{figure*}
\resizebox{\hsize}{!}{
\includegraphics{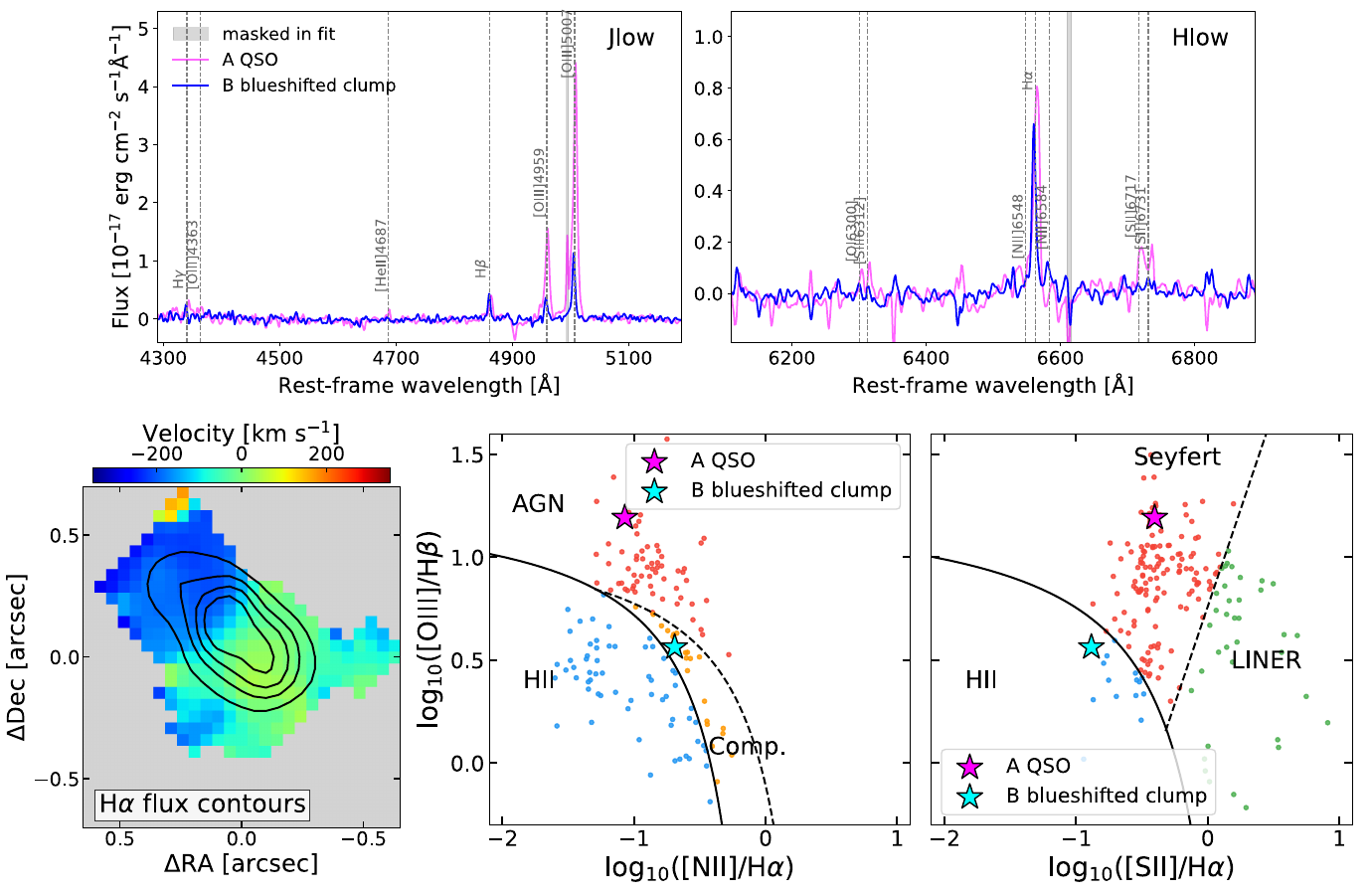}}
\caption{J0150+0809. Top panels: BLR--subtracted integrated spectra of the QSO (A, magenta) and of the \emph{blueshifted clump} (B, blue) extracted from the \jlow\ (left) and \hlow\ (right) cube, respectively. Vertical dashed black lines mark the expected positions of the most prominent emission lines at z = 1.589.
Bottom panels, from left to right: \ha\ Line-of-sight velocity field obtained after BLR subtraction, black contours are from the BLR--subtracted \ha\ flux map. Only spaxels with S/N\,$>$\,5 are shown.
[\ion{S}{ii}]- and [\ion{N}{ii}]-BPT diagrams. Coloured points represent individual spaxels, classified as H\,\textsc{ii}-region-like (blue), Seyfert (red), LINER (green), or Composite (orange) ionization. The black solid curve is the theoretical upper limit on pure star-formation \citep{Kauffmann2003}; the black dashed line in panel c) marks the \citet{Kewley2006} separation between Seyferts and LINERs; the black dashed curve in panel d) is the \citet{Kewley2001} boundary between star-forming galaxies and AGN. Magenta and cyan stars indicate the QSO component (A) and the \emph{blueshifted clump} (B), respectively. }
\label{fig:bpt}
\end{figure*}

Figure~\ref{fig:bpt} shows the ERIS--SPIFFIER data for J0150+0809: the BLR--subtracted integrated spectra of the QSO (A, magenta) and the \emph{blueshifted clump} (B, blue), the \ha\ line-of-sight velocity field obtained after BLR subtraction, and the [\ion{S}{ii}]- and [\ion{N}{ii}]-BPT diagrams.

\end{appendix}

\end{document}